\documentclass[english,reprint,aps]{revtex4-1}
\usepackage[LGR,T1]{fontenc}
\usepackage[latin9,utf8]{inputenc}
\usepackage{color}
\usepackage{babel}
\usepackage{amsmath}
\usepackage{graphicx}
\usepackage{esint}
\usepackage{subscript}
\usepackage{breakurl}
\usepackage{amsmath,amssymb}
\DeclareMathOperator{\sech}{sech}
\DeclareMathOperator{\tr}{tr}
\usepackage{dsfont}
\usepackage{algorithm,algpseudocode}

\floatname{algorithm}{}
\makeatletter

\DeclareFontEncoding{LGR}{}{}
\DeclareTextSymbol{\~}{LGR}{126}

\usepackage{braket}
\makeatother

\begin{document}
\global\long\def\tdel{q}
\newcommand{\dif}{\mathrm{d}}

\title{Lyapunov spectra of chaotic recurrent neural networks}

\author{Rainer Engelken}

\affiliation{Department of Neuroscience, Zuckerman Institute, Columbia University, New York, NY, United States of America}

\author{Fred Wolf}
\affiliation{Max Planck Institute for Dynamics and Self-Organization, G\"ottingen, Germany}
\affiliation{Bernstein Center for Computational Neuroscience, G\"ottingen, Germany}
\affiliation{Bernstein Focus for Neurotechnology, G\"ottingen, Germany}
\affiliation{Faculty of Physics, University of G\"ottingen, G\"ottingen, Germany}

\author{L. F. Abbott}
\affiliation{Department of Neuroscience, Zuckerman Institute, Columbia University, New York, NY, United States of America}
\affiliation{Department of Physiology and Cellular Biophysics, Columbia University, New York, NY, United States of America}

\begin{abstract}
	Brains process information through the collective dynamics of large neural networks. Collective chaos was suggested to underlie the complex ongoing dynamics observed in cerebral cortical circuits and determine the impact and processing of incoming information streams. While dynamic mean-field theory has uncovered key properties of recurrent network models such as the onset of chaos and their largest Lyapunov exponent, fundamental features of their dynamics remain unknown. In particular, chaotic dynamics in dissipative high-dimensional systems takes place on a subset of phase space of reduced dimension and is organized by a complex tangle of stable, neutral and unstable manifolds. Key topological invariants of this phase space structure such as attractor dimension, and Kolmogorov-Sinai entropy so far remained elusive.
	
	Here we calculate the complete Lyapunov spectrum of recurrent neural networks. We show that chaos in these networks is extensive with a size-invariant Lyapunov spectrum and characterized by attractor dimensions much smaller than the number of phase space dimensions. The attractor dimension and entropy rate increases with coupling strength near the onset of chaos but decrease far from onset, reflecting a reduction in the number of unstable directions. We find that near the onset of chaos, for very intense chaos, and discrete-time dynamics, random matrix theory provides good analytical approximations to the full Lyapunov spectrum. We show that a generalized time-reversal symmetry of the networks dynamics induces a point-symmetry of the Lyapunov spectrum reminiscent of the symplectic structure of chaotic Hamiltonian systems. Temporally fluctuating input can drastically reduce both the entropy rate and the attractor dimension. For trained recurrent networks, we find that Lyapunov spectrum analysis provides a quantification of error propagation and stability achieved by distinct learning algorithms. Our methods apply to systems of arbitrary connectivity, and we describe a comprehensive set of controls for the accuracy and convergence of Lyapunov exponents.
	
	Our results open a novel avenue for characterizing the complex dynamics of recurrent neural networks and the geometry of the corresponding high-dimensional chaotic attractor. They also highlight the potential of Lyapunov spectrum analysis as a diagnostic for machine learning applications of recurrent networks.
\end{abstract}
\maketitle

\section{Author Summary for Kids}
When you dream, think, or read this sentence, in your brain gazillions of tiny cells called neurons are talking to each other. These neurons pass on the messages coming from your five senses by sending patterns of electric pulses to each other or to your big toe if you need to run. Neuroscientists try to eavesdrop on this complex chatter and want to understand the language the neurons speak. I use math to build a super simplified imitation of this chatter. I have randomly connected thousands of neurons inside a big computer and so they form a network that looks like a giant cobweb woven by a drunken spider. I call this imitation of real neurons my 'model'.

In my model, each neuron has a very simple rule. The rule tells the neuron how messages coming from other neurons are translated into some kind of activity. As part of its activity, it sends messages on to thousands of other neurons that it is connected to. Other scientists, using pen and paper instead of computers, found out that such networks are quiet as a mouse when the connections are weak, but they start a tumult of chatter when the connections between the neurons are strong enough.

I want to better understand the space of activities and how complicated it is. It is much easier to imagine this space of possible activity patterns if you have only three neurons. You can give each neuron a number which says how active the neuron is at a particular moment in time. With three neurons, you can imagine the activity space as your bedroom. The activity of the first neuron is the direction from head to toe when you are lying in bed. The activity of the second neuron gives the position from left to right, and the activity of the third neuron gives the height above your bed. A combination of neuron activity states gives a point in space in the bedroom. Over time, the point moves around in space. Now we can ask: What patterns will form from the points of network activity if we wait long enough? Will they fill the whole room, or will they create patterns that are lying only in a small subspace? For example, any activity could lie on a thin, crumpled layer on the floor like a blanket. Or the activity points could loop around a curved line, like a hula-hoop leaning against your desk. Now think of the same questions, but with thousands of neurons whose activities span thousands of directions. It is hard to imagine this space.

I discovered that although the activity of the whole network looks like a complete mess, there is a hidden pattern in the space of activities. I reveal this using a theory called chaos theory. Chaos theory can be used to understand complex groups of many small things that interact, such as the swirling of gazillions drops of water in the clouds on a rainy day. We call such a system chaotic when a tiny poke is enough to make it do something very different from what it would have done without the poke. Actually, chaos theory started already more than a hundred years ago. Back then, people wanted to understand whether our solar system is stable. By stable, they mean whether the paths of the planets are the same after you poke them a bit. If the solar system was chaotic, Earth or Mars might take a wrong turn and end up sizzling into the sun or being catapulted out of the Milky Way one day. A Russian mathematician named Lyapunov was worried about that and thought very hard about it. There are some numbers, known as Lyapunov exponents --- after that Russian mathematician --- that measure how fast things fly apart in a chaotic system after tiny poking or tickling.

There is an almost magical link between these Lyapunov exponents and the space of all imaginable gibberish the neurons could possibly ever talk about. I use this link to show that although the activity of a chaotic network looks like a random jumble of gibberish, there is a lot more secret order than what you would expect when only listening to the neurons one by one or two at a time. 
In the space of all imaginable network activity patterns, there are lots of holes, like in Swiss cheese. Actually, in the model I studied, this space is almost empty, it is made up mostly of thin air. Therefore specific network activity patterns can never occur. So is it that the neurons have secretly agreed never to talk about certain topics? No! It's instead that the wiring of the random network (the giant cobweb of the drunken spider) and the neuron rules somehow don't allow them to chat about certain things.

In the following pages, I propose how to find out more about this secret order using other tricks from chaos theory. If you want to learn more, just write me an email$\;$:-)

\section{Introduction}

\begin{figure}[!ht]
	\includegraphics[width=1\columnwidth]{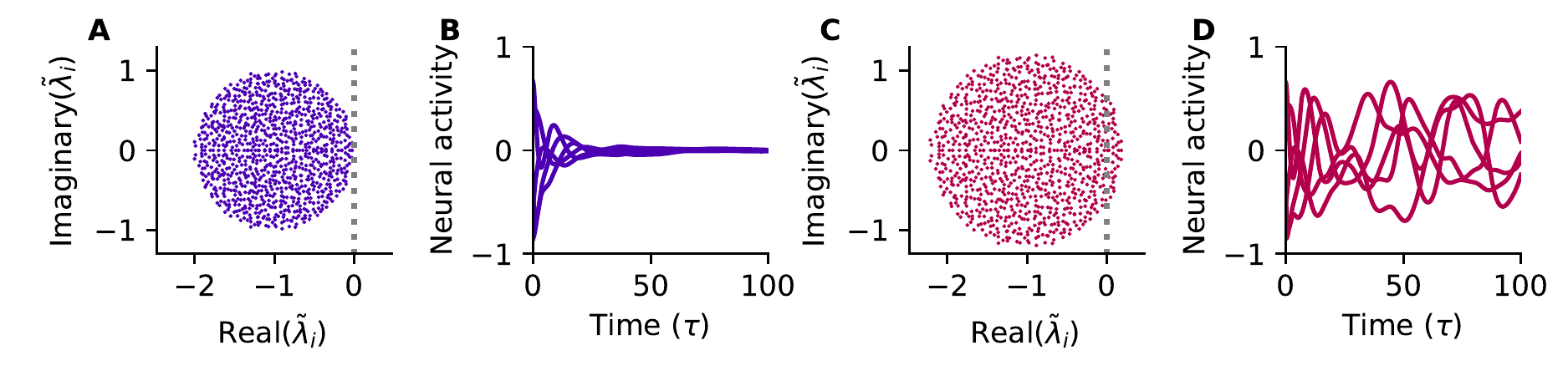}

	\caption{{\bf Transition to chaos for sufficiently strong coupling $g$ in rate networks.}
		{\bf A}~Linear stability of rate dynamics near the zero fixed point. Real vs imaginary part of eigenvalues $\hat\lambda_i$ of the stability matrix for $g=0.99$. {\bf B}~For subcritical couplings ($g=0.99$) the trivial fixed point of the system $h_i=0$ is the only stable solution. {\bf C}~In large networks the trivial fixed point loses stability at $g_{\textnormal{crit}}=1$ and chaos emerges from the nonlinear interaction of rate units where the spectral radius crosses unity (gray dotted line). {\bf D}~Rate chaos for $g=1.2$ (other parameters: network size $N=1000$, integration step $\Delta t=10^{-3}\tau$).}
	\label{fig1}
\end{figure}

A major challenge in theoretical neuroscience, statistics, and statistical physics is to develop mathematical concepts to characterize high-dimensional activity and find collective degrees of freedom and information representations of strongly interacting populations of elements, such as neurons.
Theoretical work suggested that asynchronous rate activity in neural systems may originate from chaotic dynamics in recurrent networks. A seminal study showed that large networks of randomly connected firing-rate units display a sharp transition from an inactive state to a chaotic state \cite{sompolinsky_chaos_1988} (Fig.~\ref{fig1}). In this class of models, each rate unit maps its synaptic input $h_{i}$ smoothly into a firing rate through a hyperbolic tangent input-output transfer function $\phi$. Coupling strengths are drawn independently from a Gaussian distribution with zero mean and standard deviation $g/\sqrt{N}$, where $N$ is the size of the network. A dynamic mean-field theory has been developed and is applicable in the large network limit $N\rightarrow \infty$. In this approach, the recurrent input into a typical unit is modeled by a Gaussian process whose statistics is determined self-consistently. For small coupling $g<1$, the trivial fixed point $h_i=0$ for all $i$ is the only stable solution to the mean-field theory (Fig.~\ref{fig1}A,B). For increasing coupling strength, this trivial fixed point loses stability and chaos emerges from the nonlinear interaction of unstable activity modes (Fig.~\ref{fig1}C,D). Sompolinsky, Crisanti, and Sommers showed in the large network limit $N\rightarrow \infty$ that above a critical strength $g_{\textnormal{crit}}=1$, the only stable self-consistent solution is chaotic dynamics \cite{sompolinsky_chaos_1988}. The transition to chaos occurs when the spectral radius $\hat\lambda_{\max}$ of the stability matrix obtained from linearizing the rate dynamics around the fixed point $h_i=0$ crosses unity (Fig.~\ref{fig1}A,C).

This classical work has been extended and the transition has been studied for networks with different subpopulations \cite{kadmon_transition_2015,aljadeff_transition_2015,harish_asynchronous_2015}, various input-output transfer functions \cite{kadmon_transition_2015,harish_asynchronous_2015}, bistable units \cite{stern_dynamics_2014}, adaptation \cite{muscinelli_how_2019}, sparse balanced network architectures \cite{kadmon_transition_2015,harish_asynchronous_2015,harish_network_2013} and external stimuli \cite{molgedey_suppressing_1992,rajan_inferring_2010,rajan_stimulus-dependent_2010,schuecker_optimal_2018}. For networks of spiking model neurons, quantitative agreement with a corresponding chaotic rate network in the limit of slow synaptic dynamics was found \cite{harish_network_2013,harish_asynchronous_2015} (see also \cite{shriki_rate_2003}). 

The chaotic, heterogeneous state of rate networks possesses high computational capabilities. These arise from its rich internal dynamics that can provide a substrate for complex nonlinear computations, e.g., implementing input/output maps \cite{maass_real-time_2002,jaeger_harnessing_2004,sussillo_generating_2009} and learning temporal sequences \cite{laje_robust_2013}, 
 however, it is a challenge to extend this to spiking neural networks \cite{abbott_building_2016,depasquale_using_2016,thalmeier_learning_2016,nicola_supervised_2017,ingrosso_training_2019}. 
Some studies proposed that computational features are favorable closely beyond the so-called \textit{edge of chaos} in the chaotic regime \cite{bertschinger_real-time_2004,schweighofer_chaos_2004,legenstein_edge_2007,busing_connectivity_2009,sussillo_generating_2009,toyoizumi_beyond_2011,dahmen_second_2019,schuecker_optimal_2018}. 
It was claimed and questioned much earlier in dynamical systems that the edge of chaos is computationally advantageous
\cite{kelso_dynamic_1988,langton_computation_1990,mitchell_revisiting_1993-3}.

Recent developments in machine learning, including the renaissance of deep networks, has sparked additional interest in principles of stability and information processing in recurrent rate networks \cite{saxe_exact_2013,poole_exponential_2016}. One reason for this is that recurrent networks can be unrolled in time into infinitely deep feed-forward networks with tied weights \cite{pascanu_difficulty_2012}. To avoid vanishing or exploding gradients during learning, this analogy suggests that learning in deep nonlinear networks is facilitated if the weights are initialized such that the corresponding recurrent networks are close to the edge of chaos ($g_{\textnormal{crit}} = 1$)\cite{saxe_exact_2013,hanin_products_2018,chen_dynamical_2018,gilboa_dynamical_2019,can_gating_2020}. 
Intriguingly, transient rate chaos yields exponential expressivity in deep networks \cite{poole_exponential_2016,schoenholz_deep_2016,pennington_emergence_2018}.

Here we calculate for the first time to our knowledge the full set of Lyapunov exponents of classical firing-rate networks. Previous studies only considered the largest Lyapunov exponent, which measures the average exponential rate of divergence or convergence of nearby network states. The full Lyapunov spectrum provides growth rates of volume elements along the trajectory and gives valuable additional insights into the collective dynamics of firing-rate networks. 

We use concepts from the ergodic theory of dynamical systems to further characterize the complex collective dynamics of rate networks. Often large-scale dissipative systems evolve towards a low-dimensional attractor, but it is a challenge to 
identify and characterize this lower dimensional manifold. Ergodic theory provides an estimate of the attractor dimensionality, by characterizing the diversity of collective network activity states \cite{vulpiani_chaos:_2009}. It also provides access to the dynamical entropy rate which measures the amplification of dynamical uncertainty due to sensitivity to initial conditions. The dynamical entropy rate constrains the capability of information processing. Given that the initial state is known only with finite precision, the sensitive dependence on initial conditions makes predictions of future states impossible in chaotic systems \cite{shaw_strange_1981,young_mathematical_2013}. This corresponds to a dynamical entropy rate because nearby states, which cannot be distinguished by a finite precision readout initially, are pulled apart by the chaotic dynamics and become distinguishable later on. Therefore, the dynamical entropy rate quantifies the speed at which microscopic perturbations affect macroscopic rate fluctuations \cite{shaw_strange_1981}. Sensitivity to initial conditions in cortical circuits might serve as a dynamical mechanism to pull nearby trajectories apart \cite{ashwin_nonlinear_2005,rabinovich_dynamical_2001,rabinovich_transient_2008}. If the microscopic initial state contains a relevant signal, the dynamical entropy rate measures the rate by which this information becomes accessible. From a neural coding perspective, the dynamical entropy rate can contribute to the so-called noise entropy \cite{lajoie_chaos_2013}, because the dynamic amplification of microscopic noise by chaotic dynamics can impair coding capacity.

Both the dynamical entropy rate and attractor dimensionality are invariants of dynamical systems, i.e., they do not change under diffeomorphisms of the phase space \cite{kolmogorov_new_1958,sinai_notion_1959,billingsley_ergodic_1965,kuznetsov_invariance_2016} and can be obtained from the set of Lyapunov exponents \cite{eckmann_ergodic_1985}. 
 This is the only known general way of accessing the entropy of a high-dimensional differentiable dynamical system \cite{vulpiani_chaos:_2009}. Sampling-based estimates of entropy rate and dimensionality, e.g. the Grassberger-Procaccia algorithm \cite{grassberger_characterization_1983,grassberger_estimation_1983,grassberger_generalized_1983} that estimates the correlation dimension $D_{2}$, are intractable for systems with many degrees of freedom. The data required for such sampling-based estimates of the attractor dimensionality scales exponentially in $D$ \cite{eckmann_fundamental_1992,smith_intrinsic_1988,kantz_nonlinear_2004,pikovsky_lyapunov_2016}.

Our approach is applicable for arbitrary network structures and transfer functions $\phi$. We show that both the dynamical entropy rate and the attractor dimensionality saturate with coupling strength $g$. Thus, both uncertainty amplification due to sensitivity to initial conditions and the diversity of network activity states saturates for strong coupling. We find that time-discretization increases both the entropy rate and dimensionality.
Using random matrix theory, we analytically approximate the full Lyapunov spectrum in several limiting cases. We extend the analysis to a balanced network of threshold-linear units, where entropy rate and dimensionality peak as a function of coupling strength. We find that time-varying input reduces both the entropy rate and dimensionality. Finally, we use the Lyapunov spectrum to quantify the stability of trained networks.

\section{Model}

We study the dynamics of a randomly wired network of nonlinear firing-rate units.
The dynamics of the state, $h_i$ for $i=1,2,\ldots, N$, of each firing-unit follows \cite{sompolinsky_chaos_1988, jaeger_harnessing_2004}
\begin{equation}
\tau\frac{\dif h_i}{\dif t}= F_i=-h_i +\sum_{j=1}^{N} J_{ij}\phi(h_j).
\end{equation}
Here $h_i$ is the total synaptic current received by firing-rate unit $i$ and $\tau$ is the rate-unit time constant. We draw independent identically distributed entries of the coupling matrix $J_{ij}$ from a Gaussian distribution $J_{ij} \sim \mathcal{N} (0,g^2/N)$, remove self-coupling by setting $J_{ii}=0$ 
and choose the transfer function $\phi(x)=\tanh(x)$ \cite{sompolinsky_chaos_1988}.

\section{Lyapunov spectrum of classic recurrent neural networks}
To calculate the Lyapunov spectrum, we evaluate the Jacobian of the flow of the dynamics. 
This measures how infinitesimal perturbations of the network state evolve in the tangent space along the trajectory $h_i$. 
The instantaneous Jacobian is given by 
\begin{equation}
D_{ij}(t_s) = \frac{\partial F_i}{\partial h_j}\Bigr|_{t=t_s} =-\delta_{ij}+J_{ij}\phi'(h_j(t_s)).
\label{eq:-Jacobian}\end{equation}
Thus, in our case, the Jacobian is a negative identity matrix plus the coupling matrix with columns scaled by the squared hyperbolic secant $\phi'=\sech^2$ of the network activity states $h_i$. For strong $g$, the variance of $h_i$ increases proportional to $g$ \cite{crisanti_path_2018-1} and most rates are in the saturated regime, so $\sech^2(h_i)\approx 0 $ for most $i$ and hence most columns of $D_{ij}(t_s)$ are close to zero, aside from the diagonal entries.
The full Lyapunov spectrum $\lambda_1 \geq \lambda_2 \dots \geq \lambda_N$ is obtained by a reorthonormalization procedure \cite{benettin_lyapunov_1980}, which is described in detail in Appendix \ref{Algo}, including a detailed analysis of the convergence of the Lyapunov spectra. 
 Briefly, calculating the Lyapunov spectrum involves two steps:

 First, we evolve an initially orthonormal system $\mathbf{Q}$ in the tangent space along the trajectory using the Jacobian $\mathbf{D}$. To this end, the variational equation $\tau\dot{\mathbf{Q}}=\mathbf{D}(t)\mathbf{Q}$ has to be integrated. A continuous system can be transformed to a discrete system by considering a stroboscopic representation, where the trajectory is only considered at certain discrete time points. We use here the notation of discrete dynamical systems, where this corresponds to performing the product of Jacobians along the trajectory $\widetilde{\mathbf{Q}}_{s+1}=\mathbf{D}_{s}\mathbf{Q}_{s}$. We study the discrete network dynamics in the limit of small time step $\Delta t$. The notation can be extended directly to continuous systems \cite{geist_comparison_1990}. 

Second, we extract the exponential growth rates using the QR-decomposition, $\widetilde{\mathbf{Q}}_{s+1}=\mathbf{Q}_{s+1}\mathbf{R}^{s+1}$, which uniquely decomposes $\widetilde{\mathbf{Q}}_{s+1}$
into an orthonormal matrix $\mathbf{Q}_{s+1}$
and an upper triangular matrix $\mathbf{R}^{s+1}$ with positive diagonal elements. Geometrically, $\mathbf{Q}_{s+1}$ describes the rotation of $\mathbf{Q}_{s}$ caused by $\mathbf{D}_{s}$ and the diagonal entries of $\mathbf{R}^{s+1}$
describe the stretching and shrinking of $\mathbf{Q}_{s}$, while
the off-diagonal elements describe the shearing.

The Lyapunov exponents are given by time-averaged logarithms of the diagonal elements of $\mathbf{R}^{s}$: $\lambda_{i}=\lim_{t\to\infty}\frac{1}{t}\sum_{s=1}^{t}\log\mathbf{R}_{ii}^{s} $.
Note that the QR-decomposition does not need to be performed at every simulation step, just sufficiently often, i.e., once every $t_\textnormal{ONS}$ steps
 such that $\widetilde{\mathbf{Q}}_{s+t_\textnormal{ONS}}=\mathbf{D}_{s+t_\textnormal{ONS}-1}\cdot\mathbf{D}_{s+t_\textnormal{ONS}-2}\dots\mathbf{D}_{s}\cdot\mathbf{Q}_{s}$ remains well-conditioned \cite{benettin_lyapunov_1980}. An initial transient should be disregarded in the calculation of the Lyapunov spectrum because $\mathbf h$ first has to converge towards the attractor and $\mathbf Q$ has to converge to the unique eigenvectors of the Oseledets matrix (Eq.~\ref{eq:-Oseledets}) \cite{ershov_concept_1998}. A simple example of this algorithm in pseudocode is:
\begin{algorithm}[H] \label{pseudocode}
	\caption{\bf Jacobian-based algorithm for Lyapunov spectrum}
	\begin{algorithmic}
		\State initialize $\mathbf h$, $\mathbf Q$
		\State evolve $\mathbf h$ until it is on attractor (avoid initial transient)
		\State evolve $\mathbf Q$ until it converges to the eigenvectors of the backward Oseledets matrix
		\For{$t = s \to s_\textnormal{sim}/\Delta t$} 
		\State $\mathbf h \gets \mathbf f(\mathbf h)$ 
		\State $\mathbf D \gets\frac{ \dif\mathbf f}{ \dif\mathbf h}$ 
		\State $\mathbf{ Q} \gets \mathbf D\cdot \mathbf Q $
		\If{ $ t \text{ \% } t_\textnormal{ONS}=0 $}
		\State $\mathbf Q, \mathbf R\gets \mathrm{qr}(\mathbf Q)$
		\State $\gamma_i \mathrel{+}=\log(R_{ii})$
		\EndIf
		\EndFor
		\State $\lambda_i=\gamma_i/t_\textnormal{sim}$
	\end{algorithmic}
\end{algorithm}

\subsubsection{Extensive spatiotemporal network chaos} 
In dissipative systems, arbitrary initial conditions converge towards a lower dimensional attractor. The dimensionality characterizing the diversity of collective dynamical states in this attractor, however, can be constant, grow in proportion to the size of the system, or have other more complex dependencies. If the dimensionality is proportional to the system's size, the system is called extensive, which occurs when the shape of the Lyapunov spectrum is invariant with system size. Such an invariance also implies an extensive entropy rate. 

\begin{figure}[!h]
	\includegraphics[width=1\columnwidth]{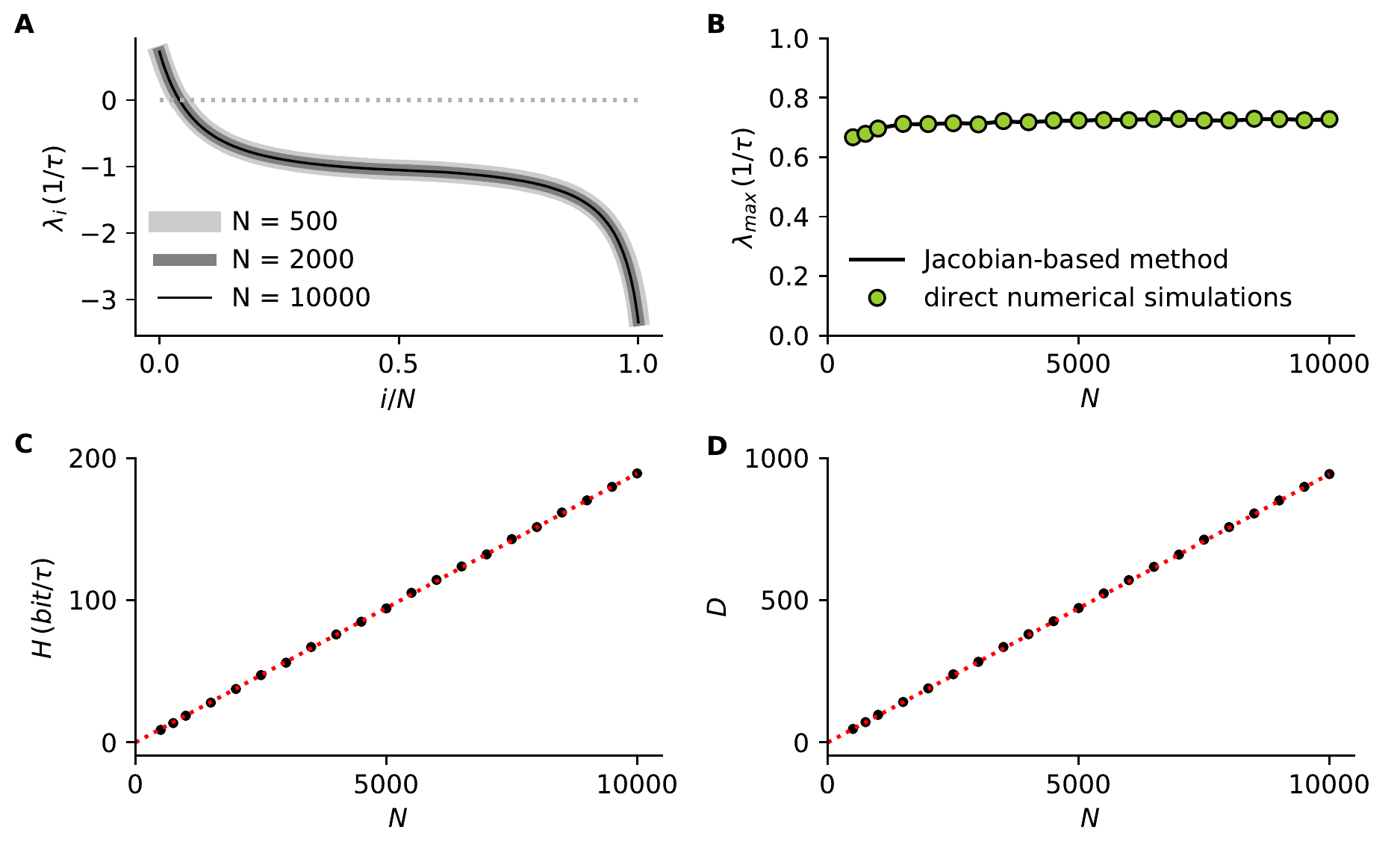}
	\caption{{\bf Extensive chaos revealed by the size-invariance of the Lyapunov spectrum }
		{\bf A}~Full Lyapunov spectra for different network sizes $N$ are on top of each other, indicating an identical shape. The Lyapunov spectrum is point-symmetric around the mean Lyapunov exponent $\bar\lambda =-1/\tau $ (See analytical derivation in Appendices \ref{LSsymmetry} and \ref{RMTmeanLE}). {\bf B}~The largest Lyapunov exponent quickly saturates with network size. {\bf C}~The Kolmogorov-Sinai entropy rate $H$ grows linearly with $N$ as shown over two orders of magnitude. {\bf D}~The same holds for the attractor dimensionality $D$ (other parameters: $g=10$, $\Delta t=0.1\tau$, $t_\textnormal{ONS}=\tau$, $t_\textnormal{sim}=10^3\tau$).}
	\label{fig3}
\end{figure}

In the case of the firing-rate networks studied here, we find extensive chaos, indicated by the invariance of the shape of the Lyapunov spectrum to network size $N$ (Fig.~\ref{fig3}A) for sufficiently large networks (although the structure of the attractor depends on the realization of the connectivity $J_{ij}$). The Lyapunov spectrum is point-symmetric around its constant mean value $-1/\tau$ (See Fig.~\ref{fig6}A). We will investigate the origin of the symmetry of the Lyapunov spectrum in section. The largest Lyapunov exponent quickly saturates as a function of network size (Fig.~\ref{fig3}B).
We investigate the finite-size effect on the largest Lyapunov exponent, its convergence to the value predicted by dynamic mean-field theory and the finite-size effect in the transition to chaos $g_{\textnormal{crit}}$ in Appendix \ref{finiteSize}.
The entropy rate $H$, also called the Kolmogorov-Sinai entropy rate, quantifies the amplification of small state differences by the chaotic dynamics. While formally defined via partitions of the phase space, it is under weak mathematical constraints given by the sum of the positive Lyapunov exponents: $H = \sum\limits _{\lambda_{i}>0}\lambda_{i}$ (See Appendix \ref{KSE_KYD}). As a consequence of the size-invariant Lyapunov spectrum, it also grows linearly, as demonstrated over two orders of magnitude in Fig.~\ref{fig3}C.
The same is true for the attractor dimensionality (Fig.~\ref{fig3}D), which is given by the interpolated number of Lyapunov exponents that sum to zero:
 \[D=k+\dfrac{\sum_{i=1}^{k}\lambda_{i}}{\left|\lambda_{k+1}\right| }\quad\text{with}\quad k=\max\limits _{n}\left\{ \sum\limits _{i=1}^{n}\lambda_{i}\ge0\right\}.\]
 Intuitively, the attractor dimension is the dimensionality of the highest dimensional infinitesimal hypersphere, whose volume does not shrink nor grow through the chaotic dynamics. In other words, on the attractor, growth along unstable manifolds is being compensated by shrinking along the stable manifolds. Thus, a $D$-dimensional hypersphere is merely deformed over time, with the volume preserved on average. 
 
The extensivity of the Lyapunov spectrum for rate networks was conjectured earlier \cite{sompolinsky_chaos_1988}, but never before demonstrated. Extensive chaos is often found in extended systems that are decomposable into weakly interacting subsystems, whose number grows linearly with system size \cite{ruelle_large_1982}. As this is not fulfilled for this fully randomly connected rate network, extensive chaos in our networks is not a trivial property. Globally coupled networks, for instance, can exhibit nonextensive chaos \cite{takeuchi_lyapunov_2009}.

\subsubsection{Strong coupling intensifies chaos} 
\begin{figure}[!ht]
	\includegraphics[width=1\columnwidth]{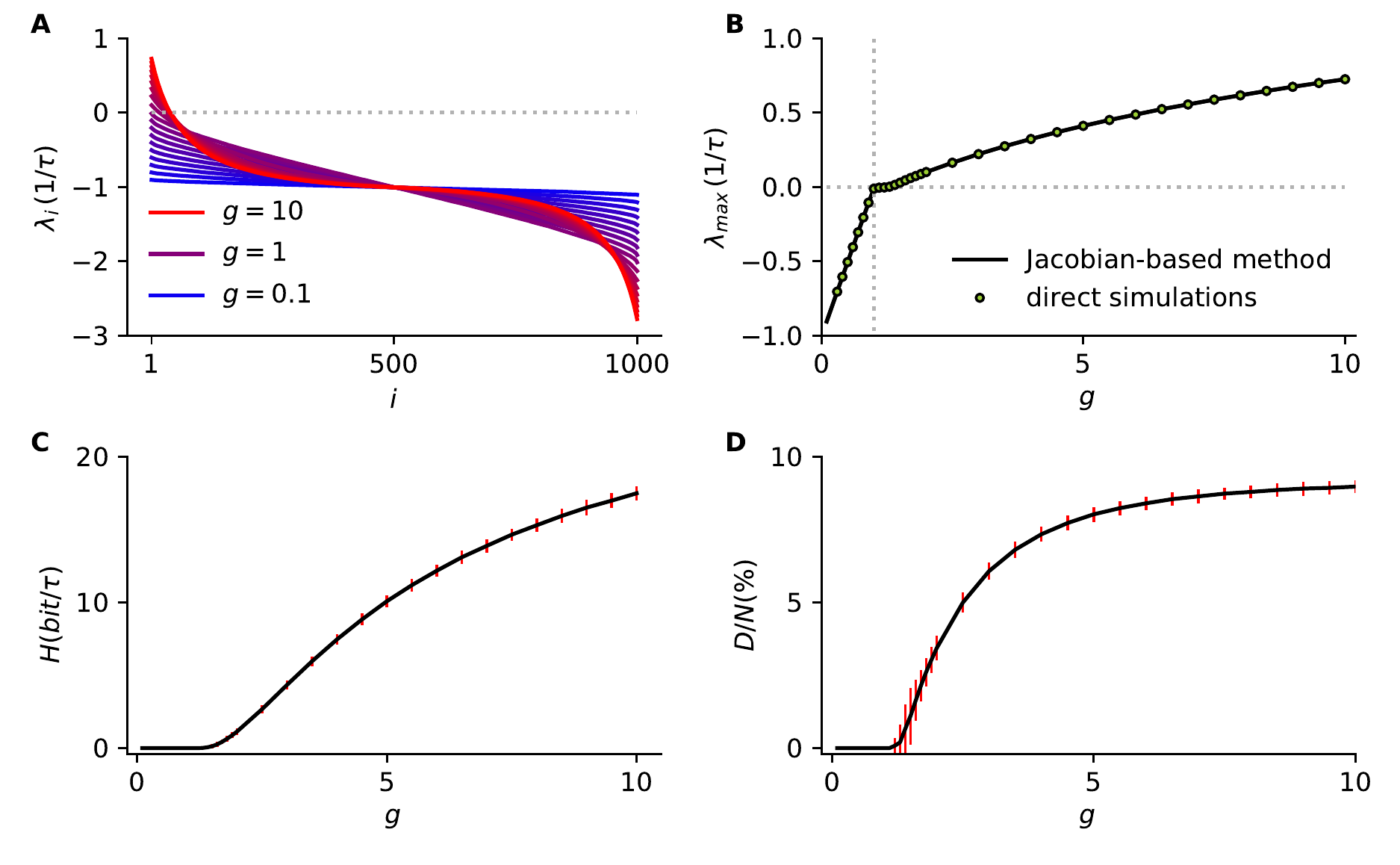}
		\caption{ {\bf Entropy rate and attractor dimensionality of firing-rate network dynamics.} {\bf A}~Full Lyapunov spectra of rate networks with different coupling strengths $g$, color-coded from blue (small $g$) to red (large $g$). 
		{\bf B}~The largest Lyapunov exponent shows the theoretically predicted linear growth for $g<1$ and first quadratic and then logarithmic growth for $g\gg 1$ as a function of $g$ \cite{sompolinsky_chaos_1988}. (Green dots: direct numerical simulations, black line: Jacobian-based method)
		{\bf C}~The dynamical entropy rate $H$ grows with $g$ but is bounded (See Appendix \ref{largeGLimit}). {\bf D}~Relative attractor dimensionality $D/N$ peaks at $D/N < 10 \%$.
		(Averages over 20 network realizations in black, red error bars indicate double std across 20 network realizations, parameters: $N=1000$, $\Delta t=10^{-2}\tau$, $t_\textnormal{sim}=10^4\tau$, $t_\textnormal{ONS}=\tau$).
		\label{fig2}}
\end{figure}

Next, we investigate the role of the synaptic coupling strength $g$ (Fig.~\ref{fig2}).  
The full Lyapunov spectrum shows an interesting dependence on $g$ (Fig.~\ref{fig2}A). For increasing $g$, the first half of the Lyapunov spectrum is increasingly curved (Fig.~\ref{fig2}A). Note that the Lyapunov spectrum is point-symmetry for all values of $g$. The largest Lyapunov exponent shows the theoretically predicted $g$-dependence in the stable regime $g<1$ (Fig.~\ref{fig2}B). In the chaotic regime $g>1$, it grows first quadratically and then logarithmically with $g$ in agreement with previous work \cite{sompolinsky_chaos_1988,crisanti_path_2018-1}. 
Note that the asymptotic large $g$ behavior $\lambda_{\max}\propto \log(g)$ is only expected when first sending $N$ and then $g\rightarrow\infty$.
The calculation of the largest Lyapunov exponent is confirmed both by tracking the amplitude of a small perturbation in direct numerical simulations and by using the Jacobian-based method \cite{benettin_lyapunov_1980} (Fig.~\ref{fig2}B). While the exponential separation rate of nearby trajectories increases for growing $g$, the overall dissipation of the system, measured by the mean Lyapunov exponent $\bar\lambda$ is independent of $g$ and only depends on the time constant $\tau$. The reasons for this are provided in Appendix \ref{RMTmeanLE}. We now focus first on the entropy rate and attractor dimensionality.

The dynamical entropy rate is zero for $g \leq 1 $ and grows monotonically for increasing values of $g$ (Fig.~\ref{fig2}C). Our numerical results suggest that for large $g$, 
the dynamical entropy rate peaks with $g$ (See Fig.~\ref{fig16LAST} and Appendix \ref{largeGLimit}). Again, this asymptotic behavior is only expected when first sending $N$ and then $g\rightarrow\infty$. In case the specific initial state of the network does not encode relevant information, the growth of the entropy rate with $g$ can be interpreted as an increasing contribution to noise entropy.

\subsubsection{Attractor dimensionality bounded for strong coupling} 
We found that the attractor dimension first increases with $g$ (Fig.~\ref{fig2}D) in the chaotic regime $g > g_{\textnormal{crit}}$ and peaks as a function of $g$ at less then 10\% of the number of phase space dimensions $N$ (Fig.~\ref{fig4}A). This suggests that despite vanishing pairwise correlations \cite{sompolinsky_chaos_1988}, rate unit activities are not independent of each other. Even for strongly chaotic networks, the strange attractor of the network dynamics does not fill the entire phase space but only a small but extensive fraction of it. Note that the geometric structure of the attractor nevertheless changes when $g$ is further increased.

\subsubsection{Comparison of attractor dimension and PCA dimension}
We compared the attractor dimensionality with a dimensionality estimate based on second-order statistics of the activity $h_i$ and $\phi (h_i)$ 
given by the effective number of principal components that account for most of the variance
(See Appendix \ref{PCAdim} for details).
Such dimensionality estimates of the network activity based on Principal Component Analysis (PCA) are commonly used in experimental and theoretical neuroscience \cite{rajan_inferring_2010,rajan_stimulus-dependent_2010,gao_simplicity_2015,cunningham_dimensionality_2014,litwin-kumar_optimal_2017,farrell_dynamic_2019,recanatesi_dimensionality_2019}, e.g., to quantify the spatiotemporal complexity of neural activity in a data set. 

PCA-based estimates of dimension are generally not invariant with respect to changes of coordinates and can be misleading if applied to limited data sets. Extensivity of the PCA-based dimension does not in general imply extensivity of the attractor dimension, nor vice versa. In addition, PCA analyses, because they are based on a 
pairwise correlation function, can miss low-dimensional structure hidden in higher-order correlations.
In general, the PCA-based dimension can both under- and overestimate the attractor dimension.

\begin{figure}[!h]
	\includegraphics[width=1\columnwidth]{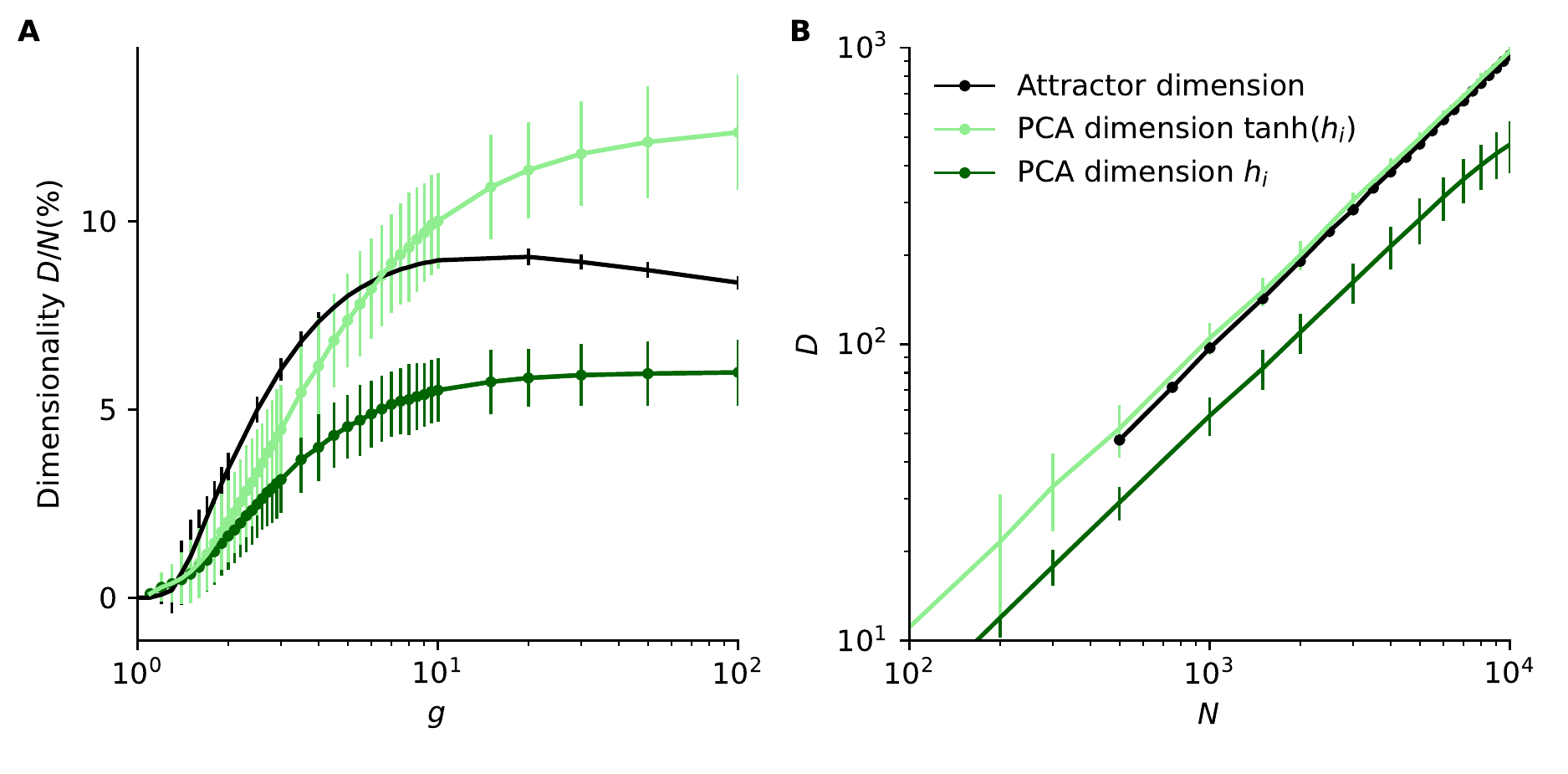}
	\caption{{\bf PCA and attractor dimensions of networks with $\tanh$ transfer function}
		{\bf A}~Principal Component Analysis (PCA)-based dimensionality estimate (green) and attractor dimension based on the Lyapunov spectrum (black) for different values of synaptic strength $g$. Both PCA dimensions saturate for $g \gg 1$ but they saturate at different levels and with distinct exponential rates 
		(error bars are double std across 20 network realizations).
		PCA dimension estimate of dynamics depends on whether $\tanh (h_i)$ or $h_i$ is considered. 
		 {\bf B}~Both PCA-based dimensionality estimates seem to be extensive, as indicated by the approximately linear growth with $N$ (other parameters: $N=1000$, $g=10$, $\Delta t=0.1\tau$, $t_\textnormal{ONS}=\tau$, $t_\textnormal{sim}=10^4\tau$).}
	\label{fig4}
\end{figure}

We found that a PCA-based dimension strongly differs depending on whether it is estimated based on the statistics of the firing rates $\phi (h{_i})$ or on $h{_i}$ (Fig.~\ref{fig4}). 
Generally, we find for all dimensionality estimates growth of dimension with $g$ in weakly chaotic networks. However, for large  $g\gg 1$, we find a peak and subsequent slight decay of the attractor dimension (Fig.~\ref{fig4}) and \ref{fig16LAST}). 
In contrast, both PCA dimensions saturate for $g \gg 1$ but they saturate at different levels and with distinct rates. 
The PCA-based dimensionality (both based on $h_i$ and $\phi(h_i)$) grows extensively with network size $N$, as does the attractor dimensionality (Fig.~\ref{fig4}B).

\subsubsection{Lyapunov spectrum of discrete-time firing-rate network} 
We next assess the effect of introducing finite temporal discretization. The dynamics of discrete-time rate networks has attracted much attention because it is mathematically simpler \cite{amari_learning_1972,parisi_asymmetric_1986,molgedey_suppressing_1992,doyon_control_1993,massar_mean-field_2013,farrell_dynamic_2019}. Here we aim to understand the impact of time-discretization on chaotic dynamics. We set $ \tau=1$ and study the evolution of the map
\begin{eqnarray*}
	h_i(t+\Delta t)&=&f_i=(1-\Delta t) h_i(t) +\Delta t \sum_{j=1}^{N} J_{ij}\phi(h_j(t)).
\end{eqnarray*}

In the limit $\Delta t\rightarrow 0$, the continuous-time dynamics \cite{sompolinsky_chaos_1988,kadmon_transition_2015,schuecker_optimal_2018}
is recovered. For $\Delta t=1$, the discrete-time network \cite{molgedey_suppressing_1992,molgedey_suppressing_1992,doyon_control_1993,massar_mean-field_2013} is obtained.

The Jacobian for the discrete-time map is 
\begin{equation}
D_{ij}(t_s)=\frac{\partial f_i}{\partial h_j}\Bigr|_{t=t_s} =(1-\Delta t)\delta_{ij} + \Delta t\cdot J_{ij}\phi'(h_j(t_s)).
\end{equation}
The full Lyapunov spectrum is again obtained by a reorthonormalization procedure of the Jacobians along a numerical solution of the map \cite{benettin_lyapunov_1980}. For details, see Appendix \ref{Algo}. 

\begin{figure}[!h]
	\includegraphics[width=1\columnwidth]{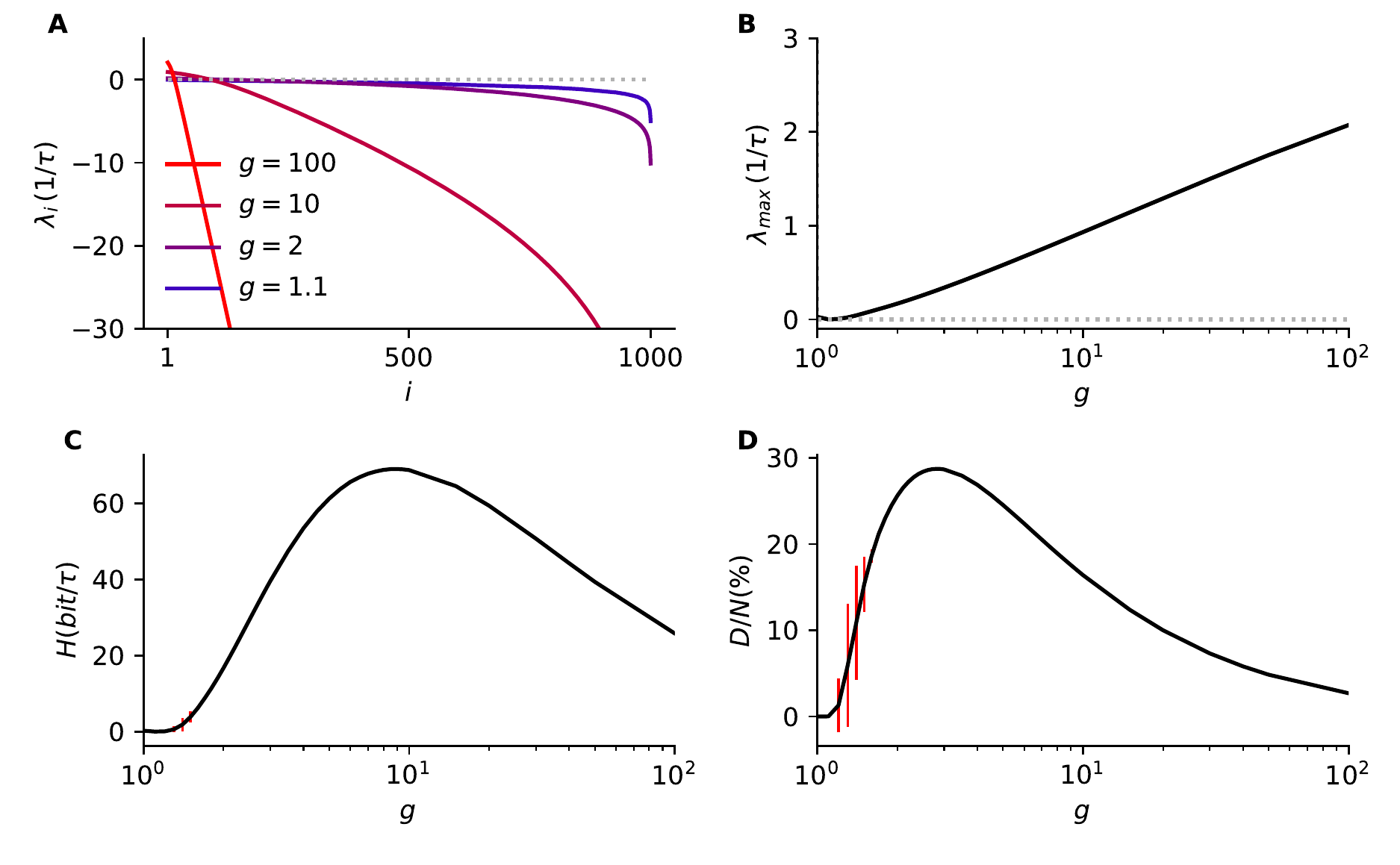}
	\caption{{\bf Lyapunov spectra of discrete-time networks.}
		{\bf A}~Full Lyapunov spectra of discrete-time rate networks with different coupling strengths $g$, color-coded from blue (small $g$) to red (large $g$). 
		{\bf B}~ The largest Lyapunov exponent grows as expected monotonically as a function of $g$ \cite{molgedey_suppressing_1992}.
		{\bf C}~The dynamical entropy rate $H$ peaks with coupling strength $g$. {\bf D}~Relative attractor dimensionality $D/N$ also peaks.
		(Averages over 10 network realizations in black, red error bars indicate double std across 10 network realizations, parameters: $N=1000$, $\Delta t=\tau$, $t_\textnormal{sim}=10^5\tau$, $t_\textnormal{ONS}=\tau$).}
	\label{fig6a}
\end{figure}

\begin{figure}[!h]
	\includegraphics[width=1\columnwidth]{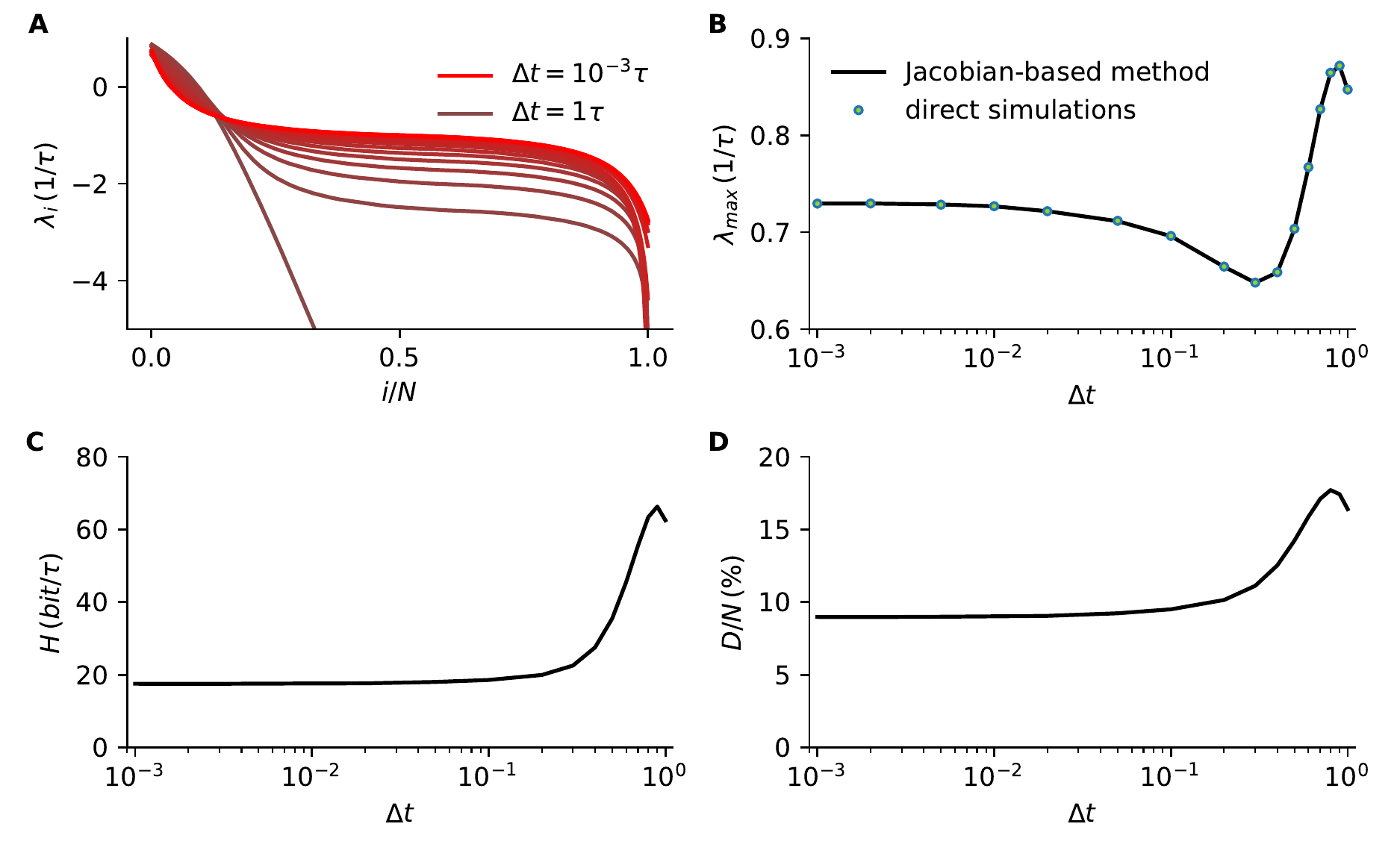}
	\caption{{\bf Full Lyapunov spectrum for different time-discretization $\Delta t$.}
		{\bf A}~The full Lyapunov spectrum reveals drastic changes for increasing $\Delta t$. For finite $\Delta t$, the Lyapunov spectrum loses its symmetry (See also Fig.~\ref{fig6} and compare with Fig.~\ref{fig2}). While the majority of Lyapunov exponents decrease for increasing $\Delta t$, the number of positive exponents increases. 
		{\bf B}~The largest Lyapunov exponent converges for small $\Delta t$. For increasing $\Delta t$, it first decreases and then increases moderately. {\bf C}~The dynamical entropy rate converges for small $\Delta t$ and increases for large $\Delta t$. {\bf D}~The attractor dimensionality behaves similar to the dynamical entropy rate, (other parameters: $N=1000$, $g=10$, $t_\textnormal{ONS}=\tau$, $t_\textnormal{sim}=10^4\tau$; averages across $10$ network realizations).}
	\label{fig5}
\end{figure}

\begin{figure}[!h]
	\includegraphics[width=1\columnwidth]{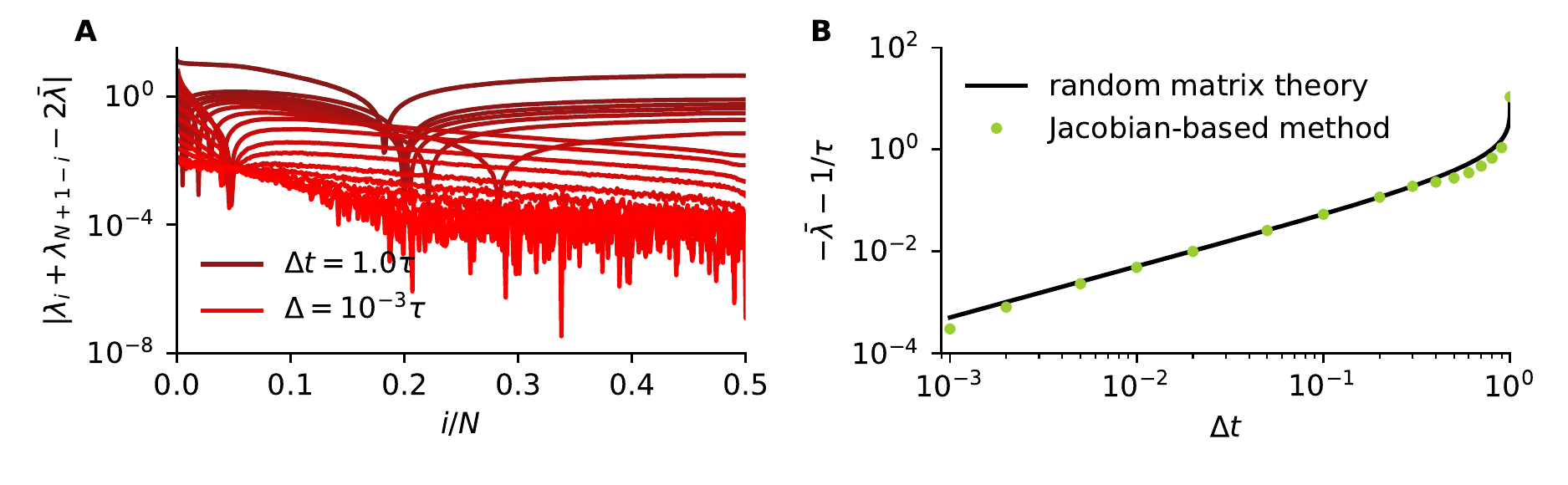}
	\caption{{\bf Point-symmetry of Lyapunov spectrum in continuous-time limit $\Delta t\rightarrow 0$ and mean Lyapunov exponent $\bar\lambda$.}
		{\bf A}~For $\Delta t\rightarrow 0$, the Lyapunov spectrum approaches point-symmetry around $i=N/2$ and $\lambda_i=-\frac{1}{\tau}$, as shown by the convergence of the Lyapunov spectrum towards point-symmetry, so $|\lambda_i+\lambda_{N+1-i}-2\bar\lambda|$ converges towards zero. The Lyapunov spectrum is only point-symmetric in the limits $N\rightarrow \infty$ and $\Delta t\rightarrow 0$. {\bf B}~The mean Lyapunov exponent $\bar \lambda$ converges for small $\Delta t$ towards $-1/\tau$. For finite $\Delta t$, the mean Lyapunov exponent can be approximated analytically (See Appendix \ref{RMTmeanLE}). (Other parameters: $N=1000$, $g=10$, $t_\textnormal{ONS}=\tau$, $t_\textnormal{sim}=10^4\tau$; averages across $10$ network realizations).}
	\label{fig6}
\end{figure}

We found a drastic effect of time-discretization on the Lyapunov spectrum (Fig.~\ref{fig6a}). In discrete-time networks ($\Delta t =1$), the Lyapunov spectrum is not point-symmetric anymore (Fig.~\ref{fig6a}A). The largest Lyapunov exponent grows slowly as a function of coupling strength $g$ (Fig.~\ref{fig6a}B), as expected from previous analytical results \cite{molgedey_suppressing_1992}. However, the slow increase of the largest Lyapunov exponent with coupling strength $g$ is overcompensated by a faster decay of the number of positive Lyapunov exponents, which results in a peak of both dynamical entropy rate H (Fig.~\ref{fig6a}C) and attractor dimensionality $D$ (Fig.~\ref{fig6a}D).
For increasing $g$, the Lyapunov spectrum bents down to strongly negative values. Very negative Lyapunov exponents can be explained by an increasing fraction of rate units in saturation, resulting in a vanishing fraction of directions that carry the gradient dynamics. In the continuous-time case for large $g$, the Lyapunov spectrum would quickly fall to the negative inverse characteristic timescale $-1/\tau$ that originates from the leak term $-h_i$ in the dynamical equation (Fig.~\ref{fig8}C). Here, in the discrete-time case, no such intrinsic timescale is present, and the last Lyapunov exponent becomes progressively more negative (Fig.~\ref{fig6a}A), indicating a quick divergence of the condition number of the long-term Jacobian $T_{t}(\mathbf{x}_{0})$. 
From a machine learning perspective, the leak term can be interpreted as a mimicking skip connection that preserves information of the network state across (unrolled) layers even if the rate units are saturated, thus ameliorating the problem of vanishing gradients \cite{bishop_pattern_2007}.
Next, we study the effect of gradually decreasing the time-discretization $\Delta t$.
 At finite $\Delta t$, the Lyapunov spectrum loses its symmetry (Fig.~\ref{fig5}A and Fig.~\ref{fig6}A), although we demonstrate that the Lyapunov spectrum again approaches point-symmetry around $i=N/2$ and $\lambda_i=-\frac{1}{\tau}$ for $\Delta t\rightarrow 0$ by showing convergence of the Lyapunov spectrum towards its point reflection, so $|\lambda_i+\lambda_{N+1-i}-2\bar\lambda| \rightarrow 0$. Even for very small $\Delta t$, however, there exists a small asymmetry because of the neutral Lyapunov exponent. Removing the neutral Lyapunov exponent, which is associated to a perturbation in the direction of the flow ($\lambda_{\textnormal{neutral}}=0$), improves the point-symmetry of the Lyapunov spectrum. Note that the symmetry of the Lyapunov spectrum originates in the approximate time-reversal symmetry of the dynamics, which only becomes exact in the limit of large $N$ (see section \ref{LSsymmetry}). Thus, the Lyapunov spectrum is only point-symmetric in the limits $N\rightarrow \infty$ and $\Delta t\rightarrow 0$. 

While the largest Lyapunov exponent changes only moderately - and non-monotonously - as the step size increases (Fig.~\ref{fig5}B), the dynamical entropy rate and attractor dimensionality both strongly grow for large $\Delta t$ (Fig.~\ref{fig5}C,D). This growth of entropy rate and dimensionality is primarily caused by an increasing number of positive Lyapunov exponents (Fig.~\ref{fig5}A). 
At the same time, the negative end of the Lyapunov spectrum decreases drastically (Fig.~\ref{fig5}A). This also strongly lowers the mean Lyapunov exponent (Fig.~\ref{fig5}A and Fig.~\ref{fig6}B).
The mean Lyapunov exponent $\bar \lambda$ converges for small $\Delta t$ towards $-1/\tau$. The dependence of the mean Lyapunov exponent on $\Delta t$ can be approximated analytically using random matrix theory by (Appendix~\ref{RMTmeanLE})
\begin{equation}
\bar \lambda(\Delta t)=\frac{\log(1-\Delta t)}{\tau\Delta t}
\end{equation}
This analytical result agrees well with numerical simulations (Fig.~\ref{fig6}B).

\section{Analytical approximations of the full Lyapunov spectrum}
The full Lyapunov spectrum is given by the eigenvalues of the Oseledets matrix \cite{oseledets_multiplicative_1968},
\begin{equation}
\boldsymbol{\Lambda}=\lim_{t\to\infty}[\mathbf{T}_{t}^{\top}{\mathbf{T}_{t}}]^{\frac{1}{2t}}.\label{eq:-Oseledets2}
\end{equation}
where $\mathbf{T}_{t}$ is the long-term Jacobian
\begin{equation}
\mathbf{T}_{t}(\mathbf{h}_{0})=\mathbf{D}_{t-1}(\mathbf{h}_{t-1})\dots\mathbf{D}_{1}(\mathbf{h}_{1})\mathbf{D}_{0}(\mathbf{h}_{0})\label{eq:long-term-Jacobian}
\end{equation}
As \textbf{$\mathbf{T}_{t}(\mathbf{h}_{0})$} is a product of generally noncommuting matrices, it is considered difficult to calculate the full Lyapunov spectrum analytically \cite{crisanti_products_1993}. However, we identified several limits where temporal correlations between subsequent Jacobians vanish and analytical random matrix approximations are justified \cite{crisanti_products_1993}. 
First, we demonstrate an approximation for the stable regime $g < g_{\textnormal{crit}}$, second in the chaotic regime just above the transition $g\rightarrow g_{\textnormal{crit}}^+$, third in the limit of large $g\rightarrow \infty$, fourth when each rate unit is driven by strong Gaussian white noise process with standard deviation $\sigma$ in the limit $\sigma\rightarrow \infty$ (see Sec. \ref{secDriven} for definition), and finally in the discrete-time case without a leak $\Delta t = \tau$. 
We numerically confirmed that in these limits the Lyapunov spectrum becomes invariant under shuffling the sequence of Jacobians (Fig.~\ref{fig8}). 

\begin{figure}[!h]
	\includegraphics[width=1\columnwidth]{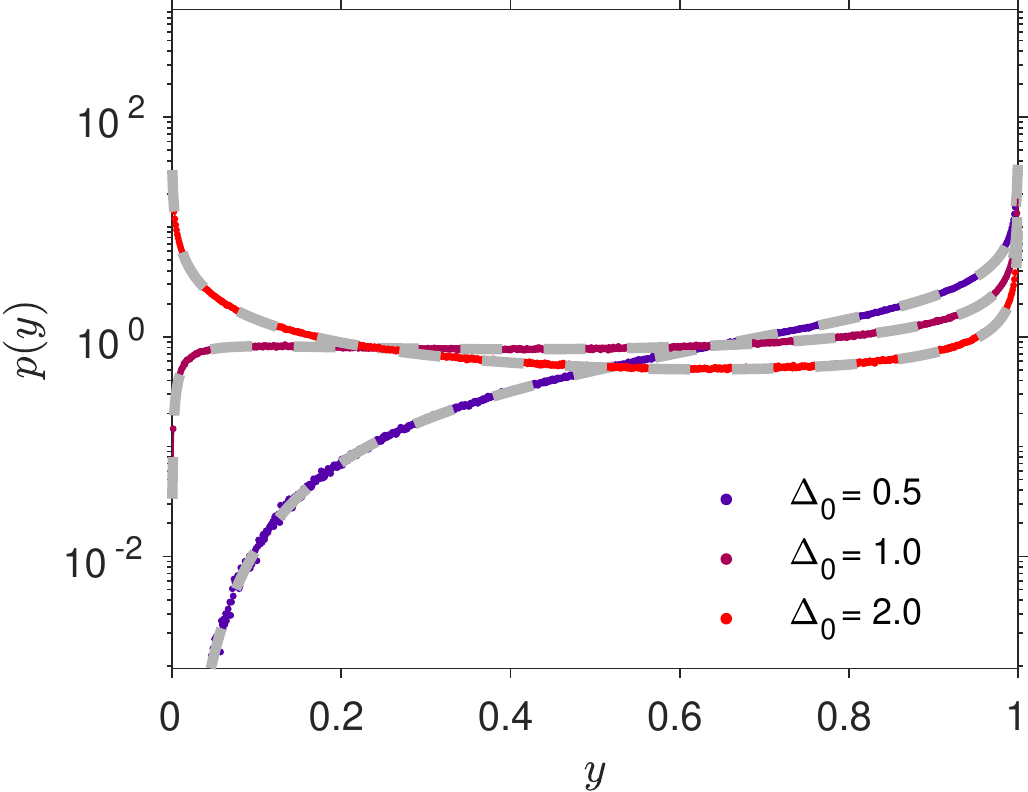}
	\caption{ {\bf Distribution of Jacobian factors $\phi(h)'$.} Colored dots are from direct numerical simulations, and grey dashed lines analytical distributions (Eq.~\ref{eq:-off-diag}) for different values of variance $\Delta_0$ of the local fields $h$. For small $\Delta_0$, most probability mass is close to 1. For large $\Delta_0$, the distribution becomes bimodal, because most units are in saturation of the nonlinearity with corresponding $y$ close to zero and few $y$ near 1.
		\label{fig7}}
\end{figure}

\begin{figure}[!h]
	\includegraphics[width=1\columnwidth]{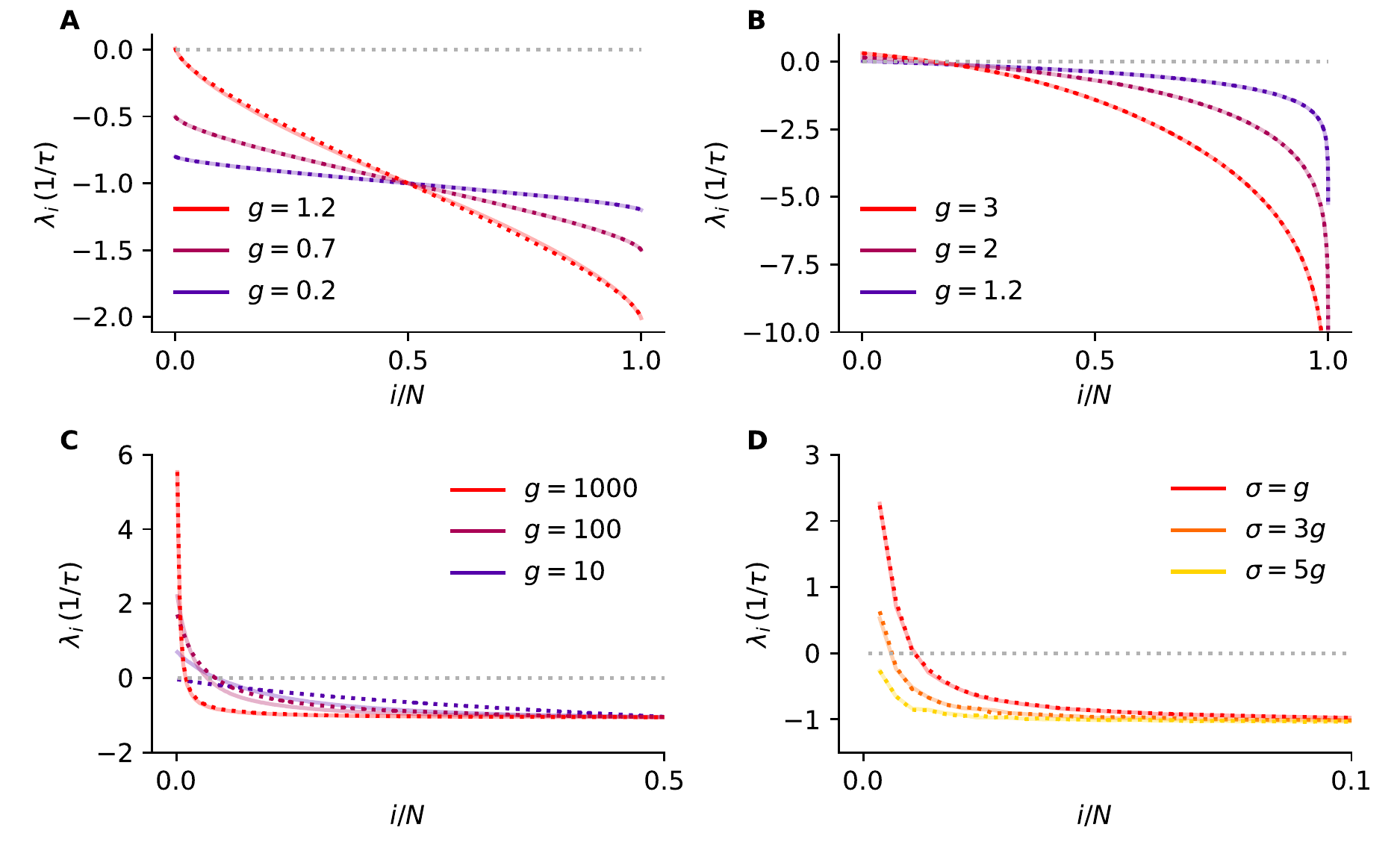}
	\caption{ {\bf Analytical approximations of the full Lyapunov spectrum.} {\bf A}~Lyapunov spectra of autonomous continuous-time rate networks for different coupling strengths $g$, where $g$ is color-coded from purple (small $g$) to red (large $g$), dashed lines are analytical results (Eq.~\ref{eq:-analyticalLyapunovSpectrumStable}) for the stable case and for $g\rightarrow g^+_{\textnormal{crit}}$ (Eq.~\ref{eq:-analyticalLyapunovSpectrumWeakChaos}), full transparent lines are numerical results.
		{\bf B}~In the discrete-time case without leak ($\Delta t = 1$), the Lyapunov spectrum is invariant under shuffling for any $g$ and can be approximated analytically by the triangle law for small $g$ (Eq.~\ref{eq:-triangleLaw}). {\bf C}~Lyapunov spectra of autonomous continuous-time rate networks for $g\gg 1$. For large $g$, the Lyapunov spectrum becomes invariant under shuffling the temporal sequence of Jacobians at fixed $\Delta t=0.1$. {\bf D}~Lyapunov spectra of driven continuous-time rate networks for different input strength $\sigma$ at fixed $g=1000$. Again, for large $g$ and $\sigma$, the Lyapunov spectrum becomes invariant under shuffling the temporal sequence of Jacobians at fixed $\Delta t=0.1$, and the full Lyapunov spectrum can be approximated by a product of random matrices with entries given by Eq.~\ref{eq:-RandomJacobian}. (Parameters if not stated differently: for $g=1.2$ in {\bf A}~$N=8000$ and $t_\textnormal{sim}=10^3\tau$, else $N=1000$, $t_\textnormal{sim}=10^4\tau$, $\Delta t=0.1\tau$, $t_\textnormal{ONS}=\tau$).
		\label{fig8}}
\end{figure}

We calculated the distribution of entries of the Jacobian analytically in the limit $N\rightarrow \infty$, where all $h_i$ follow a Gaussian distribution $h\sim \mathcal{N} (0,\Delta_0)$. First, we calculated the distribution of $y=\phi'$ 
analytically:
\begin{eqnarray}
p(y)&=&\int dh \; \delta(y-\phi'(h)) \frac{e^{-\frac{h^2}{2\Delta_0}}}{\sqrt{2\pi\Delta_0}} \\
&=&\frac{\exp\left( -\frac{\ln\left(1/\sqrt{y}\pm\sqrt{1/y-1} \right) }{2\Delta_0} \right) }{\sqrt{2\pi\Delta_0^2}|2y\sqrt{1-y}|}\label{eq:-off-diag}
\end{eqnarray}
with support $y \in [0,1]$, where $\Delta_0$ is obtained analytically from dynamic mean-field theory \cite{sompolinsky_chaos_1988}.
We can thus write the Jacobian as
\begin{equation}
D_{ij}(t_s)=(1-\Delta t)\delta_{ij} + \Delta t\cdot J_{ij}y_j,\label{eq:-RandomJacobian}
\end{equation}
where $y_j$ are random numbers drawn from the distribution in Eq.~\label{eq:-off-diag}. 
The analytically predicted distributions $p(y)$ are in excellent agreement with the results from direct numerical simulations (Fig.~\ref{fig7}). In the limits we are discussing in the following, the long-term Jacobian can be approximated by a product of random matrices of the form of Eq.~\ref{eq:-RandomJacobian},
\begin{eqnarray*}
	\textbf{$\mathbf{T}_{t}$}	
	&=&	 \prod_{s=0}^{t-1} \mathbf{D}_{s}= \prod_{s=0}^{t-1} \left[ (1-\Delta t)\mathds{1} + \Delta t\cdot \mathbf{J}\cdot \mathbf y_s \right] \\ 
\end{eqnarray*}

In the stable regime $g \leq g_{\textnormal{crit}} $, the Lyapunov spectrum is given by the real parts of the eigenvalue spectrum of the stability matrix 
\[
D_{ij}=\frac{\partial f(h_i)}{\partial h_j} =-\delta_{ij}+J_{ij}\phi'(h_j)
.\]
Because the trivial fixed point $h_i=0$ for all $i$ is the only stable solution for large $N$, this reduces to 
\[
D_{ij} =J_{ij}-\delta_{ij}.
\]
For $J_{ij}$ drawn from a Gaussian distribution $J_{ij} \sim \mathcal{N} (0,g^2/N)$, we find that the real parts of the eigenvalues of $D_{ij}$ follow the right-shifted Wigner semicircle distribution \cite{wigner_distribution_1957,crisanti_products_1993,gotze_asymptotic_2010}:
\[
p(x)=\frac{2}{2\pi g^2}\sqrt{g^2-(x+1)^2}
\] with support $x\in [-g-1, \;g-1]$. Note that this is not only expected for $J_{ij}$ drawn from a Gaussian distribution, but generally for many random matrix ensembles \cite{tao_random_2010}.
The cumulative distribution function is given by 
\[
\chi(x)=\frac{1}{2} + \frac{\arcsin\left(\frac{x+1}{g}\right)}{\pi} + \frac{(x+1)\sqrt{g^2-(x+1)^2}}{\pi g^2}.
\]
The Lyapunov spectrum in the stable regime follows from the inverse,
\begin{equation}
\lambda_i=\chi^{-1}\left(\frac{N-i+1}{N}\right).\label{eq:-analyticalLyapunovSpectrumStable}
\end{equation}
with $\frac{i}{N} \in [0, \;1]$ and $\lambda_i \in [-g-1, \;g-1]$, and with $\lambda_i$ measured in units of $1/\tau$. The analytical Lyapunov spectra are in excellent agreement with the results from direct numerical simulations (Fig.~\ref{fig8}A, purple ($g=0.2$) and maroon ($g=0.7$) lines).

Next, we consider the limit $g\rightarrow 1^+$ close to the transition to chaos. For that, we need to estimate both the distribution of Jacobian entries and its autocorrelations. The autocorrelations of the activity
\begin{equation}
\Delta_i(t') =\left\langle \delta h_i(t) \delta h_i(t+t')\right\rangle 
\end{equation}
can be solved self-consistently \cite{sompolinsky_chaos_1988,crisanti_path_2018-1}. Close to the chaotic instability $g\rightarrow g_{\textnormal{crit}}^+$, the autocorrelations are approximately \cite{sompolinsky_chaos_1988,crisanti_path_2018-1}
\begin{equation}
\Delta(t')=(g-1)\sech\left(\frac{t'(g-1)}{\sqrt{3}}\right)+\mathcal{O}\left((g-1)^2\right). 
\end{equation}

Thus, the timescale of the autocorrelations of $h_i$ diverges when approaching $g_{\textnormal{crit}}$ with $\tau_h=\sqrt{3}/(g-1)$ \cite{sompolinsky_chaos_1988}. From $\lim_{g\rightarrow 1^+}\Delta_0=g-1$ follows that $\lim_{g\rightarrow 1^+}\phi(h_i(t))=h_i(t)$. Therefore the autocorrelations of $D_{ij}$ diverge with the same time constant $\tau_D^{-1}=(g-1)/\sqrt{3}$. Consistent with these analytical considerations, numerical simulations show that for $g \gtrapprox 1 $ 
the Lyapunov spectrum obtained after shuffling the sequence of (almost identical) Jacobians is almost the same.
Thus, we conjecture that the Lyapunov spectrum is given by the logarithms of the singular values of a product of almost identical random matrices, which is still approximately given by the Wigner semicircle distribution \cite{crisanti_products_1993,gotze_asymptotic_2010}
\begin{eqnarray*}
	\chi_{\textnormal{chaos}}(x)&=& \frac{1}{2} + \frac{\arcsin\left(\frac{2(x+1)}{(x - 2) x + 3}\right)}{\pi} \\
	 &+& \frac{2(x+1)\sqrt{((g - 1)^2 - 2 x) (2 x + (g - 2) g + 5)}}{\pi ((x - 2) x + 3)^2} \label{eq:-WignerlawChaos}
\end{eqnarray*}

with support $x\in [-2\tau-\lambda_{\max}, \;\lambda_{\max}]$. Here, we used the analytical knowledge of the largest Lyapunov exponent obtained from dynamic mean-field theory \cite{sompolinsky_chaos_1988,crisanti_path_2018-1}, which behaves in the limit $g\rightarrow g^+_{\textnormal{crit}} = 1^+$ as
\begin{equation}
\lambda_{\max}(g)=\frac{1}{2}(g-1)^2 +\mathcal{O}\left((g-1)^3\right).
\end{equation}

The Lyapunov spectrum in the chaotic regime for $g\rightarrow g^+_{\textnormal{crit}}$ follows the inverse of $\chi_{\textnormal{chaos}}$:
\begin{equation}
\lambda_i=\chi^{-1}_{\textnormal{chaos
}}\left(\frac{N-i+1}{N}\right).\label{eq:-analyticalLyapunovSpectrumWeakChaos}
\end{equation}
The analytical Lyapunov spectra are in good agreement with the results from direct numerical simulations (Fig.~\ref{fig8}A, red line for $g=1.2$). The approximation breaks down if $g$ is too large and becomes more accurate as $g\rightarrow g^+_{\textnormal{crit}}$.

Next, we consider the limit of large $g~\rightarrow~\infty$. Expanding the solution to the self-consistency equation for the autocorrelation of the local fields $h$ in this limit around $t' = 0$ yields in that case
\begin{equation}
\Delta(t')=g^2\Delta_0 - g^2\left( 1-\Delta_0 \right)\frac{t'^2}{2} + \mathcal{O}\left(t'^4\right) \label{eq:Deltalarge_g} 
\end{equation}
with $\Delta_0=2(1-2/\pi)$ \cite{sompolinsky_chaos_1988,crisanti_path_2018-1}. But how can we deal with correlations between subsequent Jacobians?
We note that the autocorrelations of the Jacobians become arbitrary short in the limit of large $g$, although the autocorrelations of the activity variables $h$ approach Eq.~\ref{eq:Deltalarge_g}. For large $g$, the model behaves like the fully asymmetric Ising spin glass model \cite{crisanti_dynamics_1988,crisanti_path_2018-1}. Substituting $\Delta(t')=\Delta_0 \exp(-t'/\tau_h)$ into the self-consistency equation and taking the large $t'$ limit yield a relaxation rate for the autocorrelation equal to $\tau_h^{-1}=\sqrt{1-2/\pi}\;\tau^{-1}$ \cite{sompolinsky_chaos_1988,crisanti_path_2018-1}. 

Thus, the autocorrelation of $D_{ij}$ relaxes approximately with $\tau_D\sim \tau_h/g$. 
Intuitively, for large $g$, most rate units are in saturation, and rate units cross the non-saturated regime where they are susceptible to perturbations in shorter time windows. 
The vanishing autocorrelation time of the Jacobians $D_{ij}$ explains why the Lyapunov spectrum becomes invariant under shuffling of the sequence of Jacobians and justifies the approximation of the long-term Jacobian by a product of uncorrelated matrices drawn of the form of Eq.~\ref{eq:-RandomJacobian}. As expected, the analytical approximations of the Lyapunov spectra approach the results from direct numerical simulations when the values of $g$ increase (Fig.~\ref{fig8}C).
 
We also find that for strong uncorrelated input (see Sec. \ref{secDriven} for numerical results), the Lyapunov spectrum becomes invariant under shuffling the sequence of Jacobians. With increasing input drive $\sigma$ at fixed $g$, all Lyapunov exponents converge towards the negative inverse of the characteristic timescale $-1/\tau$ (not shown). When simultaneously increasing $g$ and $\sigma$, the Lyapunov spectrum becomes invariant under shuffling the Jacobians at finite nontrivial values of the Lyapunov exponents (Fig.~\ref{fig8}D).

Finally, in the discrete-time case $\Delta t=1$ without a leak, temporal correlations between subsequent Jacobians can be neglected for large $N$ \cite{molgedey_suppressing_1992}. For $g\rightarrow 1^+$, the Lyapunov spectrum can thus be obtained from a product of uncorrelated Gaussian matrices, whose eigenvalue distribution follows approximately a triangle law \cite{newman_distribution_1986,isopi_triangle_1992}. The full Lyapunov spectrum (Fig.~\ref{fig8}B) in this limit can thus be approximated by 
\begin{equation}
\lambda_i\approx\log\left(\exp(\lambda_{1})\cdot \sqrt{1-\frac{i}{N}}\right)=\lambda_{1}+ \frac{1}{2} \log\left(1-\frac{i}{N}\right),\label{eq:-triangleLaw}
\end{equation}
where the largest Lyapunov exponent $\lambda_{\max}$ can be obtained analytically as described earlier both in the discrete and continuous-time case with constant input and frozen white noise drive \cite{sompolinsky_chaos_1988,kadmon_transition_2015,molgedey_suppressing_1992,schuecker_optimal_2018,crisanti_path_2018-1}.

\section{Lyapunov spectrum of balanced rate network with threshold-linear transfer functions}
While odd symmetric saturated sigmoid transfer functions, e.g., $\phi(x)=\tanh(x)$ are popular because of their mathematical tractability \cite{sompolinsky_chaos_1988,crisanti_path_2018-1} and because the saturation prevents runaway activity, the firing rate of many cortical neuron types seems in a physiological operating regime not to be limited by intrinsic electrophysiological features. Evidence for this comes from the observation that artificially driven neurons can fire at much higher rates \cite{mccormick_comparative_1985} than they actually do in experimental recordings of awake behaving animals \cite{barth_experimental_2012}. 
In balanced networks, large externally incoming excitatory currents are dynamically canceled by net inhibitory recurrent currents, which yields a broad parameter regime of asynchronous irregular activity in spiking network models \cite{van_vreeswijk_chaos_1996,van_vreeswijk_chaotic_1998,van_vreeswijk_course_2005,brunel_dynamics_2000,monteforte_dynamical_2010}. Such balanced state models were recently extended from spiking networks to firing rate networks \cite{harish_network_2013,harish_asynchronous_2015,kadmon_transition_2015,mastrogiuseppe_intrinsically-generated_2017}.

\begin{figure}[!h]
	\includegraphics[width=1\columnwidth]{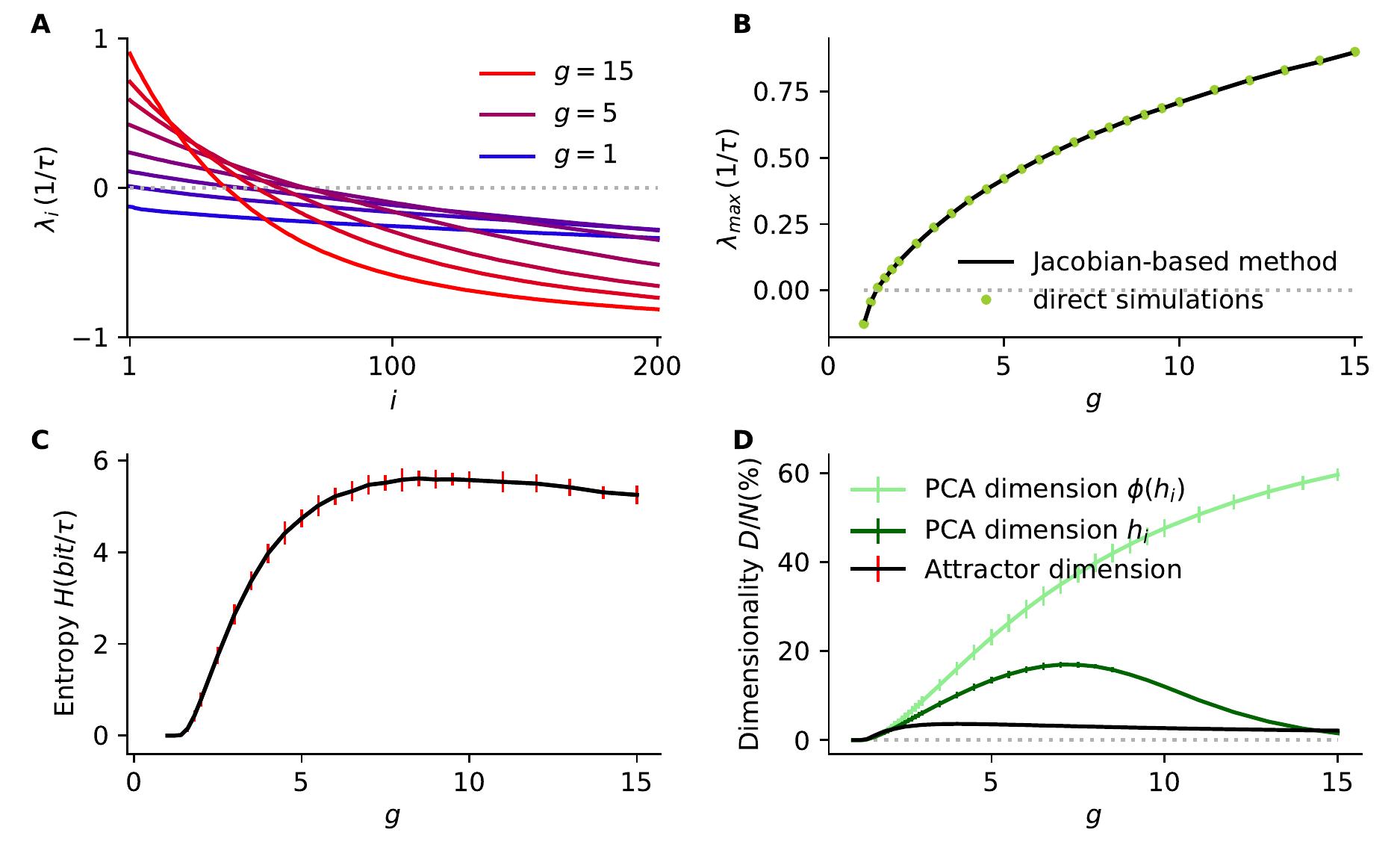} 
	\caption{ {\bf Entropy rate and dimensionality of a balanced firing-rate network with threshold-linear transfer function.} {\bf A}~Top 200 Lyapunov exponents of rate networks for different coupling strengths $g$, where $g$ is color-coded from blue (small $g$) to red (large $g$).
		{\bf B}~ The largest Lyapunov exponent grows monotonically as a function of $g$ below the divergence.
		(Green dots: direct numerical simulations, black line: Jacobian-based method)
		{\bf C}~The dynamical entropy rate $H$ peaks with coupling $g$. {\bf D}~Relative attractor dimensionality $D/N$ has a peak as a function of $g$.
		(black curves are averages over 20 network realizations, red error bars indicate double std across 20 network realizations, parameters: $N=4000$, $\bar g=300$, $I=300$, $t_\textnormal{sim}=10^3\tau$, $t_\textnormal{ONS}=\tau$).
		\label{fig9}}
\end{figure}

Here we extend our Lyapunov spectrum analysis to balanced networks with a threshold-linear transfer function and investigate the role of the synaptic coupling strength $g$ on the Lyapunov spectrum (Fig.~\ref{fig9}). Threshold-linear transfer functions are also commonly used in deep learning \cite{glorot_deep_2011,maas_rectifier_2013}. Another reason to investigate this transfer function is that experimentally measured neural nonlinearities in sensory neurons have been approximated by a power-law threshold nonlinearity \cite{hansel_how_2002,priebe_contribution_2004,priebe_direction_2005,priebe_mechanisms_2006,finn_emergence_2007}.

For simplicity, we focus on the dynamics of an inhibitory network of $N$ threshold-linear rate units that balance a constant excitatory external input. The dynamics of each firing-rate unit follows 
\begin{equation}
\tau\frac{\dif h_{i}}{\dif t}=-h_{i}+\sum_{j}J_{ij}\phi\left(h_{j}\right)+I
\end{equation}
where $I$ is a positive constant and $\phi(x)=\max(x,0)$. We draw entries of the coupling matrix $J_{ij}$ from a Gaussian distribution $J_{ij}\sim\mathcal{N}(-\mu/N,g^{2}/N)$ (similar to \cite{kadmon_transition_2015}).

As in the previously considered $\tanh$ networks, the first half of the Lyapunov spectrum is increasingly curved for increasing $g$ (Fig.~\ref{fig9}A). For large $g$ the network dynamics turns unstable and the activities $h_i$ diverge \cite{kadmon_transition_2015}, therefore there is no chaotic large $g$-limit for fixed mean coupling strength $J$. The divergence occurs at much larger $g$ than displayed in Fig.~\ref{fig9}. The largest Lyapunov exponent shows the analytically predicted behavior \cite{kadmon_transition_2015}. We confirmed the results obtained from the Jacobian-based method \cite{benettin_lyapunov_1980} by tracking the amplitude of a small perturbation in direct numerical simulations (Fig.~\ref{fig9}B). The dynamical entropy rate is zero for $g \leq g_{\textnormal{crit}} $ and first grows for increasing values of $g$ (Fig.~\ref{fig9}C) up to a peak value $g_{\textnormal{peak}}$. Beyond the peak, the growth of a small fraction of positive Lyapunov exponents is overcompensated by a decreasing number for large $g$; thus, the entropy rate decreases for large $g$. We found that the dimensionality also peaks as a function of $g$ for balanced networks with threshold-linear transfer functions (Fig.~\ref{fig9}D), but at higher values than for the entropy rate. Thus, for larger $g$, the diversity of network states as quantified by $D$ decreases. 

\section{Point-symmetry of Lyapunov spectra for continuous-time dynamics}\label{LSsymmetry}
A symmetry of Lyapunov spectra around zero is usually found in dynamical systems with a symplectic structure \cite{dressler_symmetry_1988,wojtkowski_conformally_1998}. This is given, for example, in Hamiltonian systems, where the Lyapunov spectrum is symmetric around zero. Symmetry around a negative value was previously described in a class of dissipative dynamical systems with viscous damping \cite{dressler_symmetry_1988}. 

The recurrent neuronal networks we considered have an asymmetric connectivity $J_{ij}\neq J_{ji}$. Thus, there is no conservation of energy, thus there is no time reversal symmetry. Also, a pseudo-Hamiltonian structure is not given. Moreover, our findings indicate that symmetric Lyapunov spectra are not a generic feature of recurrent neural networks. But was is the origin of the symmetry in our case of recurrent neural firing rate networks?
The symmetry of the Lyapunov spectrum of recurrent networks in the continuous-time limit originates in the approximate time-reversal symmetry of the dynamics.
This can be directly seen by a change of variables into a reference frame that contracts with time. Introducing the new variables $z=e^{\frac{t}{\tau}}h$ turns the original equation of motion
\begin{equation}
\tau\frac{\dif h_i}{\dif t}=-h_i +\sum_{j=1}^{N} J_{ij}\phi(h_j).
\end{equation}
to 
\begin{equation}
\tau\frac{\dif z_i}{\dif t}= e^{\frac{t}{\tau}} \sum_{j=1}^{N} J_{ij}\phi(e^{-\frac{t}{\tau}} z_j).
\end{equation}
Making the replacement $z(t)=\tilde z(-t)$, 
gives 
\begin{equation}
\tau\frac{\dif \tilde z_i}{\dif {t}}= e^{\frac{t}{\tau}} \sum_{j=1}^{N} (-J_{ij})\phi(e^{-\frac{t}{\tau}} \tilde z_j). 
\end{equation}
Thus, with reversed time, one obtains the same dynamics with coupling matrix $\tilde J_{ij}=-J_{ij}$. $\tilde J_{ij}$ follows the  same distribution as $-J_{ij}$. As for large $N$, the Lyapunov spectrum does not depend on the realization of the network connectivity, which is drawn from a Gaussian distribution $J_{ij} \sim \mathcal{N} (0,g^2/N)$, the Lyapunov spectrum is invariant under flipping the sign of $J_{ij}$. Thus, the dynamics is statistically invariant under  time-reversal, where 'statistically invariant' means under the statistics of the connectivity.
The Oseledets matrix in the contracting reference frame is
\begin{eqnarray}
\boldsymbol{\tilde\Lambda}(\mathbf{z}_{0})&=&\lim_{t\to\infty}[e^{-\frac{{t}}{\tau}}\mathbf{\hat T}_{t}(\mathbf{z}_{0})^{\top}\mathbf{\hat T}_{t}(\mathbf{z}_{0})e^{-\frac{{t}}{\tau}}]^{\frac{1}{2t}} \label{eq:-Oseledets-contracting}\\
&=& e^{-\frac{{1}}{\tau}} \lim_{t\to\infty}[\mathbf{\hat T}_{t}(\mathbf{z}_{0})^{\top}\mathbf{\hat T}_{t}(\mathbf{z}_{0})]^{\frac{1}{2t}}\\
&=& e^{-\frac{{1}}{\tau}} \boldsymbol{\hat\Lambda}(\mathbf{z}_{0}).
\end{eqnarray}
Growing tangent space volume elements in forward time correspond to shrinking tangent space volume elements in backward time. Because of the time-reversal symmetry, they are approximately inverse, i.e., the eigenvalues of the Oseledets matrix satisfy 
\begin{equation}
e^{-\frac{{1}}{\tau}}\mu_i^{{\boldsymbol{+}}}\approx e^{-\frac{{1}}{\tau}}\frac{1}{\mu_{N-i+1}^{{\boldsymbol{-}}}}
\end{equation}
where ${\boldsymbol{+}} ({\boldsymbol{-}})$ indicate forward (backward) time direction.
Thus, the Lyapunov exponents, given by the logarithm of the eigenvalues of the Oseledets matrix satisfy
\begin{equation}
\lambda_i^{{\boldsymbol{+}}}-\frac{1}{\tau}\approx -\frac{1}{\tau} - \lambda_{N-i+1}^{\boldsymbol{-}},
\end{equation}
where the factor $-\frac{1}{\tau}$ comes from the shrinking reference frame. Note that in contrast to Hamiltonian systems where the symplectic structure of the Hamiltonian implies an exact symmetry of the Lyapunov spectrum (around zero), here, the symmetry is only approximate for finite-size networks and around the negative inverse of the characteristic time scale that acts through the leak term as global damping on the dynamics. Also note that in autonomous systems that are not at a fixed point, there is always a zero Lyapunov exponent $\lambda_{\textnormal{neutral}}=0$ corresponding to neutral shifts in the direction of time that does not have a symmetric analogue at $\lambda=-\frac{2}{\tau}$. Finally, we note that our symmetry argument assumes that there exist statistically analogous backward trajectories, which is generally not correct but justified in our case because of the statistical mirror-symmetry of the connectivity $p(J_{ij})$ around zero. For example, this does not generally hold after training, where the negative connectivity $\tilde J_{ij}=-J_{ij}$ can yield statistically very different dynamics.

\section{Delocalization of the first covariant Lyapunov vector}
\begin{figure}[!h]
	\includegraphics[width=1\columnwidth]{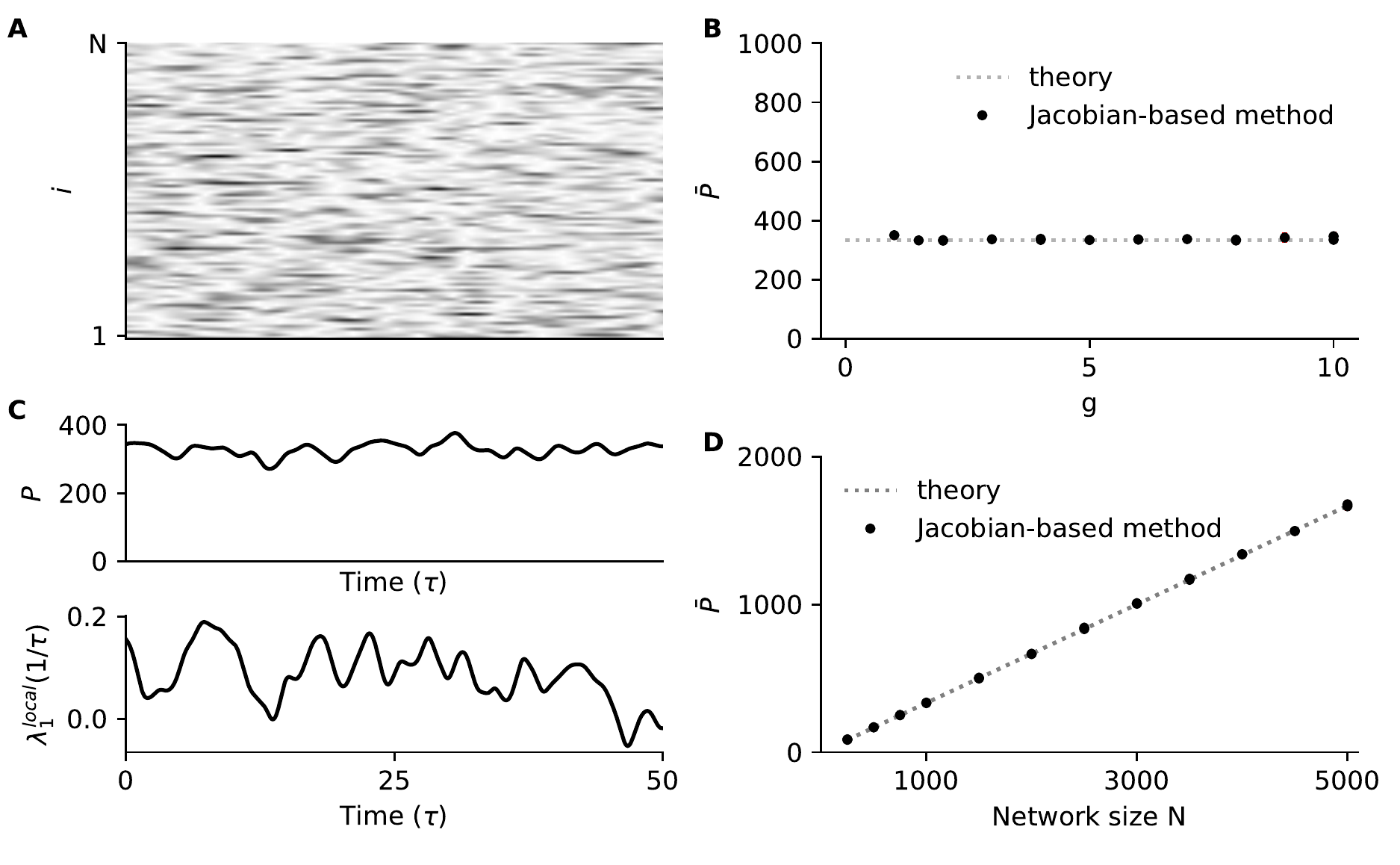}
	\caption{{\bf Spatiotemporal analysis of network chaos and localization of first covariant Lyapunov vector}
		{\bf A}~First covariant Lyapunov vector $\vec{v}(t)$ (gray-scale of $|v_i(t)|$ of a subset of 100 random directions) {\bf B}~Average participation ratio $\bar P$ vs. $g$ stays constant for $\tanh$ network and for balanced threshold-linear network, consistent with prediction of $\bar P  = N/3$. {\bf C}~Participation ratio $P(t)$ of first covariant Lyapunov vector (CLV) and corresponding local Lyapunov exponent $\lambda_{\max}^{\textnormal{local}}(t)$ {\bf D}~Average participation ratio $\bar P$ vs. network size $N$ indicates delocalized CLV $\bar P  = N/3$.
		(parameters: $N=1000$, $g=2$, $\Delta t=10^{-2}\tau$, $t_\textnormal{ONS}=\tau$, $t_\textnormal{sim}=10^3\tau$, averages across $10$ network realizations).}
	\label{fig10}
\end{figure}

To quantify how many rate units contribute to the chaotic dynamics at each moment in time, we investigated properties of the covariant Lyapunov vectors $\vec{v}^{(k)}(t)$. The first covariant Lyapunov vector gives at any point in time the direction in which almost all initial infinitesimal perturbations grow with average rate $\lambda_{\max}$. It corresponds to the first Gram-Schmidt vector and is denoted here as  $\vec{v}$ with $\sum_{i=1}^{N}v_{i}(t)^{2}=1$. The number of rate units contributing to the maximally growing direction at time $t$ can be measured by the participation ratio $P(t)=\left(\sum_{i=1}^{N} v_{i}(t)^{4}\right)^{-1}$ \cite{wegner_inverse_1980,kaneko_lyapunov_1986,cross_pattern_1993}. If all rate units contribute equally to the Lyapunov vector $\left|v_{i}(t)\right|=1/\sqrt{N}$, the participation ratio is $P(t)=1/(N/N^{2})=N$. If only one rate unit contributes to the Lyapunov vector, the participation ratio is $P(t)=1$. The Lyapunov vector of firing rate networks indicates that a temporally varying subset of rate units governs the most unstable direction (Fig.~\ref{fig10}A, C). There is only a moderate temporal fluctuation of the participation ratio (Fig.~\ref{fig10}C), which declines proportionally to $1/\sqrt{N}$ with network size (not shown).

The participation ratio $\bar{P}=\left\langle P(t)\right\rangle$ was independent of $g$ both for networks with $\tanh$ and threshold-linear input-output transfer function (Fig.~\ref{fig10}B). 

To further characterize the nature of the chaotic collective network state, we investigated the scaling of the mean participation ratio $\bar{P}$ with network size. Whether the Lyapunov vector is called localized or delocalized depends on how $\bar{P}$ scales as a function of network size $N$. A delocalized state is indicated by a linear scaling $\bar{P}\sim N$, while in the case of a localized state, the participation ratio is independent of $N$. We found a linear scaling $\bar{P}\sim N$ (Fig.~\ref{fig10}D) for both $\tanh$ and threshold-linear transfer function. This is consistent with assuming that entries of the first covariant Lyapunov vector are independent Gaussian with $v_i \sim \mathcal{N} (0,1/N)$, which yields 
\begin{eqnarray}
\bar P
&=&N\left(\int dx \; \delta(y-x^4) \frac{e^{-\frac{x^2}{2}}}{\sqrt{2\pi}}\right)^{-1}
=\frac{N}{3}.\label{gaussintegral4}
\end{eqnarray}
This is in contrast to chaos in sparse spiking neural networks, where a sublinear scaling of the participation ratio with network size has been reported for sparse networks of quadratic integrate-and-fire neurons in the balanced state \cite{monteforte_dynamical_2010}. We conclude that the direction of greatest instability in random rate networks is supported by a macroscopic number of rate units, which indicates the existence of collective Lyapunov modes that characterize the instability of the collective dynamic. This is a promising direction for future research that might link the microscopic phase space structure to macroscopic modes of activity \cite{takeuchi_collective_2013}.

\section{Lyapunov spectrum of externally driven network} \label{secDriven}
Thus far, we have analyzed the autonomous dynamics of a deterministic firing-rate network, but it is interesting to extend this to a non-autonomous system driven by time-varying input \cite{molgedey_suppressing_1992,rajan_eigenvalue_2006,rajan_stimulus-dependent_2010,massar_mean-field_2013,schuecker_optimal_2018}. We consider an input-driven network, with $\tau = 1$,
\begin{equation}
\tau\frac{\dif h_i}{\dif t}=F_i=-h_i +\sum_{j=1}^{N} J_{ij}\phi(h_j)+\xi_i(t),
\end{equation}
where in the case considered here $\xi_i $ are fixed realizations of independent Gaussian frozen white noise processes with autocorrelation function $\left\langle\xi_i(t)\xi_i(t+t')\right\rangle =\tau\sigma^2\delta(t')$.

To assess the dynamic stability of the (frozen) stochastic differential equation, we employ the theory of random dynamical systems (RDS). This theory characterizes how reliably different initial states respond to a frozen external input realization. We call a system reliable if different initial conditions converge to the same (time-dependent) trajectory, and unreliable otherwise \cite{lin_stimulus-response_2013}. More formally, the evolution of a sample measure $\mu_{\xi}^{t}$ is studied for a frozen noise realization $\xi(t)$ with $t\in(-\infty,\infty)$. This is described in more detail in Appendix \ref{LSdriven}.

The mathematical expression for the Jacobian of the flow of the dynamics is the same as in the autonomous case Eq.~\ref{eq:-Jacobian}. However, despite this similarity, an external input can have a strong effect both on the distribution of $h_i$ and on the autocorrelations $\Delta_i(\tau) =\left\langle \delta h_i(t) \delta h_i(t+\tau)\right\rangle $. First, input fluctuations increase the width of the distribution of $h_i$, meaning that more units are in the saturated regime, and the Jacobian becomes sparser, which suppresses chaos \cite{molgedey_suppressing_1992}. Second, input fluctuations temporally decorrelate network states, which destroys temporal correlations of subsequent Jacobians resulting in an independent dynamic reduction of chaos \cite{schuecker_optimal_2018}.
The full Lyapunov spectrum, which is independent of input realization $\xi$ \cite{kifer_ergodic_2012}, is again obtained by a reorthonormalization procedure of the Jacobians along a numerical solution of the stochastic differential equation integrated with the Euler-Maruyama method \cite{benettin_lyapunov_1980}. For details, see Appendix \ref{Algo}.

\begin{figure}[!h]
		\includegraphics[width=1\columnwidth]{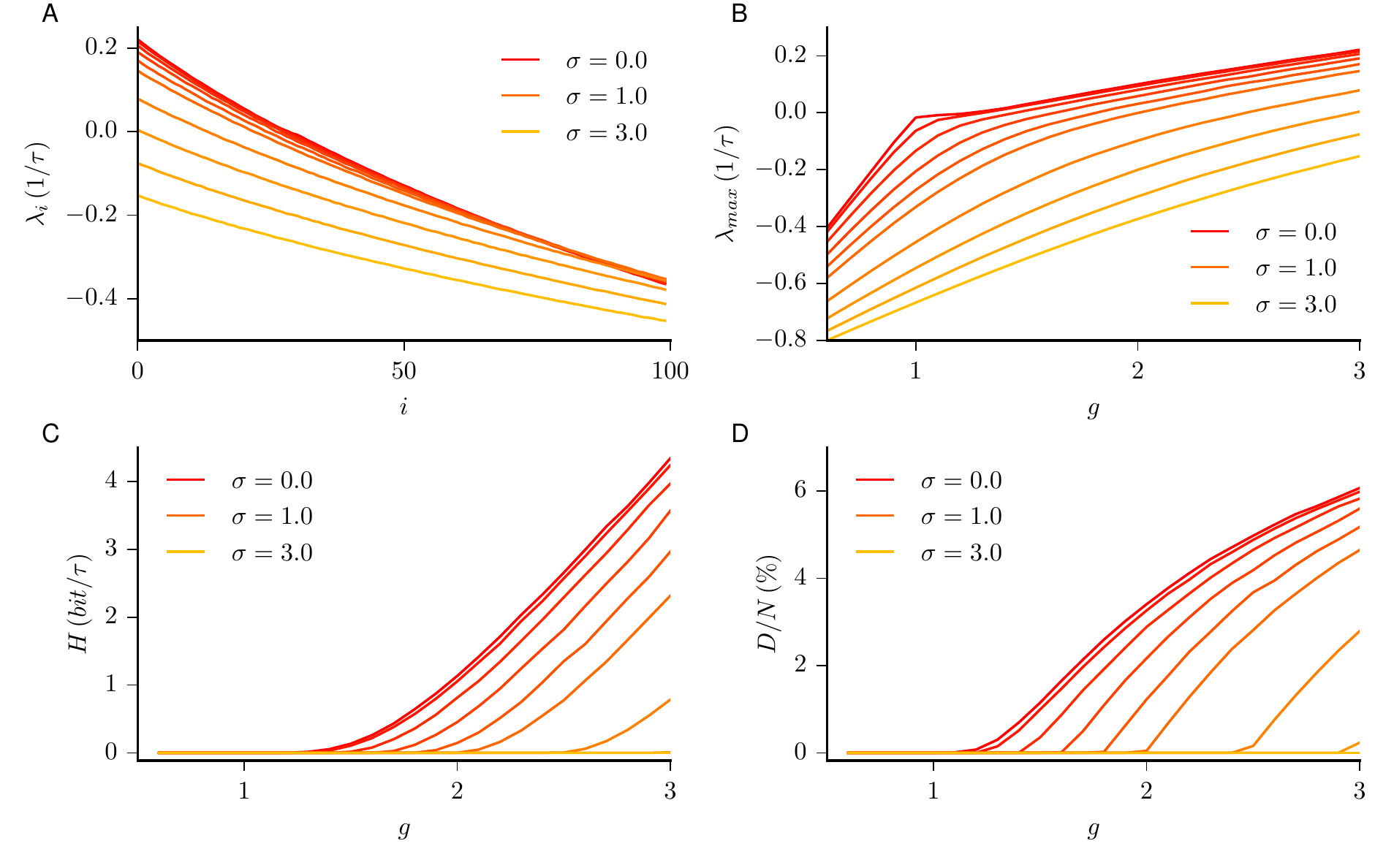}
	\caption{{\bf Time-varying stimuli reduce both the dynamical entropy rate and attractor dimensionality.}
		{\bf A}~For increasing input strength $\sigma$, the Lyapunov spectrum is increasingly pushed towards the mean Lyapunov exponent $-1/\tau$. {\bf B}~The largest Lyapunov exponent decreases, and the transition is smoothed, consistently with previous work \cite{schuecker_optimal_2018}. {\bf C}~The dynamical entropy rate $H$ is reduced. {\bf D}~The relative attractor dimensionality $D/N$ decreases for increasing $\sigma$.
		(Parameters: $N=1000$, $\Delta t=10^{-2}\tau$, $t_\textnormal{ONS}=\tau$, $t_\textnormal{sim}=10^3\tau$, averages across $10$ network realizations).}
	\label{fig11}
\end{figure}

We explored the effect of increasing input strength $\sigma$ on the Lyapunov spectrum (Fig.~\ref{fig11}). For increasing input strength $\sigma$, the Lyapunov spectrum is increasingly pushed towards the mean Lyapunov exponent $-1/\tau$ (Fig.~\ref{fig11}A). Increasing $\sigma$ monotonically reduces the largest Lyapunov exponent as previously observed in discrete \cite{molgedey_suppressing_1992,massar_mean-field_2013}
and continuous-time \cite{schuecker_optimal_2018} (Fig.~\ref{fig11}B). A similar effect has been observed in rate networks driven by periodic input \cite{rajan_inferring_2010,rajan_stimulus-dependent_2010}.

\subsubsection{Input fluctuations reduce the dynamical entropy rate and attractor dimensionality} 

The dynamical entropy rate for a given external input, which is calculated from the sum of the positive Lyapunov exponents, decreases for increasing external input strength $\sigma$ (Fig.~\ref{fig11}C). For sufficiently strong input, the entropy rate drops to zero. Thus, time-varying input impedes the flow of information from the microscopic states to the macroscopic network states. If the information in the microscopic state is considered to be noise, one can conclude that stronger external input fluctuations reduce the noise entropy arising from sensitivity to initial conditions. 
The attractor dimensionality also decreases for increasing input strength $\sigma$ (Fig.~\ref{fig11}D). Sufficiently strong input suppresses chaos, implying that the sample measure collapses on a (wandering) random sink \cite{baxendale_stability_1992,le_jan_equilibre_1987}. In other words, almost all initial conditions converge onto a set of measure zero. Thus, while the network dynamics with strong time-varying input might still seem to be high-dimensional, the attractor dimensionality given the external input can shrink drastically with a time-varying external input. Such a transition is relevant for information processing because the network loses its dependence on initial conditions, which could be a desirable feature if the network must reliably generate different output trajectories for different input patterns \cite{jaeger_harnessing_2004,sussillo_generating_2009,laje_robust_2013}.

\section{Applications to quantifying stability of trained recurrent neural networks} 
The networks we studied up to this point had random connectivity, but collective network dynamics is strongly shaped by wiring and learning algorithms for training recurrent neural networks in machine learning work by tuning connectivity. We now show that training a recurrent network to perform a task is reflected in the dynamic stability, as quantified by the Lyapunov spectrum, and show in some examples how it can affect the dimensionality and dynamic entropy rate.	

\begin{figure}
\includegraphics[width=\columnwidth]{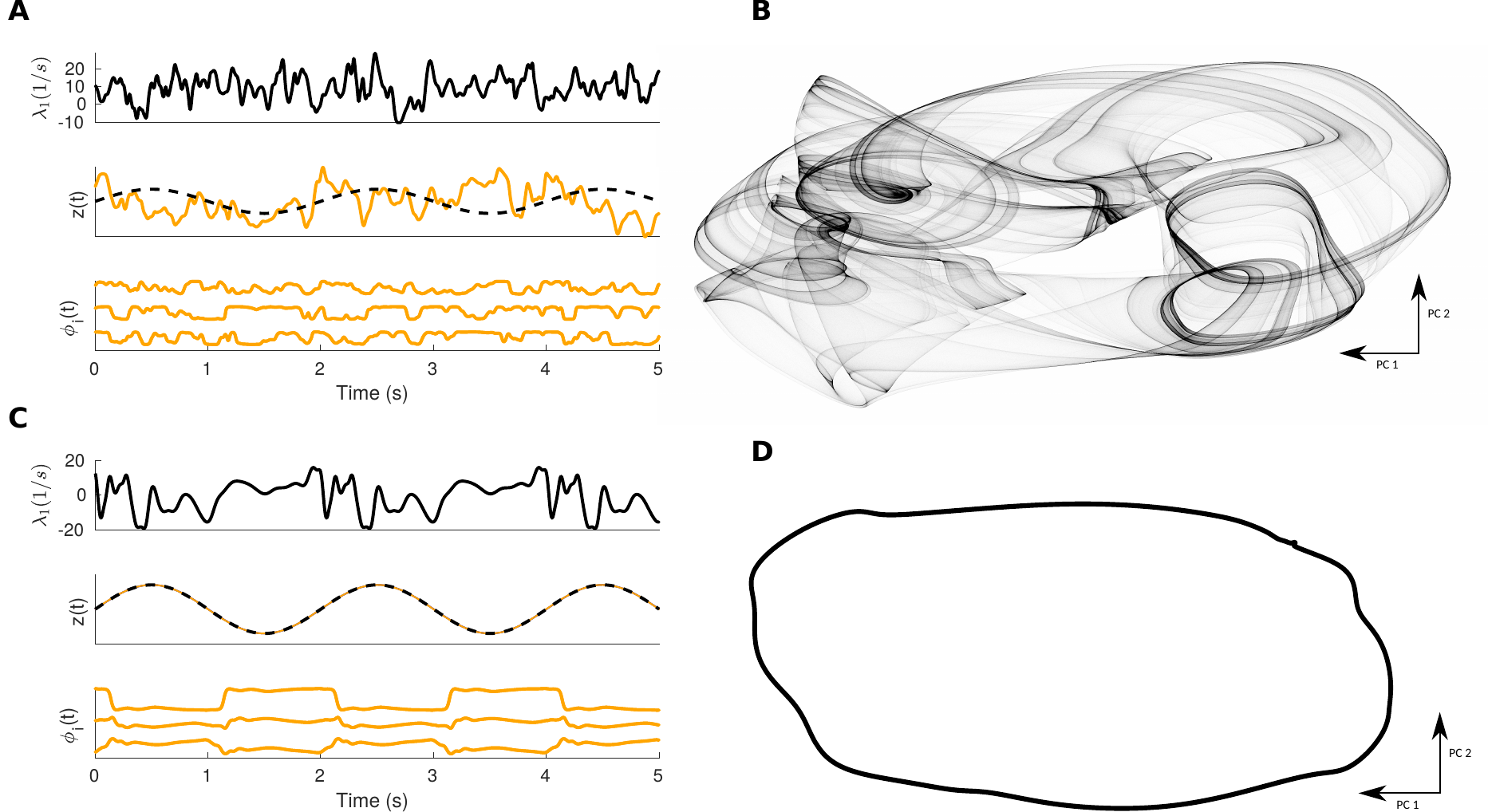}
\caption{{\bf Reorganization of rate network phase space during learning.} {\bf A} Local Lyapunov exponent ($\lambda^{\textnormal{local}}_i(t)$), output $z(t)=\mathbf{w}^{\intercal}\mathbf{\phi}(\mathbf{x}(t))$, and activity of example rate units $\phi_i(t)$ before learning in the chaotic state. {\bf B} Chaotic network activity $\mathbf{\phi}(t)$ before learning projected on the first two principal components. {\bf C} Same as {\bf A} after learning a periodic task using FORCE \cite{sussillo_generating_2009}. {\bf D} Same as {\bf B} after training.  (Parameters: $N=200$, $g=1.5$, $\Delta t=10^{-1}\tau$, $t_\textnormal{ONS}=10^{-1}\tau$, $\tau=10^{-2}$s)
		\label{fig12}}
\end{figure}

During training, network dynamics becomes confined to a low-dimensional manifold (Fig.~\ref{fig12}). 
When initializing with a random network structure in the chaotic regime ($g > 1$), the dynamics evolves on a high-dimensional attractor that spans an extensive fraction of the full $N$-dimensional phase space (Fig.~\ref{fig3}). A projection of the high-dimensional strange chaotic attractor onto the first two principal components is shown for a network of 50 rate units in Fig.~\ref{fig12}B. After training the network to perform a simple sine oscillation through a linear readout, the network dynamics is confined to a periodic orbit (Fig.~\ref{fig12}D). Note that the sine is computed by the coordinated activity of many rate units together that individually have dynamics different from the sine target (Fig.~\ref{fig12}C). For this task, the largest Lyapunov exponent becomes zero after training. This is expected for an autonomous network because all but the neutral direction along the flow become stabilized. Therefore, the dynamic entropy rate is trivially zero, and the attractor dimension is unity.

The Lyapunov spectrum can also be used as a quantification of how stable trajectories are after training. In Fig.~\ref{fig13}, we compare the result of training a recurrent rate network to output an oscillation with temporally varying frequency in response to an input pulse with three different training algorithms, backpropagation through time (BPTT), FORCE \cite{sussillo_generating_2009} and full-FORCE \cite{depasquale_full-force:_2018}. In BPTT, the full recurrent weight matrix and a readout vector are iteratively adapted by minimizing an error function using (stochastic) gradient descent {\cite{bengio_learning_1994}.
FORCE recursively updates a rank-one perturbation $u_i w_j$ to $J_{ij}$ such that the linear readout $z(t)=\sum_j w_j\phi({x_j}(t))$ matches a (potentially time-varying) target output. Full-FORCE does a full-rank recursive update of a task-performing network to match for each unit the activity to a teacher network.
 For full-FORCE, the Jacobian of the dynamics is 
\[ D_{ij}^\textnormal{fF}(t)=-\delta_{ij}+J_{ij}\phi'(x_j(t)),\] for FORCE, it is \[D_{ij}^{\textnormal{F}}(t)=-\delta_{ij}+(J_{ij}+u_i w_j)\phi'(x_j(t)).\] We obtained Lyapunov exponents by evolving an orthonormal basis along the trajectory using the analytical Jacobians as described before and in more detail in Appendix \ref{Algo}.

We find that the full-rank method full-FORCE results in a more negative largest Lyapunov exponent and thus a microscopically more stable dynamics (Fig.~\ref{fig13}). Moreover, subsequent Lyapunov exponents drop quicker towards the negative inverse of the characteristic timescale $-1/\tau$. The external input pulse makes the dynamics non-autonomous; therefore, no neutral Lyapunov exponent occurs. 
Note that convergence of infinitesimally different initial conditions does not necessarily imply stability with respect to finite-size perturbations. For example, in spiking networks there exists multistability \cite{monteforte_dynamic_2012}, and also trained firing rate networks often exhibit multistability (not shown).

\begin{figure}[!h]
	\includegraphics[width=1\columnwidth]{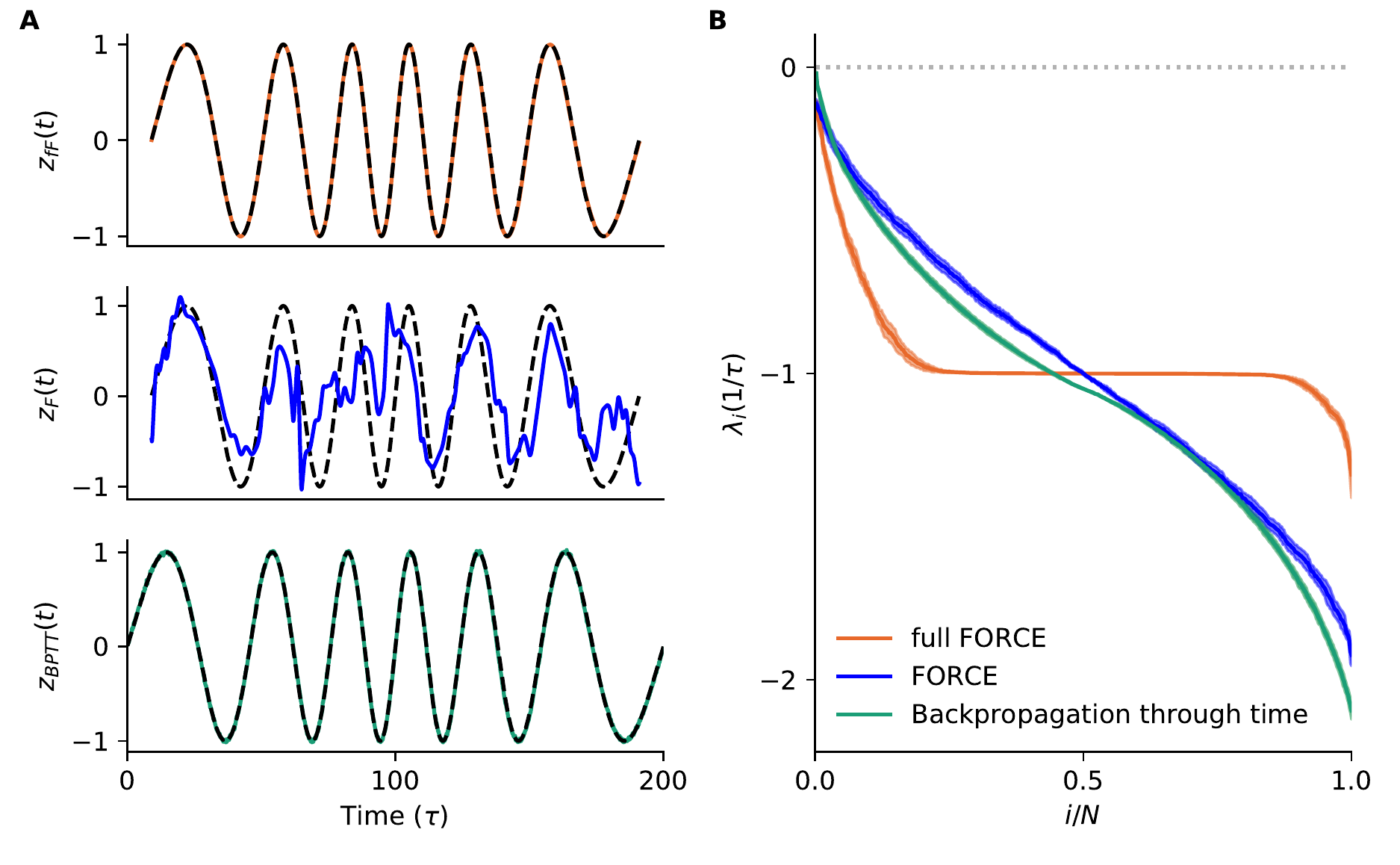}
	
	\caption{{\bf Quantification of dynamic stability after training rate networks on task using FORCE, full-FORCE, and backpropagation through time} 
		{\bf A}~Example output $z(t)=\mathbf{w}^{\intercal}\mathbf{\phi}(\mathbf{x}(t))$ of a network of 500 units trained with FORCE (blue), full-FORCE (orange), and backpropagation through time (green). {\bf B}~Lyapunov exponents calculated at the end of the training. We find that the full-rank method full-FORCE results in a more negative largest Lyapunov exponent and thus a microscopically more stable dynamics. Moreover, subsequent Lyapunov exponents drop quicker towards the negative inverse of the characteristic timescale $-1/\tau$. 
		(other parameters: $N=500$, $g=1.5$, $\Delta t=0.1\tau$, $t_\textnormal{ONS}=\tau$, $t_\textnormal{sim}=10^4\tau$, $\sigma=0$ averages across $10$ network realizations).}
	\label{fig13}
\end{figure}

\section{Lyapunov spectrum of recurrent LSTM network} 
Training recurrent neural networks on tasks that involve long time lags with gradient-based methods is hampered by the loss of gradient information. Long short-term memory (LSTM) units were introduced to ameliorate this problem of vanishing or exploding gradients by adding a latent - potentially slow  - additional degree of freedom for each rate unit with dedicated input, output, and forget gates that conspire to retain information over extended time lags \cite{hochreiter_long_1997}. 
The dynamics of the LSTM units follow \cite{hochreiter_long_1997}:
\begin{eqnarray*}%
f_t &=& \sigma_g(U_{f} h_{t-1} + W_{f} x_t +  b_f) \\
o_t &=& \sigma_g(U_{o} h_{t-1} + W_{o} x_t + b_o) \\
i_t &=& \sigma_g(U_{i} h_{t-1} + W_{i} x_t + b_i) \\
\tilde{c}_t &=& \sigma_h(U_{c} h_{t-1} + W_{c} x_t +  b_c) \\
c_t &=& f_t \odot c_{t-1} + i_t \odot \tilde{c}_t \\
h_t &=& o_t \odot \sigma_h(c_t)
\end{eqnarray*}

where $ \odot$ denotes the Hadamard product, $\sigma(x) = \frac{1}{1 + \exp(-x)}$ is the sigmoid function, and entries $U_x$ are drawn from $U_x \sim \mathcal{N} (0,g_x^2/N)$ and the bias terms $b_x $ are scalars for simplicity. The full Lyapunov spectrum is again obtained by a reorthonormalization procedure of the Jacobians along a numerical solution of the map \cite{benettin_lyapunov_1980}. For details, see Appendix \ref{Algo}.
As a proof-of-concept, we calculate Lyapunov spectra of recurrent LSTM networks in the autonomous case $W_x=0$ (Fig.~\ref{fig15new}). 
\begin{figure}[!h]
	\includegraphics[width=1\columnwidth]{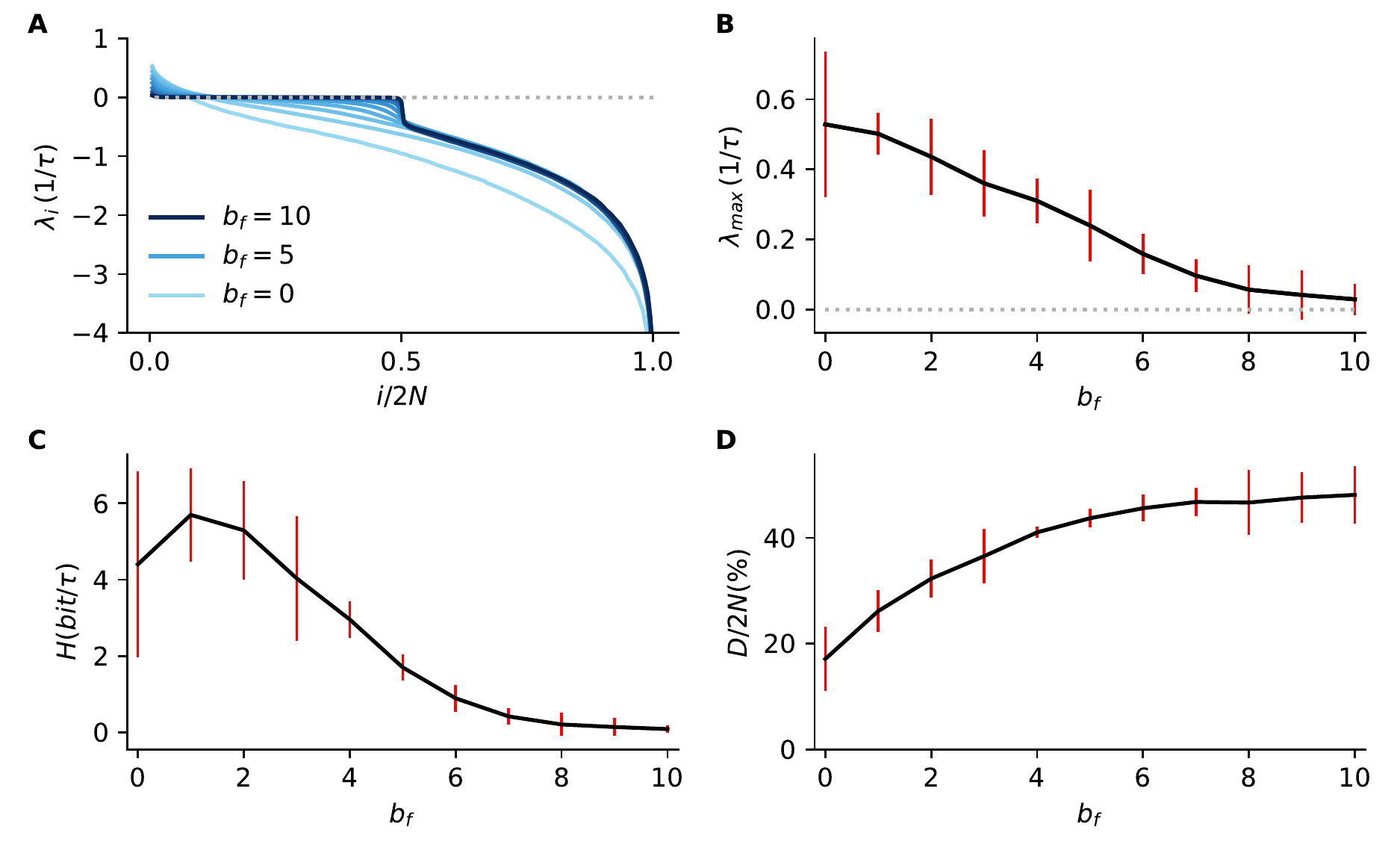}
	\caption{{\bf Lyapunov spectrum, dynamical entropy rate and dimensionality of recurrent LSTM network}
		{\bf A}~For increasing bias current in the forget gate $b_f$, the first half of the Lyapunov spectrum is increasingly pushed towards zero, concomitantly, the  autocorrelation of the latent state indicates the emergence of slow latent modes. {\bf B}~The largest Lyapunov exponent decreases for increasing $b_f$. {\bf C}~The dynamical entropy rate $H$ is also reduced for growing $b_f$. {\bf D}~In contrast, the relative attractor dimensionality $D/2N$ even increases with $b_f$, despite decreasing $\lambda_{\max}$.
		(Parameters: $N=200$, $\Delta t=\tau$, $t_\textnormal{ONS}=\tau$, $t_\textnormal{sim}=10^4\tau$, $g_f=g_o=g_i=g_c=3.0$, $b_o=b_i=b_c=0.0$, averages across $10$ network realizations).}
	\label{fig15new}
\end{figure}
We find that saturating the forget gates by increasing $b_f$ of the LSTM network results in slow latent modes and an accumulation of Lyapunov exponents close to 0 (Fig.~\ref{fig15new}A). For increasing bias current in the forget gate $b_f$, the first half of the Lyapunov spectrum is increasingly pushed towards zero, concomitantly the  autocorrelation of the latent state indicates the emergence of slow latent modes. This finding is consistent with previous theoretical work based on spectra of the state-to-state Jacobian $\mathbf{D}$ (analogous to Eq.~\ref{eq:-Jacobian} in our case) which suggested an accumulation of eigenvalues of the Jacobian close to $1$ for closed forget gates \cite{gilboa_dynamical_2019,can_gating_2020}. 
At the same time, the second half of the Lyapunov spectrum sharply drops to very negative values for increasing $b_f$, similarly to the classical $\tanh$ rate network with discrete-time dynamics (Fig.~\ref{fig6a}A), with no plateau from an intrinsic characteristic timescale (as for example the one coming from a leak term (Fig.~\ref{fig8}C), or a synaptic integration timescale, or adaptation current \cite{puelma_touzel_cellular_2016,harish_asynchronous_2015,muscinelli_how_2019}).    
Moreover, our results indicate that LSTM networks can have a high attractor dimensionality $D$ even in a weakly chaotic state when saturating the forget gates  (Fig.~\ref{fig15new}D), as a growing number of near-zero Lyapunov exponents is necessary to yield a total sum of zero. In contrast, the dynamical entropy rate $H$ is only governed by positive Lyapunov exponents, which do not reflect the large number of Lyapunov exponents close to zero for increasing $b_f$ (Fig.~\ref{fig15new}C). We find this phenomenon independent of network size $N$, which suggests extensive chaos (not shown), as in the classical rate networks (Fig.~\ref{fig3}). Driving each LSTM with independent Gaussian white noise process independent ($W_x\neq0$) into leads to a reduction of chaos, decreasing dynamical entropy rate $H$ and attractor dimensionality $D$ as expected from the case of the classical rate networks \cite{schuecker_optimal_2018} (Fig.~\ref{fig11}) (not shown). 

To what extent ou findings of the asymptotic dynamics of random LSTM networks can explain typical training scenarios with time-dependent external input, performance-optimized network structure, and finite time intervals remains to be determined in future studies.

\section{Relating gradients in backpropagation through time to the full Lyapunov spectrum} 
We found a direct mathematical link between features of the Lyapunov spectrum of the recurrent network dynamics and the problem of vanishing and exploding gradients when training with backpropagation through time.
In backpropagation through time, all connection weights are iteratively updated by stochastic gradient descent such that locally a loss is reduced \cite{werbos_beyond_1974,parker_learning-logic_1985,lecun_procedure_1985,rumelhart_learning_1986}. The gradient of the loss with respect to the weights of the recurrent network is evaluated by unrolling the network dynamics in time. The resulting expression for the gradient involves the long-term Jacobian, which is also used to calculate the Lyapunov spectrum (See Appendix \ref{gradientsBPTT_LS_fullSpectrum}). A common problem in training recurrent networks is that the gradients tend to vanish or to explode, especially in case of long temporal dependencies \cite{bengio_learning_1994,pascanu_difficulty_2012}.

The singular values of the long-term Jacobian, which determine how quickly gradients vanish or explode during backpropagation through time, are directly related to the Lyapunov exponents of the dynamics: The Lyapunov exponents are given by the logarithm of the singular values of the long-term Jacobian 
(See Appendix~\ref{gradientsBPTT_LS_fullSpectrum}). This justifies a couple of conclusions for backpropagation through time. To avoid diverging or vanishing gradients, recurrent networks should be initialized such that many singular values of the long-term Jacobian are close to one \cite{saxe_exact_2013,hanin_products_2018,schoenholz_deep_2016,chen_dynamical_2018,gilboa_dynamical_2019,can_gating_2020}, which amounts to having many Lyapunov exponents of the forward dynamics close to zero, corresponding to slowly growing/shrinking directions in tangent space. As the product of Jacobian is generally numerically ill-conditioned, we suggest using the orthonormalization procedure discussed here to quantify the stability of the tangent space.
Furthermore, the trainability of RNNs, as quantified by the maximum time difference a recurrent neural network can be trained across using BPTT before running into vanishing/exploding gradients can be quantified by Lyapunov exponents of the forward dynamics. We can thus use Lyapunov exponents to compare the effect of different initializations, nonlinearities, and optimizers on trainability.
We predict that after learning long-term dependencies, there should be some Lyapunov exponents close to zero reflecting the slow timescales.
\section{Discussion}
\subsection{Summary}

We used canonical measures from the ergodic theory of strange attractors to characterize the chaotic dynamics of randomly wired networks of firing-rate units. This is to our knowledge the first time the full Lyapunov spectrum of a continuous-time random rate network has been calculated and used to study dynamical entropy rate and attractor dimensionality.

We showed that, in the classical model, dynamical entropy rate and relative attractor dimensionality first grow and then saturate for as a function of coupling strength $g$. Thus, both the intensity and diversity of network activity states saturates for strong coupling, despite a monotonously growing largest Lyapunov exponent. We analytically approximated the full Lyapunov spectrum in several limiting cases using random matrix theory. We found that time-varying input reduces both entropy and dimensionality.

We demonstrated that the shape of the Lyapunov spectrum is size invariant and exhibits a linear growth of attractor dimensionality and entropy rate with network size~$N$. This is clear evidence of extensive chaos, which was previously conjectured in \cite{sompolinsky_chaos_1988}. We further found the Lyapunov spectrum to be point-symmetric around the mean Lyapunov exponent $-1/\tau$, which we derived analytically (Appendix \ref{RMTmeanLE}). Note that the symmetry would be around $-1$ if time is measured in units of $\tau$. A symmetry of Lyapunov spectra around zero is usually found in dynamical systems with a symplectic structure \cite{dressler_symmetry_1988,wojtkowski_conformally_1998}. Symmetry around a negative value was previously described in a class of dissipative dynamical systems with viscous damping \cite{dressler_symmetry_1988}. 

We found a strong effect of time-discretization: increasing the step size breaks the symmetry of the Lyapunov spectrum and sharply increases the entropy rate and dimensionality. This has methodological implications for further studies: The parallel update of chaotic discrete-time network dynamics has fundamentally different properties than the continuous-time limit. 
 
In balanced networks of threshold-linear units, we found that both the entropy rate and attractor dimensionality increase for small values of $g$, as in the classical model. For large values of $g$, first the attractor dimensionality and later also the entropy rate peak as a function of $g$. Different from the classical $\tanh$ model, the Lyapunov spectrum is not point-symmetric.

Time-dependent input reduced both the entropy rate and attractor dimensionality. For strong input, we found that all trajectories collapsed to a time-dependent random sink. If the input is interpreted as an input signal, this means that trajectories are reliable across repetitions of the same frozen input realization and do not depend on the initial conditions of the recurrent network.

Finally, we showed that Lyapunov spectra are a useful tool to characterize dynamic stability properties of trained networks and to analyze the solution trajectories without assuming fixed points or 'slow points' as done, for instance, in \cite{sussillo_opening_2012,rivkind_local_2017,beer_one_2018,haviv_understanding_2019}. Moreover, we show a direct link between the Lyapunov exponents of the forward dynamics and the gradient stability when training recurrent neural networks with backpropagation through time, which constrains trainability and stability.

\subsection{Relation to previous work}
Firing-rate networks can generate spontaneous rate-fluctuations by recurrent chaotic dynamics \cite{sompolinsky_chaos_1988}. Mechanisms underlying rate chaos have attracted substantial attention in studies of network heterogeneity \cite{aljadeff_transition_2015}, bistability \cite{stern_dynamics_2014}, external stimuli \cite{molgedey_suppressing_1992,rajan_inferring_2010,rajan_stimulus-dependent_2010, schuecker_optimal_2018} and the role of the single unit transfer function \cite{kadmon_transition_2015} and slow synaptic dynamics \cite{harish_network_2013,harish_asynchronous_2015} for the collective network state. See also, e.g., \cite{doyon_control_1993,ostojic_two_2014,wainrib_topological_2013,mastrogiuseppe_intrinsically-generated_2017,garcia_del_molino_synchronization_2013,cabana_large_2013,massar_mean-field_2013,engelken_reanalysis_2016}.
Our approach provides a toolkit from dynamical systems theory to analyze how these different factors shape the complex rate dynamics. 

We compared the attractor dimension with a dimensionality estimate based on principal component analysis, which is commonly used in neuroscience \cite{rajan_inferring_2010,rajan_stimulus-dependent_2010,gao_simplicity_2015,ganguli_compressed_2012,cunningham_dimensionality_2014}. We find a qualitatively similar but quantitatively different behavior of the PCA-based dimensionality and the attractor dimension: both saturate with synaptic strength for $g>1$ but they saturate at different levels and with distinct rates.
Note that Lyapunov exponents and the attractor dimension are invariant under diffeomorphisms of the phase space \cite{kuznetsov_invariance_2016}, while PCA-based dimensionality estimates are generally not invariant with respect to changes of coordinates and can be misleading for limited data sets \cite{monasson_estimating_2015}. Because the PCA-based dimensionality estimates are based on a two-point correlation function, they miss low-dimensional structure hidden in higher-order correlations. Generally, the PCA-based dimensionality can both under- and overestimate the attractor dimensionality.

Our approach allows interpolation from continuous-time to discrete dynamics. Discrete-time dynamics of rate networks has previously been studied in random diluted networks \cite{doyon_control_1993}, noise-driven networks \cite{molgedey_suppressing_1992} and on a ring network \cite{bauer_lyapunov_1991,aljadeff_low-dimensional_2016}. 

Chaotic rate dynamics provide a substrate for complex nonlinear computations, such as learning input-output relations \cite{maass_real-time_2002,sussillo_generating_2009,toyoizumi_beyond_2011,rivkind_local_2017,abbott_building_2016,barak_fixed_2013}
and learning temporal sequences \cite{laje_robust_2013}. Intriguingly, transient rate chaos yields exponential expressivity in deep networks, which has been explained by transient chaos across layers \cite{poole_exponential_2016}. Our tools facilitate the quantification of the reorganization of the collective network dynamics during learning and the underlying mechanisms of different computing strategies.

A suppression of chaos by time-dependent input was studied previously, both with white noise input in discrete-time \cite{molgedey_suppressing_1992} and continuous-time networks \cite{ schuecker_optimal_2018} and with sinusoidal input \cite{rajan_stimulus-dependent_2010}. Such a transition has relevance for information processing because the network loses its dependence on initial conditions, which is expected to affect the ability of a network to generate controlled output trajectories in response to certain input patterns after learning \cite{jaeger_harnessing_2004,sussillo_generating_2009,laje_robust_2013}.
A transition to complete control by an external stimulus and concomitant independence of initial conditions was previously studied in rate networks in the context of echo state networks for reservoir computing and termed the echo state property \cite{jaeger_echo_2001,buehner_tighter_2006,jaeger_echo_2007,manjunath_echo_2012}.

An accumulation of Lyapunov exponents close to zero is consistent with previous theoretical work based on spectra of the state-to-state Jacobian $\mathbf{D}$ (analogous to Eq.~\ref{eq:-Jacobian} in our case) which suggested an accumulation of eigenvalues of the Jacobian close to $1$ for closed forget gates \cite{gilboa_dynamical_2019,can_gating_2020}.

\subsection{Outlook}
We are only beginning to use ergodic theory to understand neural computation. By employing these concepts in large-scale rate networks, we have laid a foundation for further investigation. Computational ergodic theory of firing-rate networks is currently the only way to measure information-theoretic quantities in large recurrent circuits. It is an important challenge to obtain a more comprehensive understanding of how different factors shape collective network dynamics.

The link between firing-rate networks and spiking neural networks has been studied by investigating networks in the limit of very slow synaptic dynamics. In this limit, the synaptic input current integrates over a long time, and the network dynamics is analogous to a rate network \cite{harish_network_2013} with quantitatively similar activity fluctuations. An interpolation from spiking to rate dynamics with increasing $\tau_{s}$ and a comparison of the associated Lyapunov spectra of rate and spiking networks might improve our understanding of chaos both in spiking and rate networks.

Collective network dynamics is expected to be strongly shaped by wiring, and learning algorithms operate by modifying connectivity. Investigating how features of connectivity shape the dynamics is therefore important and could also be investigated with these tools. The role of an excess of bidirectional connections \cite{sommers_spectrum_1988}, other second-order motifs \cite{zhao_synchronization_2011} 
and strong self-coupling \cite{stern_dynamics_2014} could all be examined. 
A time-resolved analysis of dynamic stability of the recurrent network dynamics using covariant Lyapunov vectors and local Lyapunov exponents can also help to understand the mechanisms of learning, and under what conditions the training of recurrent networks fails.

\begin{acknowledgments}
We thank A. Crisanti, F. Fumarola, S. Goedeke, J. Kadmon, R. Khajeh, G. Lajoie, J. Liedtke, L. Logiaco, U. Parlitz, M. Schottdorf, M. Stern, H. Sompolinsky and M. Timme for fruitful discussions. Research supported by NSF NeuroNex Award (DBI-1707398), the Gatsby Charitable Foundation (GAT3708), the Simons Collaboration for the Global Brain (542939SPI), and the Swartz Foundation (2019-5).
\end{acknowledgments}

\appendix

\section*{Appendix: Overview}\label{secMM}
We first give a brief summary of the mathematical foundations of Lyapunov spectra (\ref{LSmath}) and our concrete implementation for rate networks (\ref{Algo}). We check the convergence of the Lyapunov spectra with various system parameters (\ref{LSconvergence}). We extend the approach to random dynamical systems and discuss the implementation of Lyapunov spectra for non-autonomous networks with time-dependent input (\ref{LSdriven}). Then we give details about the PCA-based dimensionality estimate (\ref{PCAdim}). Finally, we approximate the mean Lyapunov exponent analytically (\ref{RMTmeanLE}).

\section{Lyapunov spectrum of a dynamical system\label{LSmath}}
An autonomous dynamical system is usually defined by a set of ordinary
differential equations $\dif \mathbf{x}/\dif t=\mathbf{F}(\mathbf{x}),\;\mathbf{x}\in\mathbb{R}^{N}$
in the case of continuous-time dynamics, or as a map $\mathbf{x}_{s+1}=\mathbf{f}(\mathbf{x}_s)$
in the case of discrete-time dynamics. In the following, the theory is presented for discrete-time dynamical systems for ease of notation, but everything directly extends to continuous-time systems \cite{geist_comparison_1990}. In our specific case, we study the discrete network dynamics for small $\Delta t$. This corresponds to the usual Euler method in the autonomous case or to the Euler-Maruyama method \cite{kloeden_numerical_1992} in the non-autonomous case with stochastic input drive. We confirm our autonomous results for small $\Delta t$ using the Tsitouras 5/4 Runge-Kutta method, the Dormand-Prince 5/4 Runge-Kutta method, the Bogacki-Shampine 5/4 Runge-Kutta method, and Verner's 9/8 Runge-Kutta method \cite{tsitouras_rungekutta_2011,dormand_family_1980,bogacki_efficient_1996,verner_numerically_2010} employing the implementation provided by the DifferentialEquations.jl package in the programming language Julia \cite{rackauckas_differentialequations.jl_2017,bezanson_julia:_2017}.
Together with an initial condition $\mathbf{x}_{0}$, the map forms a trajectory. As a natural extension of linear stability analysis, one can ask how an infinitesimal perturbation
$\mathbf{x}'_{0}=\mathbf{x}_{0}+\epsilon\mathbf{u}_{0}$ evolves in
time. Chaotic systems are sensitive to initial conditions; almost all infinitesimal perturbations $\epsilon\mathbf{u}_{0}$ of
the initial condition grow exponentially $|\epsilon\mathbf{u}_{t}|\approx \exp(\lambda_{\max} t)|\epsilon\mathbf{u}_{0}|$. Finite-size perturbations therefore may lead to a drastically different subsequent behavior. The largest Lyapunov exponent $\lambda_{\max}$ measures
the average rate of exponential divergence or convergence of nearby
initial conditions:

\begin{equation}
\lambda_{\max}(\mathbf{x}_{0})=\lim_{t\to\infty}\frac{1}{t}\lim_{\epsilon\to0}\log\frac{||\epsilon\mathbf{u}_{t}||}{||\epsilon\mathbf{u}_{0}||}
\end{equation}
In dynamical systems that are ergodic on the attractor, the Lyapunov exponents do not depend on the initial conditions, as long as the initial conditions are in the basins of attraction of the attractor. Note that it is crucial to first take the limit $\epsilon\to0$ and then $t\to\infty$,
as $\lambda_{\max}(\mathbf{x}_{0})$ would be trivially zero for a
bounded attractor if the limits are exchanged, as $\lim_{t\to\infty}\log\frac{||\epsilon\mathbf{u}_{t}||}{||\epsilon\mathbf{u}_{0}||}$
is bounded for finite perturbations even if the system is chaotic.
To measure \textbf{$m$} Lyapunov exponents, one has to study the
evolution of \textbf{$m$} independent infinitesimal perturbations $\mathbf{u}_{s}$
spanning the tangent space:

\begin{equation}
\mathbf{u}_{s+1}=\mathbf{D}_{s}\mathbf{u}_{s}
\end{equation}
where the $N\times N$ Jacobian \textbf{$\mathbf{D}_{s}(\mathbf{x_{s}})=\dif \mathbf{f}(\mathbf{x_{\mathbf{s}}})/\dif \mathbf{x}$}
characterizes the evolution of generic infinitesimal perturbations
during one step. Note that this Jacobian along the trajectory is equivalent to a stability matrix only at a fixed point, i.e., when $\mathbf{x}_{s+1}=\mathbf{f}(\mathbf{x}_s)=\mathbf{x}_s$. 

We are interested in the asymptotic behavior, and therefore we study the long-term Jacobian

\begin{equation}
\mathbf{T}_{t}(\mathbf{x}_{0})=\mathbf{D}_{t-1}(\mathbf{x}_{t-1})\dots\mathbf{D}_{1}(\mathbf{x}_{1})\mathbf{D}_{0}(\mathbf{x}_{0}).\label{eq:long-term-Jacobian}
\end{equation}
Note that \textbf{$\mathbf{T}_{t}(\mathbf{x}_{0})$} is a product
of generally noncommuting matrices. 
The Lyapunov exponents $\lambda_{1}\geq\lambda_{2}\dots\geq\lambda_{N}$
are defined as the logarithms of the eigenvalues of the Oseledets matrix
\begin{equation}
\boldsymbol{\Lambda}(\mathbf{x}_{0})=\lim_{t\to\infty}[\mathbf{T}_{t}(\mathbf{x}_{0})^{\top}\mathbf{T}_{t}(\mathbf{x}_{0})]^{\frac{1}{2t}},\label{eq:-Oseledets}
\end{equation}
where $\top$ denotes the transpose operation. The expression inside
the brackets is the Gram matrix of the long-term Jacobian $\mathbf{T}_{t}(\mathbf{x}_{0})$.
Geometrically, the determinant of the Gram matrix is the squared volume
of the parallelotope spanned by the columns of $\mathbf{\mathbf{T}_{t}(\mathbf{x}_{0})}$. Thus, the exponential volume growth rate is given by the sum of the logarithms of its first $m$ (sorted) eigenvalues. 
Oseledets' multiplicative ergodic theorem guarantees the existence of the Oseledets matrix $\boldsymbol{\Lambda}(\mathbf{x}_{0})$ for almost all initial conditions\textbf{ $\mathbf{x}_{0}$} \cite{oseledets_multiplicative_1968}. In ergodic systems, the Lyapunov exponents $\lambda_{i}$ do not depend on the initial condition $\mathbf{x}_{0}$. However, for a numerical calculation of the Lyapunov spectrum, Eq.~\ref{eq:-Oseledets} cannot be used directly because the long-term Jacobian $T_{t}(\mathbf{x}_{0})$ quickly becomes ill-conditioned, i.e., the ratio between its largest and smallest singular value diverges exponentially with time.

\section{Algorithm for calculating Lyapunov spectrum of rate networks\label{Algo}}
For calculating the first $m$ Lyapunov exponents, we exploit the fact that the growth rate of an $m$-dimensional infinitesimal volume element is given by $\lambda^{(m)}=\sum_{i=1}^{m}\lambda_{i}$. Therefore, $\lambda_{1}=\lambda^{(1)}$, $\lambda_{2}=\lambda^{(2)}-\lambda_{1}$, $\lambda_{3}=\lambda^{(3)}-\lambda_{1}-\lambda_{2}$, \dots \cite{benettin_lyapunov_1980}. The volume growth rates can be obtained via QR-decomposition. 

First, one needs to evolve an orthonormal basis $\mathbf{Q}_{s}=[\mathbf{q}_{s}^{1},\,\mathbf{q}_{s}^{2},\dots\mathbf{q}_{s}^{m}]$
in time using the Jacobian $\mathbf{D}_{s}$:
\begin{figure*}
	\includegraphics[width=2\columnwidth]{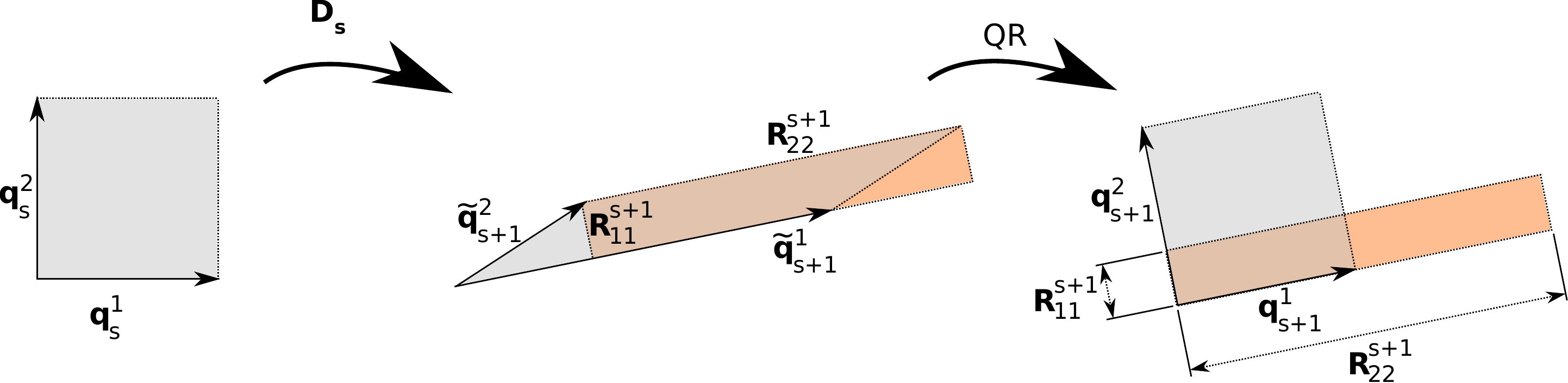} 	
	\vspace{0.3cm}\caption{\label{fig:fig14}\textbf{Geometric illustration of Lyapunov spectrum calculation.} An orthonormal matrix $\mathbf{Q}_{s}=[\mathbf{q}_{s}^{1},\,\mathbf{q}_{s}^{2},\dots\mathbf{q}_{s}^{m}]$, whose columns are the axes of an $m$-dimensional cube, is rotated
		and distorted by the Jacobian $\mathbf{D}_{s}$ into an $m$-dimensional parallelotope $\widetilde{\mathbf{Q}}_{s+1}=\mathbf{D}_{s}\mathbf{Q}_{s}$
		embedded in $\mathbf{\mathbb{R}^{N}}$. The figure illustrates this for $m=2$, in which case the columns of $\widetilde{\mathbf{Q}}_{s+1}$ span a parallelogram, which can be divided into a right triangle and a trapezoid and rearranged into a rectangle. Thus, the area of the gray parallelogram is the same as that of the orange rectangle. The QR-decomposition reorthonormalizes $\widetilde{\mathbf{Q}}_{s+1}$ by decomposing it into the product of an orthonormal matrix $\mathbf{Q}_{s+1}=[\mathbf{q}_{s+1}^{1},\,\mathbf{q}_{s+1}^{2},\dots\mathbf{q}_{s+1}^{m}]$
		and the upper-triangular matrix $\mathbf{R}^{s+1}$. $\mathbf{Q}_{s+1}$
		describes the rotation of $\mathbf{Q}_{s}$ caused by $\mathbf{D}_{s}$.
		The diagonal entries of $\mathbf{R}^{s+1}$ gives the stretching/shrinking
		along the columns of $\mathbf{Q}_{s+1}$, thus the volume of the parallelotope
		formed by the first $m$ columns of $\widetilde{\mathbf{Q}}_{s+1}$
		is given by $V_{m}=$$\prod_{i=1}^{m}\mathbf{R}_{ii}^{s+1}$. The time-averaged
		logarithms of the diagonal elements of $\mathbf{R}^{s}$ give the
		Lyapunov spectrum: $\lambda_{i}=\lim_{t_{\textnormal{sim}}\to\infty}\frac{1}{t_{\textnormal{sim}}}\log\prod_{s=1}^{t}\mathbf{R}_{ii}^{s}=\lim_{t_{\textnormal{sim}}\to\infty}\frac{1}{t}\sum_{s=1}^{t}\log\mathbf{R}_{ii}^{s}$.}
	\label{fig14}
\end{figure*}
\begin{equation}
\widetilde{\mathbf{Q}}_{s+1}=\mathbf{D}_{s}\mathbf{Q}_{s}
\end{equation}
Second, the volume growth rates are obtained by applying a QR-decomposition 
\begin{equation}
\widetilde{\mathbf{Q}}_{s+1}=\mathbf{Q}_{s+1}\mathbf{R}^{s+1}
\end{equation}
As a result of this, the non-orthonormal matrix $\widetilde{\mathbf{Q}}_{s+1}$
is uniquely decomposed into an orthonormal matrix $\mathbf{Q}_{s+1}$
of size $N\times m$ so $\mathbf{Q}^{\top}_{s+1}\mathbf{Q}_{s+1}=\mathds{1}_{m\times m}$
and to an upper triangular matrix $\mathbf{R}^{s+1}$ of size $m\times m$
with positive diagonal elements. 

Geometrically, $\mathbf{Q}_{s+1}$ describes the rotation of $\mathbf{Q}_{s}$
caused by $\mathbf{D}_{s}$ and the diagonal entries of $\mathbf{R}^{s+1}$
describe the stretching or shrinking of $\mathbf{Q}_{s}$, while
the off-diagonal elements describe the shearing. Fig.~\ref{fig14}
visualizes $\mathbf{D}_{s}$ and the QR-decomposition for $m=2$.
The Lyapunov exponents are given by time-averaged
logarithms of the diagonal elements of $\mathbf{R}^{s}$:

\begin{equation}
\lambda_{i}=\lim_{t\to\infty}\frac{1}{t}\log\prod_{s=1}^{t}\mathbf{R}_{ii}^{s}=\lim_{t\to\infty}\frac{1}{t}\sum_{s=1}^{t}\log\mathbf{R}_{ii}^{s}
\end{equation}
Note that the QR-decomposition does not need to be performed in every simulation
step, just sufficiently often that $\widetilde{\mathbf{Q}}_{s+w}=\mathbf{D}_{s+w-1}\cdot\mathbf{D}_{s+w-2}\dots\mathbf{D}_{s}\cdot\mathbf{Q}_{s}$
is well-conditioned \cite{benettin_lyapunov_1980}. An appropriate reorthonormalization interval
$w_{\textnormal{ONS}}=t_\textnormal{ONS}/\Delta t$ thus depends on the condition number, 
the ratio of the smallest and largest singular value:
\begin{equation}
\kappa_{2}(\widetilde{\mathbf{Q}}_{s+w})=\kappa_{2}(\mathbf{R}^{s+w})=\frac{\sigma_{1}(\mathbf{R}^{s+w})}{\sigma_{m}(\mathbf{R}^{s+w})}=\frac{\mathbf{R}_{11}^{s+w}}{\mathbf{R}_{mm}^{s+w}}.
\end{equation}
Therefore, the condition number can be estimated based on the ratio
of the largest and smallest Lyapunov exponent that is calculated:
$\kappa_{2}(\widetilde{\mathbf{Q}}_{s+w})\approx\exp\left(\lambda_{1}-\lambda_{m}\right)$.
Thus, an appropriate reorthonormalization interval is given by $t_\textnormal{ONS}=\mathcal{O}\left(\log(\hat{\kappa}_{2})/(\lambda_{1}-\lambda_{m})\right)$,
where $\hat{\kappa}_{2}$ is some acceptable condition number. The
acceptable condition number depends on the desired accuracy of the
entries of $\mathbf{R}^{s+w}$. As the dynamical system first has to converge onto the attractor and the initially random orthonormal basis is not aligned, i.e., the first vector does not point in the direction of the first covariant Lyapunov vector, and so on, an initial transient should be discarded. It is guaranteed that under general conditions initially random orthonormal systems will exponentially converge towards a unique basis that is given by the eigenvectors of the Oseledets matrix Eq.~\ref{eq:-Oseledets} \cite{ershov_concept_1998}. A minimal example of this algorithm in pseudocode is shown in the main text (see \ref{pseudocode}). A feasible strategy to determine $t_\textnormal{ONS}$ is to get first a rough estimate of the Lyapunov spectrum using a short simulation time $t_{\textnormal{sim}}$ and a small $t_\textnormal{ONS}$ and repeat with a longer simulation time and a $t_\textnormal{ONS}$ based on the Lyapunov spectrum of the rough estimate of the Lyapunov spectrum. Another strategy is, to first iteratively adapt $t_\textnormal{ONS}$ on a short simulation run to get a condition number that is acceptable.

\section{Convergence of the Lyapunov spectrum\label{LSconvergence}}
We checked the convergence of the Lyapunov spectrum as a function of different simulation parameters. 
First, the Lyapunov exponents were checked to converge with simulation time $t_\textnormal{sim}$ (Fig.~\ref{figConvergenceT}). Figure \ref{fig9} shows the temporal convergence of selected Lyapunov exponents for ten random network realizations for different values of $g$ and $\sigma$. The Lyapunov spectra were independent of initial conditions but showed some variability across different realizations of the random network structure. There are two main contributions to the variability of numerically calculated Lyapunov spectra, finite-time sampling noise and quenched fluctuations. Indeed, Lyapunov exponents are asymptotic properties numerically estimated from finite time calculations. Variability also arises from the quenched disorder in different random network realization. The first contribution would vanish in the limit of long simulations for ergodic systems. The second contribution is expected to vanish in the large network limit due to self-averaging. Quantities that are self-averaging converge in the limit of large system size to the ensemble average.
\begin{figure}[!h]
	\includegraphics[width=1\columnwidth]{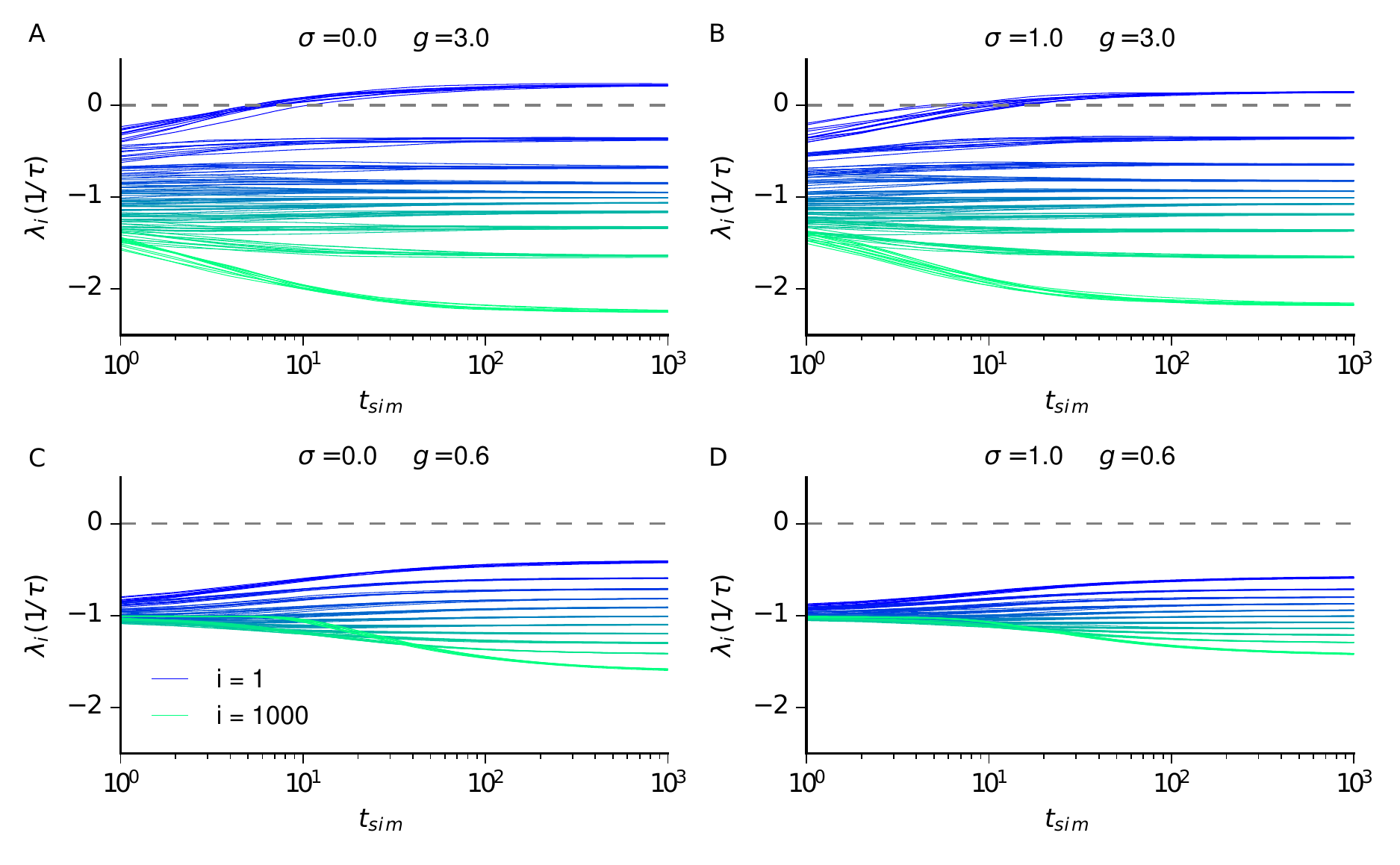}
	\caption{{\bf Convergence of Lyapunov spectrum with simulation time $t_\textnormal{sim}$.} 
		{\bf A}~Convergence of selected Lyapunov exponents $\lambda_i$ for ten different network realizations with simulation time (in units of $\tau$)($i = 1,100,200,\dots1000$) for $\sigma=0$ and $g=3$. {\bf B}~Same as top left, but for $\sigma=1$ and $g=3$. {\bf C}~$\sigma=0$ and $g=0.6$. {\bf D}~$\sigma=1$ and $g=0.6$.
		(Other parameters: $N=1000$, $\Delta t=0.01\tau$, $t_\textnormal{sim}=10^4\tau$, $t_\textnormal{ONS}=\tau$).}
	\label{figConvergenceT}
\end{figure}

Second, we confirmed that the orthonormalization interval was chosen sufficiently small (Fig.~\ref{fig16}A). If the reorthonormalization is not carried out sufficiently often, the long-term Jacobian $\mathbf T_{t}(\mathbf{x}_{0})$ becomes ill-conditioned. As a consequence, the orthonormalization becomes numerically unstable, and errors start to accumulate. This results in a flattening of the Lyapunov spectrum beginning at small Lyapunov exponents (Fig.~\ref{fig16}A, D). As described above, a suitable orthonormalization interval inversely scales with the difference between smallest and largest Lyapunov exponent that is calculated $|\lambda_{\max}-\lambda_k|$. Therefore, it is no surprise that for large $\Delta t$, the errors in the Lyapunov spectrum grow faster with $t_\textnormal{ONS}$ (Fig.~\ref{fig16}C, D), because the difference $|\lambda_1-\lambda_k|$ is larger (Fig.~\ref{fig16}A).

Third, we checked convergence with the integration time step $\Delta t$ (Fig.~\ref{fig5}A). For large $g$, the integration time step $\Delta t$ has to be chosen smaller, because the autocorrelation of the Jacobians become very short ($\tau_{AC}\ll \tau$), although the autocorrelation of the dynamics variables $h_i$ stays finite even for $g\rightarrow \infty$ \cite{sompolinsky_chaos_1988,crisanti_path_2018-1}.

Fourth, we confirmed the convergence of the shape of the Lyapunov spectrum for large network size $N$ (Fig.~\ref{fig3}B). Note that even for very small $\Delta t$, there exists a small asymmetry in the Lyapunov spectrum because of the neutral Lyapunov exponent ($\lambda_i=0$). Thus, the Lyapunov spectrum is only symmetric in the limits $N\rightarrow \infty$ and $\Delta t\rightarrow 0$.

\begin{figure}[!h]
	\includegraphics[width=1\columnwidth]{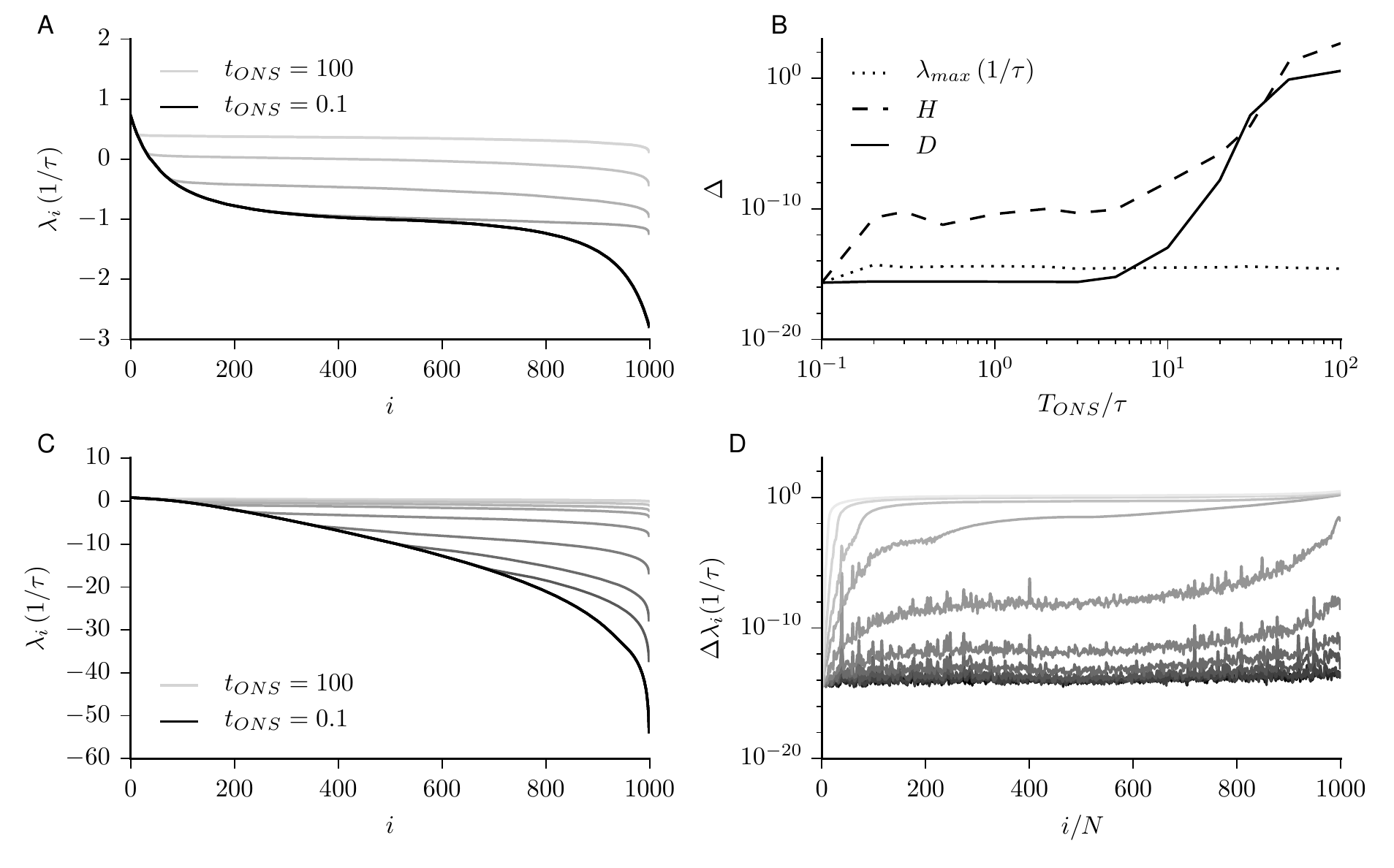}
	\caption{{\bf Convergence of Lyapunov spectrum with reorthonormalization interval $t_\textnormal{ONS}$.} 
		If the reorthonormalization is not performed sufficiently often, the Lyapunov spectrum is flattening from the end for large $t_\textnormal{ONS}$.
		{\bf A}~Lyapunov spectra for $t_\textnormal{ONS}\in \{0.1, 0.2, 0.3, 0.5, 1,2,5,10,20,50,100\}\tau$ for $\Delta t =0.01$. {\bf B}~$\Delta \lambda_{\max}$ shows the deviation of the largest Lyapunov exponent for different $t_\textnormal{ONS}$ from the smallest $t_\textnormal{ONS}=0.1\tau$. The same is shown for $H$ and $D$. For our typical parameter sets, an orthonormalization interval of $t_\textnormal{ONS}= 1\tau$ is sufficient to keep errors in $H$ and $D$ orders of magnitudes smaller than the deviations across network realizations due to quenched fluctuations. {\bf C}~Same as {\bf A}~for $\Delta t =1$. {\bf D}~Deviations of full Lyapunov spectra for different $t_\textnormal{ONS}$ from the smallest $t_\textnormal{ONS} =0.1$ for $\Delta t =0.01$. (Other parameters: $N=1000$, $\Delta t=0.01\tau$, $g=10$, $t_\textnormal{sim}=10^4\tau$, averages across 10 network realizations).}
	\label{fig16}
\end{figure}

Fifth, we confirmed numerically that the neutral Lyapunov exponent ($\lambda_i=0$) associated to a perturbation in the direction of the flow converges towards zero in the limit of small $\Delta t$ (not shown).

Sixth, we confirmed numerically that the Lyapunov spectrum does not depend on the realization of the initially random orthonormal system. Dependence in the realization of the orthonormal system would indicate that the ONS did not converge to the eigenvectors of the Oseledets matrix Eq.~\ref{eq:-Oseledets} \cite{ershov_concept_1998} (not shown).

Seventh, for large $N$, the numerical estimate of the largest Lyapunov exponent can be compared to one calculated analytically using dynamic mean-field theory \cite{sompolinsky_chaos_1988,kadmon_transition_2015,molgedey_suppressing_1992,schuecker_optimal_2018,crisanti_path_2018-1} (Fig.~\ref{fig16a}A,C).

Eighth, we confirmed that the Lyapunov spectrum does not systematically change when increasing the floating-point precision by using arbitrary precision floating point arithmetic in spot checks (not shown). 

\section{Finite network size effects on the transition to chaos and Lyapunov exponents}\label{finiteSize}

We described so far chaos in large firing rate networks. Here, we investigated the finite network size effect on the largest Lyapunov exponent and the critical coupling strength $g_{\textnormal{crit}}$, where the transition to chaos occurs. We calculated for the largest Lyapunov exponent of the classical random rate networks with $\tanh$-nonlinearity as a function of network size $N$ for $100$ network realizations per size. We found that the largest Lyapunov exponent for small networks exhibits a large diversity across network realizations (Fig.~\ref{fig16a}A). For increasing network size, the median Lyapunov exponent increases and approaches an asymptotic limit for large $N$. At the same time, the diversity as quantified by the 20\% and 80\% percentile across the network realizations vanishes. This indicates that for large network size $N$, the variability of Lyapunov exponents coming from the quenched disorder of different network realizations vanishes, and the Lyapunov exponent becomes independent of network realization.

Complementary, we calculated for different realizations and different network size $N$ the critical coupling strength $g_{\textnormal{crit}}$ where the network turns chaotic as indicated by the largest Lyapunov exponent using a noisy bisection method. 
For small networks, we found a broad diversity of $g_{\textnormal{crit}}$ (Fig.~\ref{fig16a}B). For many small random networks, we could not find a chaotic regime at all. For increasing values of $N$, the median $g_{\textnormal{crit}} -1 $ across 100 realizations decreased $\propto 1/\sqrt{N}$ from a median $g_{\textnormal{crit}}\approx 4$ for $N=50$ to $g_{\textnormal{crit}}\approx 1.1$ for $N=10^4$, and the diversity of $g_{\textnormal{crit}}$ as quantified by the $20\%$ and $80\%$ percentile across the network realizations shrank $\propto 1/\sqrt{N}$. This indicates that for large $N$, the coupling strength $g_{\textnormal{crit}} $ converges to 1 and the variability arising from quenched fluctuations disappears. Note that for small networks, there exist not necessarily a unique critical coupling strength $g$, so details of Fig.~\ref{fig16a}B may depend on the bisection scheme utilized.

\begin{figure}[!h]
\includegraphics[width=\columnwidth]{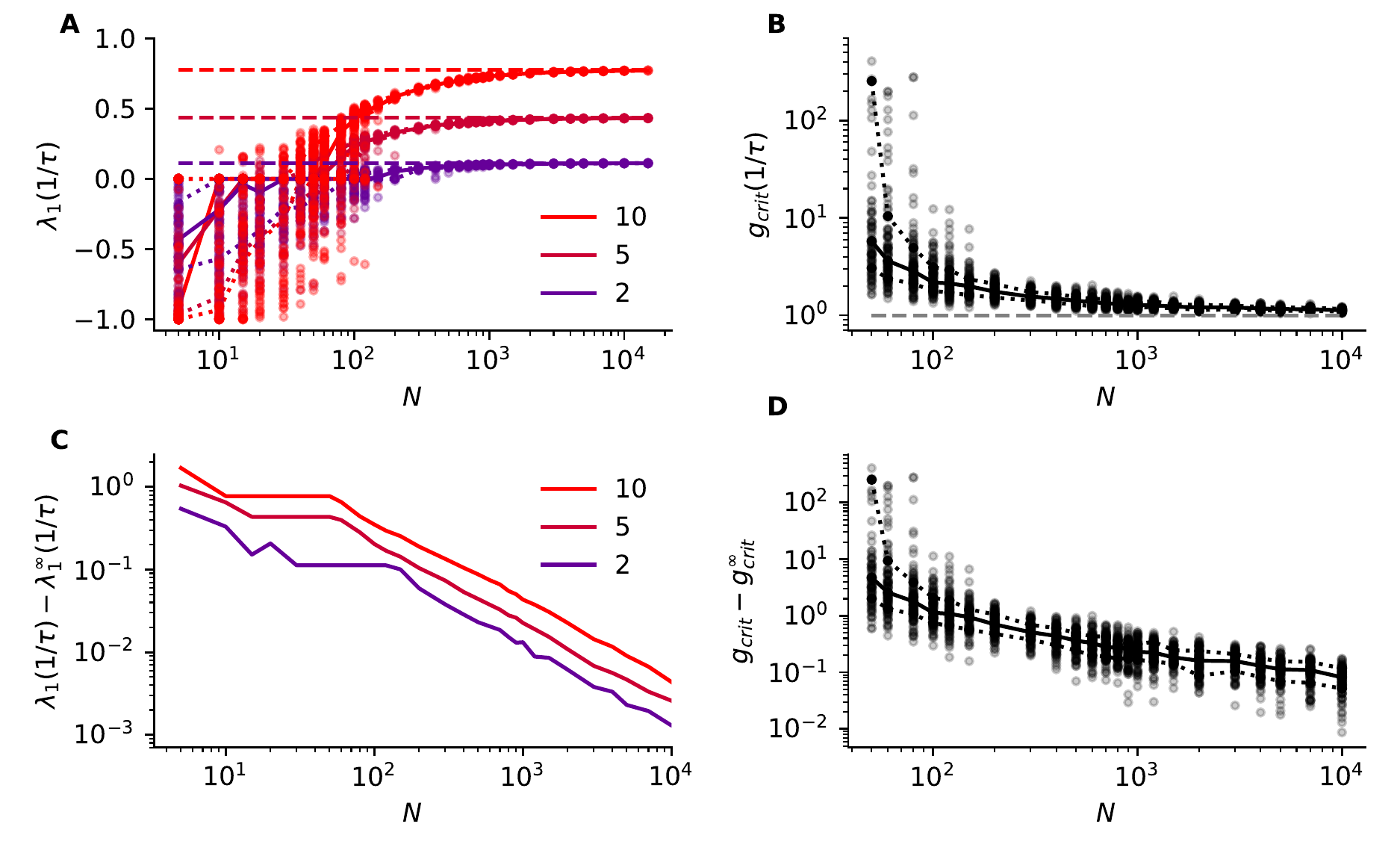}
		
	\caption{{\bf Finite-size effect on the largest Lyapunov exponent and transition to chaos}
		{\bf A}~Largest Lyapunov exponent across 100 network realizations as a function of network size $N$ for $g\in \{2, 5, 10\}$. Dots indicate individual realizations, full line are median, dotted curves are 20\% and 80 \% percentile, dashed lines are the prediction obtained from dynamic mean-field theory where $g$ is color-coded from blue (small $g$) to red (large $g$). {\bf B}~Critical coupling strength $g_{\textnormal{crit}}$ as a function of network size $N$ for 100 network realizations obtained by bisection method. Dots indicate individual realizations, full line are median, dotted curves are 20\% and 80 \% percentile, dashed line the analytical prediction. {\bf C}~Difference between mean-field theory prediction and median across 100 realizations as a function of network size. {\bf D}~Same data as {\bf B}~but $g_{\textnormal{crit}}-1$ depicted on $\log$-scale (Other parameters: $\textnormal{relative tolerance}=10^{-10}$, $t_\textnormal{sim}=10^4\tau$, $t_\textnormal{ONS}=\tau$, median across 100 network realizations).}
	\label{fig16a}
\end{figure}

\section{Kolmogorov-Sinai entropy rate and Kaplan Yorke attractor dimensionality}\label{KSE_KYD}
\paragraph{Entropy rate}
Chaos of a dynamical system is always associated with a dynamical entropy rate because nearby states, which could not be distinguished by a finite precision readout, are pulled apart by the sensitive dependence on initial conditions \cite{shaw_strange_1981}. This concept was formalized by Kolmogorov and Sinai in 1959 and termed metric entropy (also called Kolmogorov-Sinai entropy or dynamical entropy rate) \cite{eckmann_ergodic_1985,vulpiani_chaos:_2009,sinai_kolmogorov-sinai_2009,young_mathematical_2013,pikovsky_lyapunov_2016}.

Ruelle showed that the sum of the positive Lyapunov exponents gives an upper bound to the Kolmogorov-Sinai entropy \cite{ruelle_inequality_1978-1}:\[h_{\textnormal{KS}}\leqslant\sum_{\lambda_{i}>0}\lambda_{i}\]
Equality holds if and only if the system is endowed with an SRB (Sinai-Ruelle-Bowen) measure (Pesin entropy formula) \cite{ledrappier_metric_1985}. An $f$-invariant Borel probability measure $\mu$ is an SRB measure if the conditional probability of $\mu$ on smooth manifolds is absolutely continuous \cite{kuehn_multiple_2015}.$f$-invariant means here that $ \mu \left( f^{-1} (\mu) \right) = \mu (A)$. The Pesin entropy formula thus says that uncertainty in the prediction of future states comes from positive Lyapunov exponents, or more precisely from the expanding manifolds with smooth densities \cite{young_mathematical_2013}. In several classes of dynamical systems, the existence of an SRB measure was proved \cite{young_what_2002}. The angles between unstable and stable manifolds can be used to test numerically whether a system is hyperbolic. If a dynamical system is hyperbolic, there is always a finite angle between stable and unstable manifolds. In this case, the existence of an SRB measure is guaranteed \cite{sinai_gibbs_1972,bowen_ergodic_1975,ruelle_inequality_1978,eckmann_ergodic_1985}.

\paragraph{Attractor dimensionality}
The trajectory of a dissipative chaotic system with $N$ degrees of freedom does not cover the whole phase space. After a transient period, it relaxes onto an attractor, which has a dimensionality $D\leq N$. This can be a zero-dimensional fixed point, a one-dimensional periodic orbit, a higher-dimensional quasi-periodic orbit, or a strange attractor with typically non-integer dimensionality in case of a chaotic system. Such a strange attractor is often a fractal set and one classical approach to measuring its dimensionality is box counting. The idea is to count the number $M$ of $N$-dimensional boxes of side length $a$ that are necessary to cover the attractor. The box-counting dimension is then defined as $D=-\lim\limits _{a\to0}\frac{\log(M(a))}{\log(a)}$. For increasing dimension, one runs into the curse of dimensionality, because the data necessary for the box counting scales exponentially with the dimensionality.

 A more generalized concept of dimensionality of fractals is given by the R\'{e}nyi dimension (also called generalized dimension) \cite{grassberger_generalized_1983}. The R\'{e}nyi dimension of order $\alpha$ is given by \[D_\alpha = \lim_{\varepsilon \to 0} \frac{1}{\alpha-1}\frac{\log(\sum_{i} p_i^\alpha)}{\log\varepsilon}
\] 
For $\alpha = 0$ the capacity dimension (box-counting dimension) is obtained. $\alpha = 1$ gives the information dimension and $\alpha = 2$ the correlation dimension \cite{grassberger_characterization_1983}. 
Besides box-counting, there exist other sampling-based techniques to obtain entropies and dimensionalities directly from data, e.g., the Grassberger-Procaccia algorithm \cite{grassberger_estimation_1983,grassberger_characterization_1983}, which estimates the correlation dimension $D_{2}$. Similar to the case of box counting, a strict lower bound on the data required to estimate the attractor dimensionality with a fixed desired accuracy scales exponentially in the degrees of freedom $D$ \cite{eckmann_fundamental_1992,smith_intrinsic_1988}. It is well understood in nonlinear dynamics that such direct approaches of measuring dimensionality are inappropriate for high-dimensional dynamical systems \cite{ruelle_claude_1990}.

A more tractable way to quantify the attractor dimension and thus the number of degrees of freedom of a strange chaotic attractor can be obtained based on the Lyapunov spectrum if the equations of motion of the dynamical system are known and differentiable. The attractor dimension is then given by the interpolated number of Lyapunov exponents that sum to zero:
 \[D_{\textnormal{KY}}=k+\dfrac{\sum_{i=1}^{k}\lambda_{i}}{\left|\lambda_{k+1}\right| }\quad\text{with}\quad k=\max\limits _{n}\left\{ \sum\limits _{i=1}^{n}\lambda_{i}\ge0\right\}.\]
 The attractor dimension has been conjectured to be `in general' equivalent to the information dimension $D_1$ \cite{kaplan_preturbulence:_1979,frederickson_liapunov_1983,alexander_fat_1984,eckmann_ergodic_1985}. While there exists no proof in general, it has been proven for several low-dimensional systems \cite{ledrappier_relations_1981,young_dimension_1982} and for other systems supporting numerical evidence has been found \cite{russell_dimension_1980}. The following bound on the capacity dimension has been proven: $D_{0}\leq D_{KY}$ \cite{ledrappier_relations_1981,ruelle_chaotic_1989}. 
 
 Intuitively, the attractor dimension is the dimensionality of the highest dimensional infinitesimal hypersphere, whose volume does not shrink nor grow by the chaotic dynamics. In other words, on the attractor, growth along unstable manifolds is being compensated by shrinking along the stable manifolds and any $D$-dimensional hypersphere is merely deformed and the volume is preserved on average.

 Remembering the inequalities
 $h_{\textnormal{KS}}\leq h$ and $D_{0}\leq D_{\textnormal{KY}}$, we will call $h=\sum_{\lambda_{i}>0}\lambda_{i}$ the
 entropy rate and $D=D_{\textnormal{KY}}$ 
 the attractor dimension throughout this paper.
 
\section{Random Dynamical Systems and trial-to-trial variability\label{LSdriven}}
The extension of concepts of the ergodic theory of dynamical systems to input-driven systems was done in the theory of Random Dynamical Systems \cite{arnold_random_1995}. This can be useful for neuroscience to better understand trial-to-trial variability, controllability and input-driven chaos (see e.g., \cite{lin_reliability_2009,lajoie_chaos_2013}).
Consider a stochastic differential equation of the form:
\[
\dif x_{t}=a(x_{t})\dif t+\sum_{i=1}^{N}b_i(x_{t})\circ \dif W_{t}^{i}
\]
where $ \dif W_{t}^{i}$ are independent Brownian motions. An associated
\emph{stochastic flow map} is a solution for the dynamics, i.e. $F_{t_{1},\, t_{2};\zeta}(\mathbf{x}_{t_{1}})=\mathbf{x}_{t_{2}}$ maps the state $x$ from $t_1$ to $t_2$, where $\zeta$ denotes the realization of the stochasticity.
Instead of studying the temporal evolution of some initial measure
$\mu$, where each initial condition receives ``private'' noise,
as it is usually done in a Fokker-Planck approach, the theory of random
dynamical systems studies the evolution of a \emph{sample measure
}$\mu_{\zeta}^{t}$, defined as
\[
\mu_{\zeta}^{t}=\lim_{s\rightarrow\infty}(F_{-s,\,t;\zeta})_{*}\mu
\]
where the propagator $(F_{-s,\,t;\zeta})_{*}$ transports the initial
measure $\mu$ for some fixed white noise realization $\zeta(t)$
defined for all $t\in(-\infty,\infty)$ along the flow $F_{-s,\,t;\zeta}$.
In other words, the sample measure $\mu_{\zeta}^{t}$ is the conditional
measure at time $t$ given the infinite past history of
$\zeta(t)$. Note that in general, while $\mu_{\zeta}^{t}$ depends both
on time $t$ and the noise realization $\zeta$, it posses invariant
properties, characterizing its structure. For example, the Lyapunov
exponents $\lambda_{1}\geqslant\lambda_{2}\geqslant\ldots\geqslant\lambda_{N}$
are independent of the input realization $\zeta$ \cite{kifer_ergodic_2012}.

Two theorems for random dynamical systems link sample measure $\mu_{\zeta}^{t}$ and Lyapunov spectrum in chaotic and stable systems, respectively. First, Ledrappier and Young proved that if $\lambda_{1}>0$, then
$\mu_{\zeta}^{t}$ is a random SRB (Sinai-Ruelle-Bowen) measure \cite{ledrappier_entropy_1988}. As a consequence, in contrast to autonomous systems, for random dynamical systems, the Pesin identity $H=\sum\limits _{\lambda_{i}>0}\lambda_{i}$ is guaranteed to hold. Note that in contrast
to SRB measures of autonomous systems, random SRB measures are time-dependent.
However, they have a similar meaning: systems with SRB measure have
smooth conditional measures along the unstable manifolds.

In addition, Baxendale and Le Jan showed that if $\lambda_{1}<0$ and
the stationary measure is ergodic and some nondegeneracy conditions
on the measure are fulfilled \cite{baxendale_stability_1992}, then
$\mu_{\zeta}^{t}$ is a random sink, which means $\mu_{\zeta}^{t}(\mathbf{x})=\delta(\mathbf{x}-\mathbf{x}_{t})$,
where $\mathbf{x}_{t}$ is a solution of the stochastic dynamics for
a given noise realization $\zeta$ \cite{le_jan_equilibre_1987,baxendale_stability_1992}. This means that any trajectory of a stable rate network driven by white noise will after finite time be absorbed into one single trajectory, which is independent of the initial condition but depends only on the noise realization. Equally, any smooth initial measure will asymptotically coalesce into a time-dependent random sink. Note that the theorems by Baxendale and Le Jan do not say when the globally attracting random sink will be reached, which means that for very long transients, its asymptotic existence might have no practical relevance on biologically relevant timescales \cite{young_mathematical_2013}.

\section{Principal component-based dimensionality estimate\label{PCAdim}}
We compared the attractor dimension to a principal component-based dimensionality estimate. 
Principal component analysis (PCA) has been widely used as a dimensionality reduction technique both in experimental and theoretical neuroscience \cite{rajan_inferring_2010,rajan_stimulus-dependent_2010,gao_simplicity_2015,cunningham_dimensionality_2014}.

For a given data set, PCA provides the succeeding orthogonal directions that account for most of the variance in the data and the associated fraction of variance explained.
Mathematically, PCA is given by the eigenvalue decomposition of the covariance matrix.  
The number of principal components necessary to account for the majority of the total variance gives an estimate of the number of degrees of freedom of the underlying dynamics. If a few principal components explain most of the variance, the dynamics is mostly constrained to a hyperellipsoid with few long axes. If many principal components are necessary, no such localized structures in the second-order statistics of the collective dynamics are detected.
To avoid choosing an arbitrary threshold of variance (e.g., 95 \%), one can use a participation ratio, commonly used in physics to quantify, e.g., localization of collective activity modes \cite{bell_atomic_1970}, Anderson localization of waves in a disordered medium \cite{bauer_correlation_1990} or localized Lyapunov vectors \cite{ginelli_characterizing_2007,monteforte_dynamical_2010}.
We calculated PCA-based dimensionality estimates both based on the covariance of the total synaptic currents $h_i$ and of the rates $\phi_i=\tanh (h_i)$. For instance, for $h_i$, we compute the covariance matrix $C^h_{ij}$:
\begin{equation}
C^h_{ij}=\left\langle (h_{i}-\left\langle h_i\right\rangle )(h_{j}-\left\langle h_j\right\rangle )\right\rangle
\end{equation}
A PCA-based dimensionality estimate is then given by the participation ratio
\begin{equation}
D^h_{\textnormal{PCA}}=\frac{(\sum_{n=1}^N \mu_n^h)^2}{\sum_{n=1}^N \mu_n^{h2}} 
\end{equation}
where $ \mu_n^h$ is the $n$\textsuperscript{th} 
eigenvalue of the covariance matrix $C^h_{ij}$. If all eigenvalues contribute equally (i.e. $\frac{ \mu^h_n}{\sum_i \mu^h_n}=1/N$), the dimension estimate is $D^h_{\textnormal{PCA}} = N$. Conversely, if only one eigenvalue contributes then $D^h_{\textnormal{PCA}} = 1$ \cite{monteforte_dynamical_2010,gao_simplicity_2015,rajan_inferring_2010}.
$D_{\textnormal{PCA}}^{\tanh h}$ was calculated the same way, but for the covariance matrix $C^{\tanh h}$ of the firing rates.

\section{Random matrix theory of mean Lyapunov exponent}\label{RMTmeanLE} 
From the Jacobian, we derive a random matrix approximation of the mean Lyapunov exponent $\bar{\lambda}=\frac{1}{N}\sum_{i=1}^{N}\lambda_{i}$. The mean Lyapunov exponent describes the average dissipation rate of phase space compression, captured by the determinant of the long-term Jacobian $\mathbf{T}_t=\mathbf{D}_{t}\cdots\mathbf{D}_0$.
In the discrete-time case, the Jacobian is given by:
\begin{equation}
D_{ij}(t_s)\!=\!\frac{\partial f (h_i(t))}{\partial h_j(t)}\Bigr|_{t=t_s}\!=\!(1\!-\!\Delta t)\delta_{ij} \!+\!\Delta t\cdot J_{ij}\sech^2(h_j(t_s)).
\end{equation}
It is known that in the chaotic regime for large $N$, the activity variables $h_i$ approximately follow a Gaussian distribution both in discrete and continuous time, $h\sim \mathcal{N} (0,\Delta_0)$, where for large $N$, $\Delta_0$ solely depends on $g$ \cite{sompolinsky_chaos_1988,kadmon_transition_2015,molgedey_suppressing_1992, schuecker_optimal_2018}. The variance of $h_i$ grows with $g$, thus the squared hyperbolic secant of $h_i$ is close to zero for most $i$. For this reason, in the case of strong $g$, most columns of $D_{ij}(t_s)$ are, aside from the diagonal entries, close to zero and $D_{ij}$ becomes sparse. 

The long-term Jacobian \textbf{$\mathbf{T}_{t}(\mathbf{h}_{0})$} is
\begin{eqnarray*}
\textbf{$\mathbf{T}_{t}(\mathbf{h}_{0})$}	&=&\mathbf{D}_{t-1}(\mathbf{h}_{t-1})\dots\mathbf{D}_{1}(\mathbf{h}_{1})\mathbf{D}_{0}(\mathbf{h}_{0})\\ 
&=&	 \prod_{s=0}^{t-1} \mathbf{D}_{s}\\
&=& \prod_{s=0}^{t-1} \left((1-\Delta t)\mathds{1} + \Delta t\cdot \mathbf{J}\cdot \sech^2(\mathbf{h}(t_s)) \right) \\
\end{eqnarray*}
Thus, the mean Lyapunov exponent for large $N$ is

\begin{eqnarray*}
\bar{\lambda}&=&\left[\frac{1}{N}\sum_{i=1}^{N}\lambda_{i}\right]=\left[\frac{1}{N}\sum_{i=1}^{N}\ln{\mu_{i}}\right]=\left[\frac{1}{N}\ln{\prod_{i=1}^{N}\mu_{i}}\right]\\
&=&\left[\frac{1}{N}\ln{\big(\det\boldsymbol{\Lambda}\big)}\right]=\left[\frac{1}{N}\ln{\big(\det \lim_{ t\to\infty} [\mathbf{T}_{t}(\mathbf{x}_{0})^{\top}\mathbf{T}_{t}(\mathbf{x}_{0})]^{\frac{1}{2t}} \big)}\right]\\
&=&\left[\frac{1}{N\tau}\lim_{ t\to\infty}\frac{1}{t}\ln\big(\det\mathbf{T}_t\big)\right]\\
&=&\frac{1}{N\tau\Delta t}\left[\lim_{n\to\infty}\frac{1}{n}\sum_{s=0}^{n-1}\ln\left(\det \left( (1-\Delta t)\mathds{1} + \Delta t \mathbf{J} \phi'(\mathbf{h}(t_s))\right)\right)\right]\\
&=&\frac{1}{N\tau\Delta t}\left[\left<\ln(\det ( (1-\Delta t)\mathds{1})\det(\mathds{1} + \tfrac{\Delta t}{1-\Delta t} \mathbf{J} \phi'(\mathbf{h}(t_s))))\right>\right]\\
&=&\frac{1}{N\tau\Delta t}\left[\left<\ln(\det ( (1-\Delta t)\mathds{1})(\mathds{1} + \tfrac{\Delta t}{1-\Delta t} \tr( \mathbf{J} \phi'(\mathbf{h}(t_s)))))\right>\right] \\
 &+&\mathcal{O}\left((\Delta t)^2\right)\\
&=&\frac{1}{\tau\Delta t}\ln\left( 1-\Delta t \right)+\frac{1}{N\tau\Delta t}\left[\left<\ln(1 + \tfrac{\Delta t}{1-\Delta t} \tr( \mathbf{J} \phi'(\mathbf{y}))))\right>\right] \\
 &+&\mathcal{O}\left((\Delta t)^2\right)\\
&=&\frac{1}{\tau\Delta t}\ln\left( 1-\Delta t \right)+\mathcal{O}\left((\Delta t)^2\right)\\
\end{eqnarray*}

where $\mathbf{y}$ follows the distribution of Eq. \ref{eq:-off-diag},$\left<\dots \right>$ denotes the time average and $\left[\dots \right]$ denotes the ensemble average. For small $\Delta t$, we find excellent agreement with numerical simulations (See Fig.~\ref{fig6}B). In the limit $\Delta t\rightarrow0$, the mean Lyapunov exponent becomes $-\frac{1}{\tau}$.

\section{Dynamic mean-field theory}
We used dynamic mean-field theory to obtain the autocorrelations and the largest Lyapunov exponent. Briefly, following \cite{sompolinsky_chaos_1988,crisanti_path_2018-1,schuecker_optimal_2018}, we solved the autocorrelations $\Delta(\tau)$ self-consistently. We first obtained the variance $\Delta_0=\Delta(0)$, integrated them to obtain $\Delta(\tau)$. We calculated the largest Lyapunov exponent by calculating the ground-state  energy via $\lambda_{\max} = -1+\sqrt{1-\epsilon_0}$, where the $\epsilon_0$ is obtained from the smallest eigenvalue of the time-independent Schr\"odinger eigenvalue equation, where the quantum potential $W(\tau)=-V^{\prime\prime}(c(\tau))=1-g^{2}f_{\phi^{\prime}}(c(\tau),c_{0})$ is evaluated based on the self-consistent solution of the autocorrelation $\Delta(\tau)$ \cite{sompolinsky_chaos_1988,crisanti_path_2018-1,schuecker_optimal_2018}. We compared the solution of the dynamic mean-field theory with the previously proposed explicit expressions for autocorrelations and the largest Lyapunov exponent in the limits $g\rightarrow g_{\textnormal{crit}}^+$ and $g\rightarrow \infty$ (Fig.~\ref{fig16x}).

\begin{figure}[!h]
	\includegraphics[width=\columnwidth]{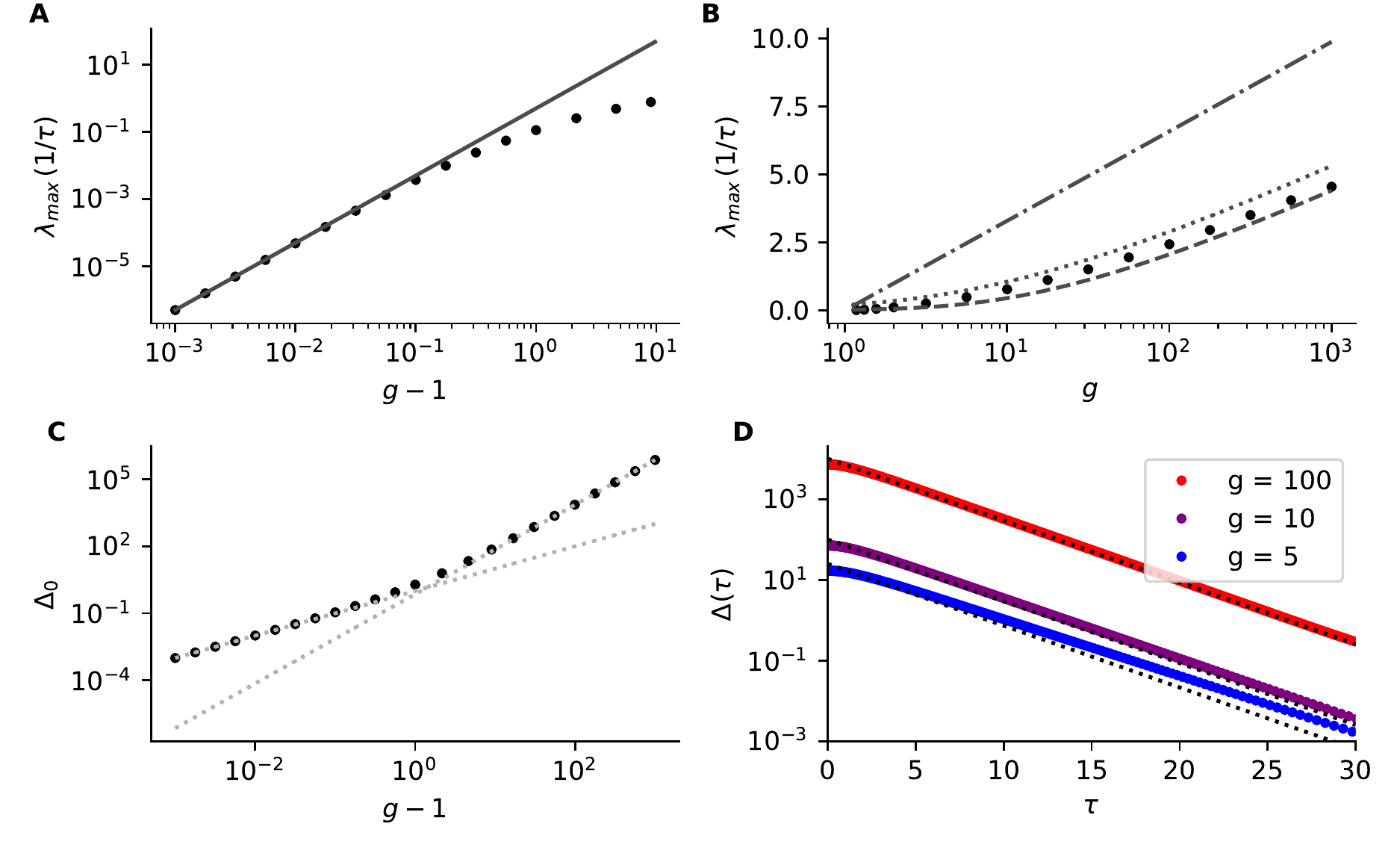}
	
	\caption{{\bf Dynamic mean-field theory of autocorrelations and the largest Lyapunov exponent in the limits $g\rightarrow g_{\textnormal{crit}}^+$ and $g\rightarrow \infty$}
		{\bf A}~Largest Lyapunov exponent for $g\rightarrow g_{\textnormal{crit}}^+$. Dots are the solution obtained from dynamic mean-field theory \cite{sompolinsky_chaos_1988,crisanti_path_2018-1,schuecker_optimal_2018}. Full line is $\lambda_(g)=\frac{1}{2}(g-1)^2$ \cite{sompolinsky_chaos_1988,crisanti_path_2018-1}. {\bf B}~Largest Lyapunov exponent for $g\rightarrow \infty$. Dots are solution obtained from dynamic mean-field theory, dash-dotted line is the explicit approximation $\lambda_{\max}(g)=C \log(g)$ with
		$C = \frac{2}{\pi}/\sqrt{\Delta_0(1-\Delta_0)}$ and $\Delta_0 = 2(1-2/\pi)$ \cite{crisanti_path_2018-1}, dotted line is $\lambda_{\max}(g)=-1 + \sqrt{1+C^2(W(\frac{g}{d}))^2}$, where $W$ is the Lambert $W$ function and $d = 6/((4-\pi)\sqrt{\pi-2})$ \cite{rozzi_exact_2006}.
		{\bf C}~$\Delta_0$ obtained from dynamic mean-field theory, dashed line are explicit limits for small $g$($\Delta_0=g-1$) and large $g$( $\Delta_0=2(1-2/\pi)g^2$) \cite{sompolinsky_chaos_1988,crisanti_path_2018-1}. {\bf D}~Autocorrelations $\Delta(\tau)$ for $g \in \{5,\; 10,\; 100\}$. For large $\tau$  and $g$ the autocorrelations decay exponentially with time constant $\sqrt{(\pi-3)/(\pi-2)})$ (dashed lines)	 (Other parameters: $\textnormal{for {\bf A-C} relative tolerance}=10^{-9}$, for {\bf D}~rel. tol. $10^{-11}$).}
	\label{fig16x}
\end{figure}

\section{Mathematical link of gradients in BPTT and Lyapunov spectrum\label{gradientsBPTT_LS_fullSpectrum}}
To train recurrent networks using backpropagation through time, the gradient of the loss $E$ with respect to the weights of the recurrent network has to be evaluated. This is done by unrolling the network dynamics in time \cite{pascanu_difficulty_2012}:

\begin{eqnarray}
\frac{\partial E_t}{\partial \mathbf{J}} 
&=& \frac{\partial E_t}{\partial \mathbf{o}_t}  \frac{\partial \mathbf{o}_t}{\partial \mathbf{h}_t} \sum_\tau \frac{\partial \mathbf{h}_t}{\partial \mathbf{h}_\tau}\frac{\partial \mathbf{h}_\tau}{\partial \mathbf{J}}\\
&=& \frac{\partial E_t}{\partial \mathbf{o}_t}  \frac{\partial \mathbf{o}_t}{\partial \mathbf{h}_t} \sum_\tau \left(\prod_{\tau'=\tau}^{t-1}\frac{\partial \mathbf{h}_{\tau'+1}}{\partial \mathbf{h}_{\tau'}}\right)\frac{\partial \mathbf{h}_\tau}{\partial \mathbf{J}} \\
&=& \frac{\partial E_t}{\partial \mathbf{o}_t}  \frac{\partial \mathbf{o}_t}{\partial \mathbf{h}_t} \sum_\tau \left(\prod_{\tau'=\tau}^{t-1}\mathbf{J}\phi'(\mathbf{h}_{\tau'})\right)\frac{\partial \mathbf{h}_\tau}{\partial \mathbf{J}} \\
&=& \frac{\partial E_t}{\partial \mathbf{o}_t}  \frac{\partial \mathbf{o}_t}{\partial \mathbf{h}_t} \sum_\tau \left(\prod_{\tau'=\tau}^{t-1}\mathbf{D}_{\tau'}\right)\frac{\partial \mathbf{h}_\tau}{\partial \mathbf{J}} \\
&=& \frac{\partial E_t}{\partial \mathbf{o}_t}  \frac{\partial \mathbf{o}_t}{\partial \mathbf{h}_t} \sum_\tau \mathbf{T}_{t}(\mathbf{h}_\tau)\frac{\partial \mathbf{h}_\tau}{\partial \mathbf{J}} \\
\end{eqnarray} 
where  $\mathbf{D}_{\tau'}$ is the Jacobian Eq.~\ref{eq:-Jacobian} that we already considered when calculating the Lyapunov spectrum. The recursive dependence of the gradient on the previous network state results in a product of Jacobians, which takes the form of the long-term Jacobian $\mathbf{T}_{t}(\mathbf{h})$ (Eq.~\ref{eq:long-term-Jacobian}) whose inner product gives the Oseledets matrix (Eq.~\ref{eq:-Oseledets}). 

The singular values of the long-term Jacobian $\mathbf{T}_{t}(\mathbf{h}_\tau)$, which determine how quickly gradients vanish or explode during backpropagation through time, are directly related to the Lyapunov exponents of the forward dynamics: 
The Lyapunov exponents of the forward dynamics are given by the logarithm of the singular values of the long-term Jacobian \cite{geist_comparison_1990}. Thus, our results on how the global coupling strength $g$,  simulation parameters (e.g., time-discretization $\Delta t$), time-dependent input, and nonlinearity $\phi$ (e.g., threshold-linear vs. $\tanh$) shape the Lyapunov spectrum can directly be translated into predictions on the gradient instability during backpropagation through time.
As pointed out previously \cite{schoenholz_deep_2016,chen_dynamical_2018,gilboa_dynamical_2019}, the trainability of recurrent networks is constrained by the condition number $\kappa$ of the long-term Jacobian $\mathbf{T}_{t}(\mathbf{h}_\tau)$. The condition number can be approximated by the Lyapunov spectrum: 
$\kappa_{2}(\mathbf{T}_{t}(\mathbf{h}_\tau))=\frac{\sigma_1({\mathbf{T}_{t}(\mathbf{h}_\tau)})}{\sigma_N({\mathbf{T}_{t}(\mathbf{h}_\tau)})}\approx(t-\tau)\exp\left(\lambda_{1}-\lambda_{N}\right)$, where $t-\tau$ is the time to be bridged by backpropagation through time.

\section{Lyapunov spectra for huge coupling $g$}
spectrum\label{largeGLimit}}
\begin{figure}[!h]
	\includegraphics[width=\columnwidth]{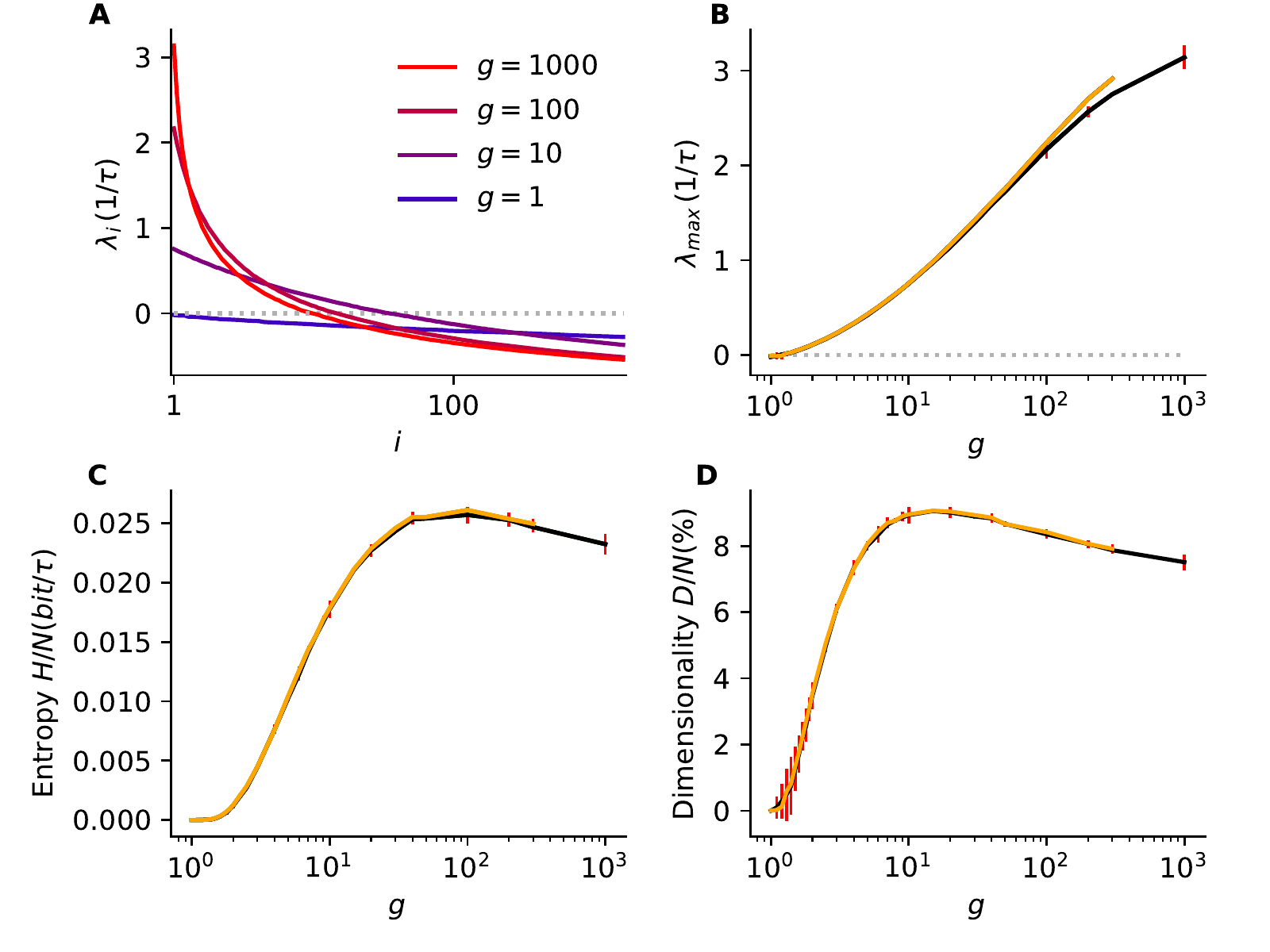}
	\caption{{\bf Peak in dynamical entropy rate and attractor dimensionality for large $g$}
		{\bf A}~For increasing coupling strength $g$, the Lyapunov spectrum is increasingly bent upward with a decreasing fraction of positive Lyapunov exponents ($N=2000$). {\bf B}~The largest Lyapunov exponent grows monotonically for increasing values of $g$ as predicted analytically (Fig.~\ref{fig16x}B). For very large $g$, the largest Lyapunov exponents flattens (black lines, $N=2000$, orange lines, $N=3000$). Increasing $N$ reduces the flattening, indicating a finite $N$ effect on $\lambda_{\max}$. 
		{\bf C}~The dynamical entropy rate $H$ peaks as a function of $g$. {\bf D}~The relative attractor dimensionality $D/N$ also peaks as function of $g$. Both positions of the peak in $D$ and $H$ do not shift with $N$ indicating that the peak is not a finite $N$ effect. 
		(Parameters: $\textnormal{relative tolerance}=10^{-10}$, $t_\textnormal{ONS}=\tau$, $t_\textnormal{sim}=10^3\tau$, averages across $3$ network realizations).}
	\label{fig16LAST}
\end{figure}
For very large values of $g$, we observe that both attractor dimensionality $D$ and dynamical entropy rate $H$ peak as a function of $g$ (Fig.~\ref{fig16LAST}). For increasing network size $N$, this peak did not vanish or shift, indicating that it is not merely an finite-size effect. The peak in both $H$ and $D$ can be explained by a growing fraction of rate units in saturation and consequently, increasingly sparse Jacobian $D_{ij}(t_s)$, thus fewer active units at each moment and therefor fewer unstable space directions as indicated by the decreasing number of positive Lyapunov exponents. Our seemingly contradictory claim of decreasing dynamical entropy rate $H$ despite growing largest Lyapunov exponent for large $g$ are consistent for large $N$ and finite $g$, where relative contribution of the largest Lyapunov exponent to the sum of the positive Lyapunov exponents vanishes. We note that the analytical argument for point-symmetry in the Lyapunov spectrum around $i=N/2$ and $\lambda_i=-\frac{1}{\tau}$ together with the fact that $D$ is always bounded by $N$ already imply that $H$ has to be bounded: $H<D<N$, as \[H\leqslant\sum_{i=1}^p\lambda_{i}=-\sum_{i=p+1}^D\lambda_{i}\leqslant -\sum_{i=p+1}^D -\frac{1}{\tau}\leqslant D-p-1 \leqslant D\leqslant N,\] where $p$ is the number of positive Lyapunov exponents. For notational simplicity, we assumed here an integer dimensionality $D$, but the argument holds generally.

\section{Supporting Information}

\paragraph{S1 Code}
\label{S1_Code}
{\bf Source code for Lyapunov spectrum of rate networks. } We provide all necessary code to calculate the full Lyapunov spectrum written in Julia \cite{bezanson_julia:_2017}. The efficient implementation is parallelized using level-3 matrix-matrix operations from BLAS (Basic Linear Algebra Subprograms) called via LAPACK (Linear Algebra PACKage). The code also provides an alternative estimate of the largest Lyapunov exponents by tracking the evolution of a small but finite initial perturbation and resizing it iteratively \cite{eckmann_ergodic_1985}. 
Furthermore, the program provides bootstrapped 95 percentile confidence intervals for the first and the last Lyapunov exponent, the Kolmogorov-Sinai entropy rate, and the attractor dimensionality. Optionally, a principal component-based dimensionality estimate can also be calculated. Finally, the program provides the convergence of the Lyapunov spectrum in time.
Input variables are network size $N$, coupling strength $g$, time-discretization $\Delta t$, simulation time $t_\textnormal{sim}$, number of Lyapunov exponents to be calculated, $nLE$, orthonormalization time interval $t_\textnormal{ONS}$, seed for initial conditions $seed_{\textnormal{IC}}$, seed for random network realization $seed_{\textnormal{net}}$, seed for orthonormal system $seed_\textnormal{ONS}$ and finally the subdirectory where the results are stored.
Code written in MATLAB\textsuperscript{\textregistered}/Octave/Python is available on github.

\paragraph{S2 Code}
\label{S2_Code}
{\bf Source code for Lyapunov spectrum of input-driven rate networks.}
We also provide Julia code to obtain the full Lyapunov spectrum of a noise-driven rate network by a reorthonormalization procedure \cite{benettin_lyapunov_1980}. This is done along a numerical solution of the stochastic differential equation obtained with the Euler-Maruyama method \cite{kloeden_numerical_1992}. The noise strength $\sigma$ is now an additional input parameter. Code written in MATLAB\textsuperscript{\textregistered}/Octave/Python is available on github.
\label{S3_Code}
\bibliographystyle{apsrev}

\end{document}